\documentclass[aps,pr,twocolumn]{revtex4-2}
\usepackage{mathtools}
\usepackage{graphicx}
\usepackage{mathrsfs}
\usepackage{amssymb}
\usepackage{xcolor}
\usepackage{amsmath}
\usepackage{bm}
\usepackage{amsmath}
\usepackage{physics}
\makeatletter\AtBeginDocument{\let\@elt\relax}\makeatother
\usepackage{amsmath}
\usepackage[normalem]{ulem}
\usepackage[unicode=true,colorlinks=true,citecolor=blue,urlcolor=blue]{hyperref}

\def\be {\begin{equation}}
\def\ee {\end{equation}}
\def\bea {\begin{align}}
\def\eea {\end{align}}

\def\bee{\begin{eqnarray}}
\def\eee{\end{eqnarray}}
\def\BC {\begin{cases}}
\def\EC {\end{cases}}

\makeatother

\begin{document}
\title{Photocurrents in bulk tellurium}

\author{M. D. Moldavskaya$^1$, L. E. Golub$^1$, S. N. Danilov$^1$, V. V. Bel'kov$^2$, D.~Weiss$^1$ and S.~D.~Ganichev$^{1}$
}

\affiliation{$^1$Terahertz Center, University of Regensburg, 93040 Regensburg, Germany}

\affiliation{$^2$Ioffe Institute, 194021 St. Petersburg, Russia}

\begin{abstract}	
We report a comprehensive study of polarized infrared/terahertz photocurrents in bulk tellurium crystals. We observe different photocurrent contributions and show that, depending on the experimental conditions, they are caused by the trigonal photogalvanic effect, the transverse linear photon drag effect, and the magnetic field induced linear and circular photogalvanic effects. All observed photocurrents have not been reported before and are well explained by the developed phenomenological and microscopic theory. We show that the effects can be unambiguously distinguished by studying the polarization, magnetic field, and radiation frequency dependence of the photocurrent. At frequencies around 30 THz, the photocurrents are shown to be caused by the direct optical transitions between subbands in the valence band. At lower frequencies of 1 to 3 THz, used in our experiment, these transitions become impossible and the detected photocurrents are caused by the indirect optical transitions (Drude-like radiation absorption). 
\end{abstract}

\maketitle
\section{Introduction}
\label{intro}

Tellurium is an elementary semiconductor that has been studied from the very beginning of the history of semiconductor physics. In the 60's-70's such phenomena as quantum effects in cyclotron resonance~\cite{Button1969}, {Shubnikov-de~Haas effect~\cite{Bresler1970},} Nerst-Ettingshausen and Seebeck effects~\cite{Saffert1974}, surface quantum states~\cite{Klitzing1971b,Englert1977}, and natural optical activity~\cite{Nomura1960,Ivchenko1974,Fukuda1975,Stolze1977} were detected in Te crystals. In addition, several novel photoelectric phenomena, including the circular photogalvanic effect~\cite{Ivchenko78p640,Asnin1978}, the circular photon drag effect~\cite{Shalygin2016}, and the electric current-induced optical activity~\cite{Ivchenko78p640,Vorobjev79p441}, were discovered for the first time in bulk Te, for a recent review see~\cite{Shalygin2023}. These effects arise from spin splitting of the valence band at the boundary of the first Brillouin zone (''camel back'' structure).

In recent years, studies of Te have experienced a renaissance due to the possibility of fabricating 2D Te crystals (tellurene) which exhibit unique material properties, for reviews see e.g. Refs.~\cite{Wang2018,Wu2018b,Shi2020,Zhang2020t,Xu2020,Yan2022}, and theoretical proposals for closing the energy gap in Te and the appearance of Weyl points near the Fermi level with Fermi arcs at the surface by applying proper strain, see e.g. Refs.~\cite{Agapito2013,Hirayama2015,Murakami2017,Ideue2019,Zhang2020,Oliveira2021,Glazov2022}.

Experimental access to the properties of tellurene as well as to the specific properties of Weyl fermions should allow studies of photoelectric effects excited by infrared/THz radiation. The power of the method has already been demonstrated for other 2D materials, surface states of topological insulators and Weyl semimetals, see e.g. Refs.~\cite{Glazov2014,Otteneder2020,35,Plank2018,Ishizuka2016,Golub2017,Juan2017,Chan2017,Ma2017,Ji2019}.
Recently, a transverse CPGE has been observed in bulk unstrained Te at oblique incidence and has been attributed to Weyl fermions~\cite{Ma2022}. However, no features specific to Weyl fermions have been detected and these results can be explained alternatively by
considering optical transitions in the conventional Te band structure
without involving Weyl bands, which are not expected without a significant strain~\cite{Agapito2013,Hirayama2015,Murakami2017,Ideue2019,Zhang2020,Oliveira2021,Glazov2022}. 
In view of the increasing interest in photoelectric effects excited in 2D and on the surface of 3D Te crystals, it becomes important to understand the photocurrents excited in bulk Te that are not related to 2D states or the topological charges of the Weyl points.

In the present work, we report the observation of three photoelectric phenomena in Te crystals, which have not been previously addressed either experimentally or theoretically. These phenomena are: (i) trigonal linear photogalvanic effect (LPGE) due to intersubband optical transitions in the valence band,
(ii) transverse linear photon drag effect,
and (iii) circular (radiation helicity driven) magnetophotocurrents due to intersubband optical transitions in the valence band.
While the effect (ii) is detected for Drude absorption (THz frequencies) only, the effects (i) and (iii) are observed at both infrared frequencies and Drude absorption.
The observed phenomena are characterized by different dependences on radiation frequency and polarization. Furthermore, the linear and circular magnetophotocurrents depend linearly on an external magnetic field $B$ and vanish for $B=0$. The qualitatively different functional behavior allows us to clearly distinguish and study all these individual effects. The results are well described by the developed phenomenological and microscopic theories. It is shown that all phenomena are excited in the bulk of the material and are caused by the displacement of electrons in real space due to direct intersubband optical transitions (trigonal LPGE at IR frequencies), asymmetric scattering of carriers at Drude absorption (trigonal LPGE at THz frequencies), transfer of linear photon momentum to free carriers (photon drag effect at THz frequencies), and magnetic field assisted asymmetric scattering (magnetic field induced LPGE and CPGE). 

The paper is structured as follows.  In Sec.~\ref{samples-methods} we describe the investigated samples and the experimental technique. In Sec.~\ref{Experiment} we discuss the experimental results. In the following Secs.~\ref{Phenomenological_theory} we perform a symmetry analysis of the photocurrent excited by radiation propagating along the $c$-axis and identify different mechanisms of the observed photocurrents excited by linearly (Sec. \ref{linear_polarization}) and elliptically (Sec.~\ref{elliptical_polarization}) polarized radiation. In the following section we discuss possible optical transitions in the studied experimental arrangement 
and some details of the band structure. 
Next, we present the developed theory and corresponding model pictures for the photogalvanic effects excited at the \textit{inter} subband transitions (Secs.~\ref{trigonal_intersubband} and~\ref{MPGE_intersubband}) and the \textit{intra} subband Drude-like transitions (Secs.\ref{trigonal_intraband} and~\ref{MPGE_intraband}). After discussing the mechanisms of photogalvanic currents, we consider the linear 
photon drag effect (PDE) and the linear magnetic field induced PDE excited at intraband (Drude like) transitions 
(Sec.~\ref{Drag_intraband}). In Sec.~\ref{discussion} we compare the experimental and theoretical results. Finally, in Sec.~\ref{summary} we summarize the results.   
We have also included four appendices. In Appendix~~\ref{appendix2} we cover the complete phenomenology of PGE and PDE currents at normal incidence. In Appendix~\ref{Abs_intersubband} we present the equations that describe
the absorption coefficient at direct intersubband
transitions, and Appendices~\ref{PDE_micro} and~\ref{App_MPDE} contain the  microscopic derivation of the
PDE and MPDE currents.

\begin{figure}
	\centering \includegraphics[width=\linewidth]{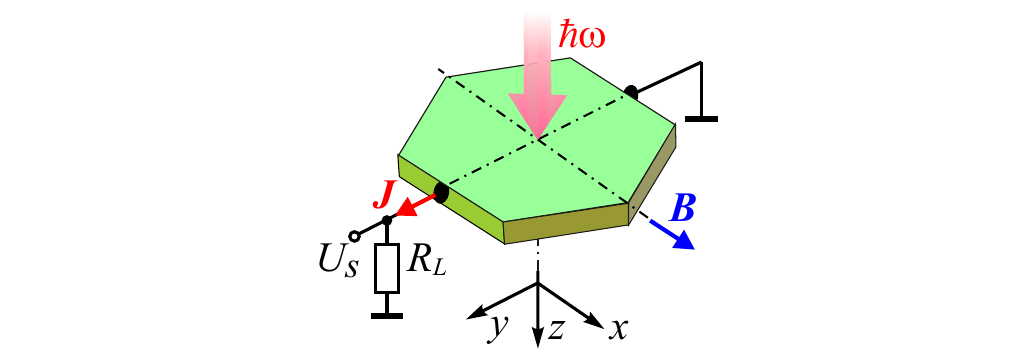}
	\caption{{Experimental setup. {Here $z$ is the $c$-axis of Te crystal, and $x$ is parallel to one of three $C_2$ rotation axes.}} }
	\label{Fig1setup}
\end{figure}

\section{Samples and experimental setup}
\label{samples-methods}

The measurements were carried out on an $p$-type tellurium single crystal grown by the Czochralski method in a hydrogen atmosphere. The inset in Fig.~\ref{Fig1setup} shows the sample and the experimental setup. 
A plate with thickness 0.8~mm was cut perpendicular to the  to $c$-axis. A pair of Ohmic contacts was fabricated on opposite sides of the hexagon shaped plate~\footnote{Note that due to growth conditions the hexagon was slightly distorted.}. This allowed us to measure the photocurrents along the $y$-direction.  
Note that we use the coordinate system $(x,y)$ where $x$ is parallel to one of three $C_2$ rotation axes.  
The contacts were made of an alloy of tin, bismuth, and antimony with a low melting temperature (Sn\,:\,Bi\,:\,Sb\,=\,50\,:\,47\,:\,3)~\cite{Shalygin2016}.  {The magnetotransport measurements were performed in the van der Pauw geometry on a sample cut from the same tellurium crystal as the sample used for the photocurrent measurements. The room temperature carrier density was $p=7\times10^{16}$ cm$^{-3}$ and the hole mobility $\mu = 700~$ cm$^2$/(V$\cdot$s). For the effective mass $m \approx 0.2~m_0$ (see Refs.~\cite{Bauer1973,Ribakovs1977}) this results in a momentum relaxation time $\tau \approx 8\times10^{-14}$~s.
Note that the same parameters were obtained for similar Te crystals used in Ref.~\cite{Shalygin2016}.
}


\begin{figure}
	\centering \includegraphics[width=\linewidth]{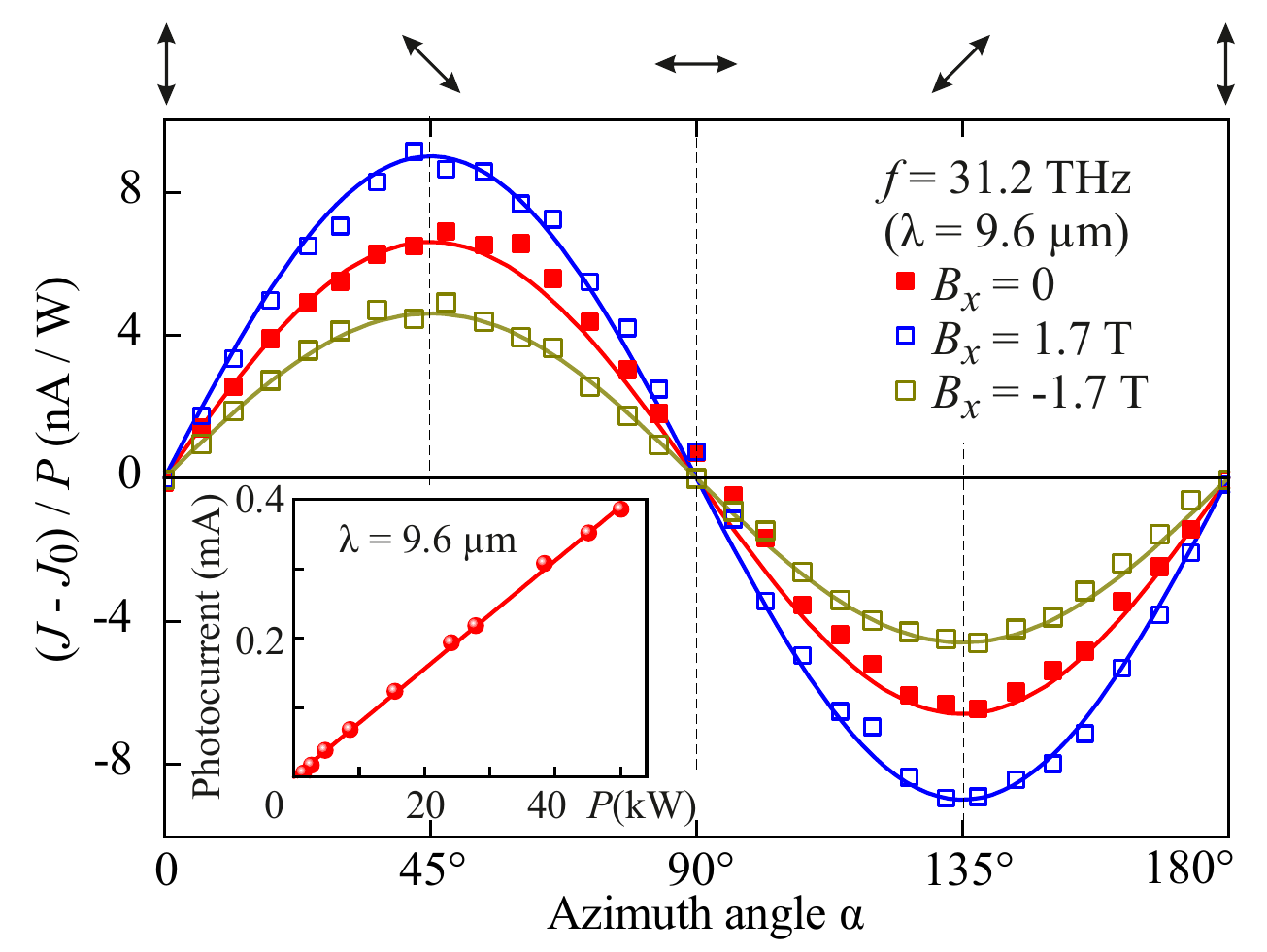}
	\caption{
	Azimuth angle dependence of the normalized photocurrent $J_y/P$ measured at zero magnetic field (red squares) and $B_x=\pm 1.7$~T (blue and dark yellow squares). {Note that the small polarization independent offset $J_0 \ll J_y$ is subtracted for clarity.}  The data are obtained at $f = 31.2$~THz ($\lambda  = 9.6 \,\,\mu$m). Solid lines are fits to	$(J_y-J_0)/P \propto \sin 2\alpha$. The double arrows at the top illustrate the state of the polarization for different values of the azimuth angle $\alpha$.  { The inset shows the intensity dependence of the photocurrent. Solid line is the linear fit.}  }
	\label{Fig1}
\end{figure}


To study the photocurrent in a wide frequency range we used two pulsed laser systems: a TEA CO$_2$ laser and an optically pumped molecular terahertz laser~\cite{Ganichev2006}. The lasers operated at single frequencies in the range from $f\approx 1$ to 30~THz (corresponding photon energy range from $\hbar \omega = 4.4$ to 132\,meV, where $\omega = 2\pi f$ is the angular frequency). Radiation with frequencies of about 30~THz was obtained by a line-tunable TEA CO$_2$ laser~\cite{Ganichev2003a,Ganichev2003}.  We used four frequencies of the laser radiation from 28.3~THz (wavelength $\lambda = 10.6~\mu$m, $\hbar\omega =117$~meV) to 32.2~THz ($\lambda = 9.3~\mu$m, $\hbar\omega =133.3$~meV)~\footnote{The number of studied frequencies was defined by the laser tunability and availability of the polarizers operating at specific frequencies.}.
 The laser generated single pulses with a duration of about 100\,ns and a repetition rate of 1~Hz. The radiation power on the sample surface {$P$} was about 50~kW. For the low frequencies (from 1 to 3.3~THz) we used a line-tunable  pulsed molecular laser with NH$_3$ as active media~\cite{Ganichev1993,Ganichev1995,Ganichev1998}. The laser operated at $f = 1.07$~THz ($\lambda = 280~\mu$m, $\hbar\omega =4.4$~meV) and 3.3~THz ($\lambda = 90.5~\mu$m, $\hbar\omega =13.7$~meV). The operation mode of the NH$_3$ laser was similar to that of the TEA CO$_2$ laser. The radiation power on the sample surface was about 5~kW. The peak power of the radiation was monitored by infrared and terahertz photon-drag  detectors~\cite{Ganichev2006,Ganichev1985}, as well as by a pyroelectric   power meters. The beam positions and profiles were checked with pyroelectric camera or thermally sensitive paper. The radiation was focused onto spot sizes of about 0.5 to 3~mm diameter, depending on the radiation frequency.

Photocurrents were  measured at room temperature by applying polarized radiation at normal incidence, see the inset in Fig.~\ref{Fig1}. In experiments with linearly polarized radiation the in-plane radiation electric field vector $\bm E$  was rotated counter-clockwise with respect to the $y$-axis. The orientation of the vector $\bm E$ is defined by the azimuth angle $\alpha$ ($\bm E \parallel y$  corresponds to $\alpha = 0$) and was varied by rotation of a $\lambda/2$-plate. To study photocurrents sensitive to the radiation helicity we used $\lambda/4$ plates. By rotating the $\lambda/4$ plate we varied the THz radiation helicity according to $P_{\rm circ} \propto \sin{2 \varphi}$~\cite{Belkov2005}, where $\varphi$ is the angle between the laser polarization plane and the optical axis of the plate. Note that for $\varphi=0$ the radiation is linearly polarized along the $y$-direction.

The induced photocurrents were detected as a voltage drop  across load resistors $R_L=50$~Ohm.  The signals were recorded  using digital oscilloscopes. In experiments on magneto-photocurrents an external in-plane magnetic field $B$ up to 1.7~T was applied along $x$-direction using an electromagnet.

\section{Experiment}
\label{Experiment}

We begin by describing the experimental results obtained under various experimental conditions. The phenomenological theory and identification of the photocurrent mechanisms are given in the next section.

\begin{figure}
	\centering \includegraphics[width=\linewidth]{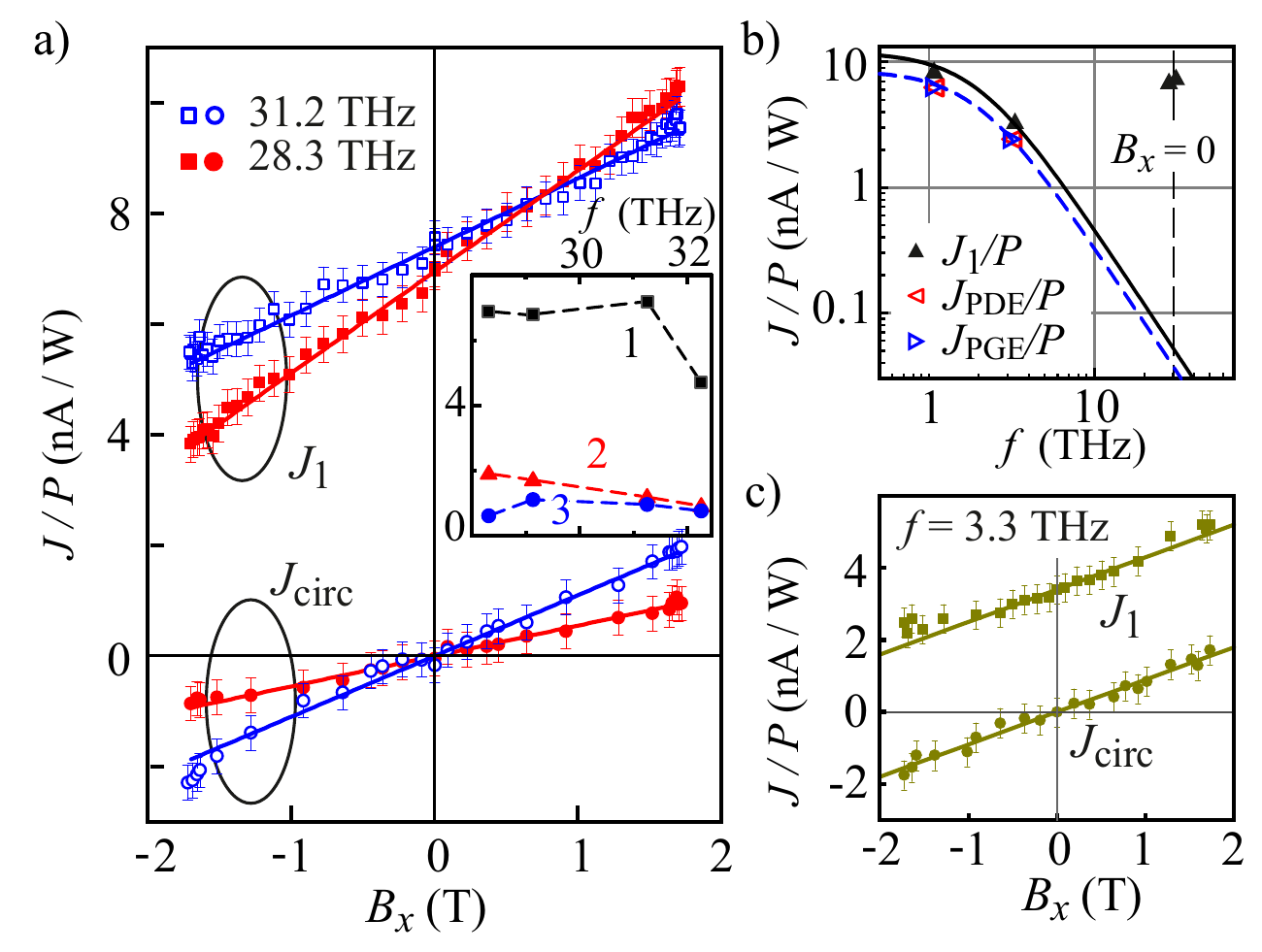}
	\caption{(a) Magnetic field dependencies of the amplitudes of linear ($J_1$, squares) and circular ($J_{\rm circ}$, circles)  photocurrents normalized on the radiation power $P$. The red and blue symbols correspond to data obtained at $f = 28.3$~THz ($\lambda = 10.6 \,\,\mu$m) and $f = 31.2$~THz ($\lambda = 9.6 \,\,\mu$m), respectively. 
Inset: 1 -- $A^{\rm LPGE}$ in nA/W, 2 -- $D^{\rm MLPGE}$ in nA/(WT), 3 -- $D^{\rm MCPGE}$ in nA/(WT). 
	(b) Frequency dependencies of  photocurrents $J_1/P$ and its components due to the photon drag $J_{\rm PDE}$ and photogalvanic $J_{\rm PGE}$ effects measured at zero magnetic field in the THz frequency range. Lines correspond to the spectral dependence of the Drude absorption. (c) Magnetic field dependencies of the amplitudes of linear (squares) and circular (circles)  photocurrents $J/P$ measured at
$f = 3.3$~THz ($\lambda = 90 \,\,\mu$m).
}
	\label{Fig3}
\end{figure}

First we present the photocurrent excited at frequencies about $30$~THz.
 Figure~\ref{Fig1} shows the dependence of the normalized photocurrent $J_y/P$ excited by linearly polarized radiation 
 as a function of the orientation of the electric field vector.
 The data obtained at $f=31.2$~THz are shown for zero magnetic field and for magnetic fields $B_x=\pm 1.7$~T. 
 All three traces can be well fitted by
\begin{equation}
\label{jy}
 J_y = J_1(B_x) \sin2\alpha + J_0\,\,\,\,,
\end{equation}
where coefficients $J_1(B_x)$ and $J_0$ are fit parameters~\footnote{{While in magnetic fields up to about $\pm~1.2$~T the current linearly depends on the magnetic field at higher fields, it possibly tends to deviate from this behavior. To justify this tendency experiments at higher magnetic fields are needed, which was out of scope of the present work.}}. Note that the polarization-independent offset $J_0$ is more than an order of magnitude smaller than the parameter $J_1(B_x)$ and will not be discussed below. 
The magnetic field  dependence of the coefficient $J_1(B_x)$ 
{measured for $f=30.2$ and 28.3~THz are shown in Fig.~\ref{Fig3}(a). 
It demonstrates that the photocurrent depends linearly on $B$ and has a substantial magnitude at zero magnetic field. 
Solid lines in Fig.~\ref{Fig3}(a) show that the coefficient $J_1(B)$ within the error bars are well described by the function
\begin{equation}
\label{j1coeff}
 J_1 = A^{\rm LPGE} P+ D^{\rm MLPGE} P B_x \,\,\,\,,
\end{equation}
where $A^{\rm LPGE}$ and $D^{\rm MLPGE}$ are fitting parameters, which, as we show below, describe the linear photogalvanic and linear magneto-photogalvanic effects, respectively.
The magnitudes of these coefficients measured for four laser frequencies in the range between 28.3 and 32.2~THz are shown in the inset in Fig.~\ref{Fig3}(a).
}
The slope of the straight line in the inset in Fig.~\ref{Fig3}(a) depends on the radiation frequency. 
Measuring the photocurrent as a function of the radiation power $P$, we found that it scales linearly with $P$, i.e. depends quadratically on the radiation electric field $E$, {see the inset in Fig.~\ref{Fig1}.}

\begin{figure}
	\centering \includegraphics[width=\linewidth]{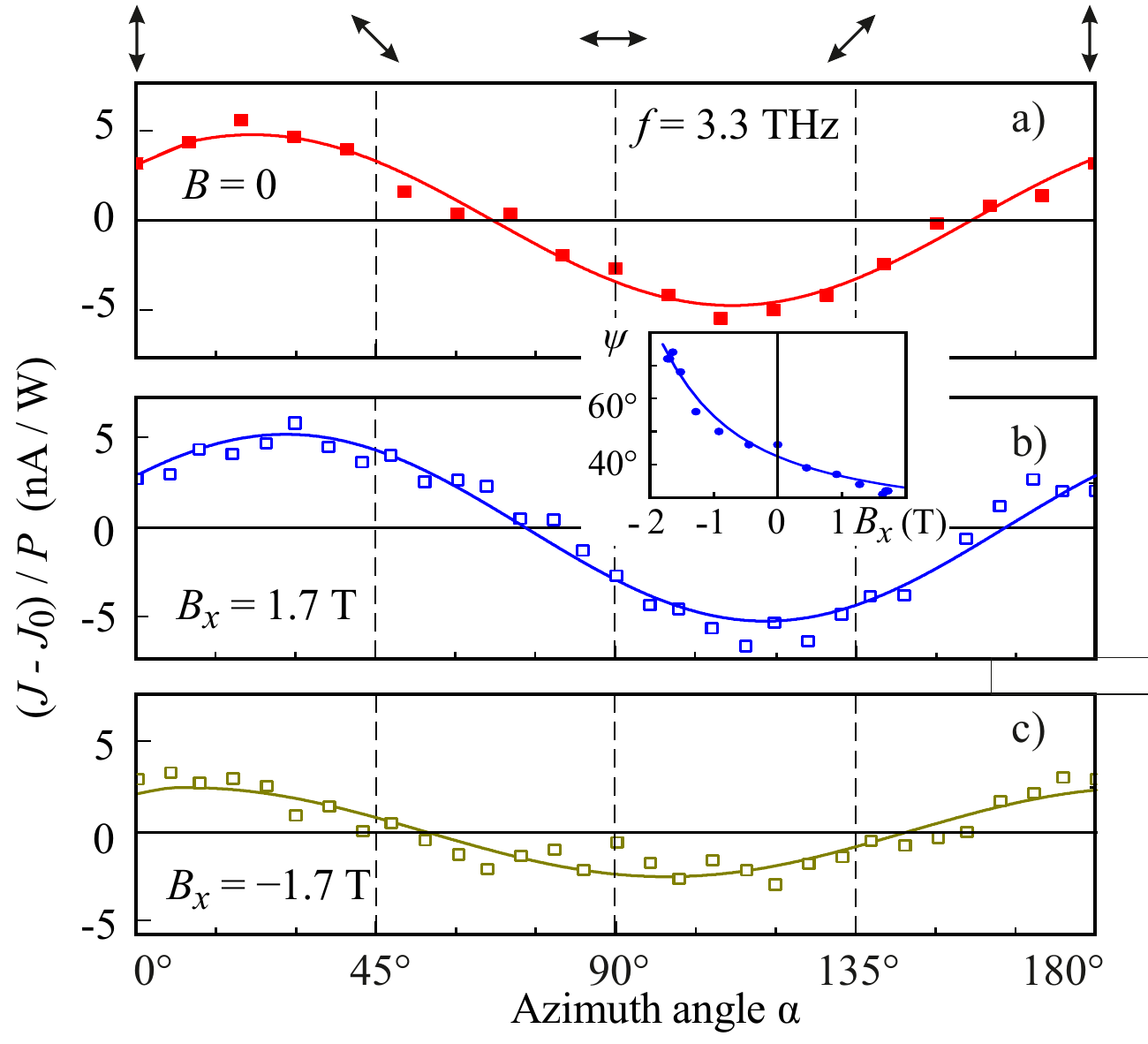}
	\caption{Azimuth angle dependence of the normalized photocurrent $(J-J_0)/P$ measured  at zero magnetic field (red squares) and $B_x=\pm 1.7$~T (blue and olive squares). The data are obtained applying radiation with $f = 3.3$~THz ($\lambda = 90\,\,\mu$m). Solid lines are fits after
$(J-J_0)/P\propto \sin[2\alpha-\psi(B_x)]$. The double arrows at the top illustrate the state of the polarization for different values of the azimuth angle $\alpha$. Inset shows magnetic field dependence of phase $\psi$ (circles). Solid line is fit after Eq.~(\ref{phase}). }
	\label{Fig4}
\end{figure}

For radiation at lower frequencies ($f= 1.07$ and 3.3~THz) the photocurrent still varies sinusoidally with the double angle $\alpha$ but the experimental traces become phase shifted, see Fig.~\ref{Fig4}.
Now the data can be well fitted by
%
\begin{equation}
\label{jyTHz}
J_y = J_1(B_x) \sin[2\alpha - \psi(B_x)] + J_0\,.
\end{equation}
The magnetic field dependencies of the amplitude $J_1(B_x)$ and the phase $\psi(B_x)$ are shown in Fig.~\ref{Fig3}(c) and the inset in Fig.~\ref{Fig4}, respectively.
{Figure~\ref{Fig3}(c)  reveals that, alike at high frequencies, the coefficient  $J_1(B_x)$ depends linearly on magnetic field $B_x$ and has a substantial amplitude at zero magnetic field.}

Figure~\ref{Fig3}(b) shows the frequency dependence of the photocurrent magnitude $J_1(0)$ at zero magnetic field.  For the low frequencies used in the experiments,  $f= 1.07$ and 3.3~THz, which correspond to the photon energies of 4.4 and 13.7~meV, optical transitions in  Te can only be due to Drude absorption. It varies with the radiation frequency as $K_D \propto 1/(1 +\omega^2 \tau^2)$, where $K_D$ is the absorption coefficient. The solid line in Fig.~\ref{Fig3}(b) shows the frequency dependence of $J_1 \propto K_D$ calculated for a momentum relaxation time $\tau = 8\times10^{-14}$~s determined from magneto-transport measurements. It shows that Drude absorption describes the data well at low frequencies. At high frequencies, however, the signal magnitudes are by more than two orders of magnitude larger than would be expected from the calculated dependence. This observation shows that at frequencies of about 30~THz, i.e. photon energies of about 125~meV, the photocurrent is not related to the indirect optical transitions in the valence band, but stems from direct optical transitions between subbands of the valence band. Below we discuss this in more detail. 
{Comparing the slopes of the magnetic field dependencies of $J_1(B_x)$ measured at high frequencies [28.3 and 31.2~THz, see Fig.~\ref{Fig3}(a)] with that measured at low frequencies [3.3~THz, see Fig.~\ref{Fig3}(c)], we obtain that they have comparable magnitudes. 
Using the same arguments as above for the zero magnetic field photocurrent we conclude that also the magneto-photocurrent at high frequencies $\sim 30$~THz stems from direct intersubband transitions and not the Drude absorption.}

\begin{figure}
	\centering \includegraphics[width=\linewidth]{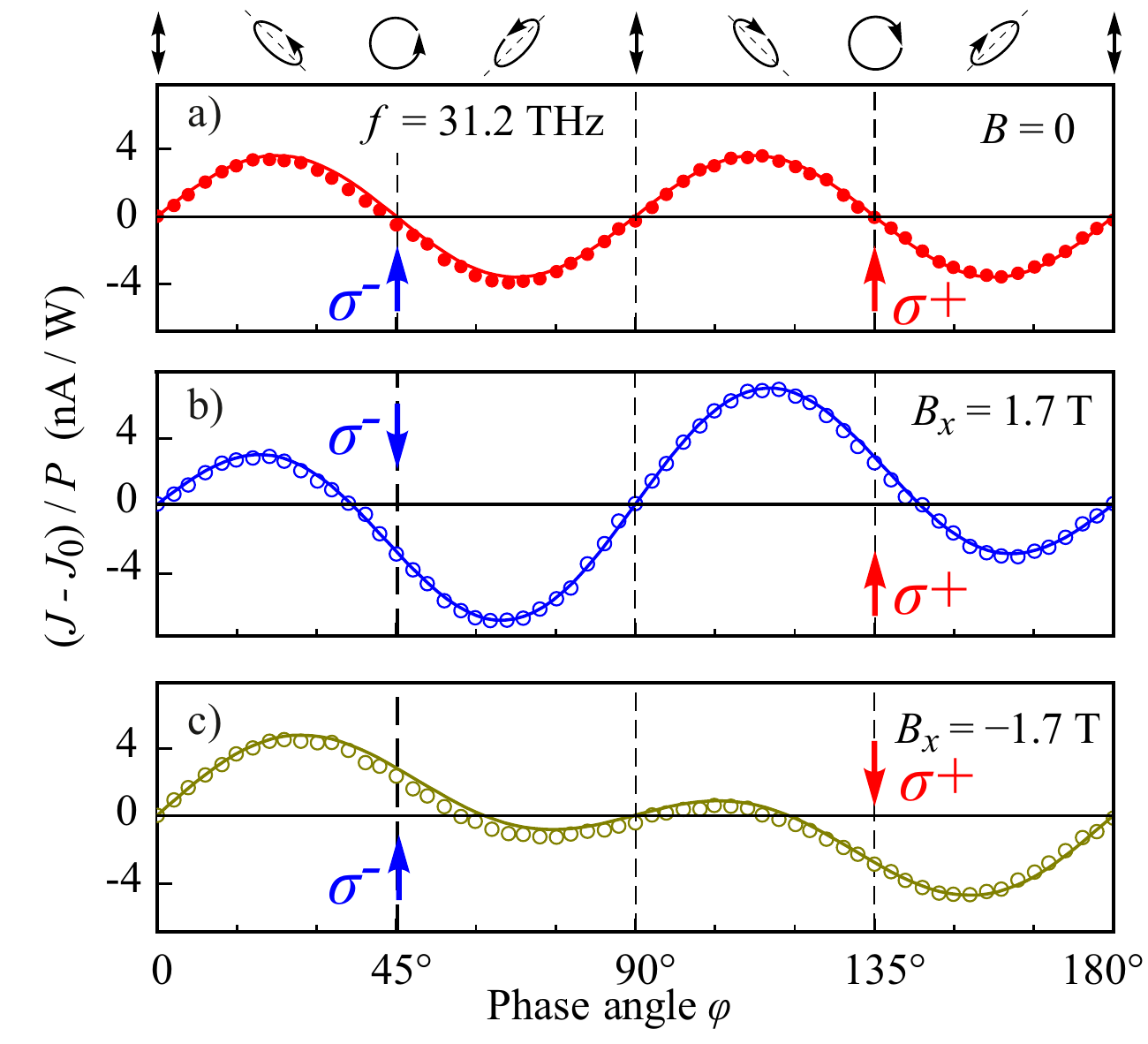}
	\caption{Helicity dependences of the normalized photocurrent $(J-J_0)/P$ obtained at zero magnetic field (red circles) and $B_x=\pm 1.7$~T (blue and dark yellow circles).  The data are measured at $f = 31.2$~THz ($\lambda  = 9.6 \,\,\mu$m). Solid lines are fits after Eq.~(\ref{jycirc}). The ellipses on top illustrate the polarization states at several angles $\varphi$. Vertical blue and red arrows in the panels indicate left- and right-handed circularly polarized radiation.}
	\label{Fig2}
\end{figure}

Using circularly polarized radiation we also detected a magnetic field induced circular photocurrent whose direction reverses when the radiation helicity reverses.  Figure~\ref{Fig2} shows the dependencies of the photocurrent on the angle $\varphi$ measured at $f=31.2$~THz. The curves can be  fitted by
\begin{equation}
	\label{jycirc}
	J_y = {1\over 2}J_1(B_x) \sin4\varphi + J_{\rm circ}(B_x) \sin2\varphi + J_0\,\,\,,
\end{equation}
where $J_1(B_x)$ is the same as used in Eq.~(\ref{jy})
and $J_{\rm circ}(B_x)$ is a fitting parameter that corresponds to the magnitude
of the circular photocurrent,
which is proportional to $P_{\rm circ}$ and reverses its sign by changing polarization from $\sigma^-$ to $\sigma^+$
\footnote{Note that in the geometry applying $\lambda$-quarter plate the $\sin4\varphi/2$ corresponds to the Stokes parameter describing the degree of linear polarization and the polarization ellipse orientation, which in experiments applying $\lambda$-half plate is given by $\sin 2\alpha$, see below and Refs.~\cite{Saleh2019,Belkov2005}.
}. 
{The magnetic field dependencies of the circular photocurrent, measured at two different radiation frequencies, are shown in Figs.~\ref{Fig3}(a) and (c). These plots reveal that the sign of 
\begin{equation}
\label{J_MCPGE}
J_{\rm circ}(B_x) = D^{\rm MCPGE} P B_x
\end{equation}
changes by reversing the magnetic field direction; at $B_x=0$ this contribution vanishes. 
}

\begin{figure}
	\centering \includegraphics[width=0.8\linewidth]{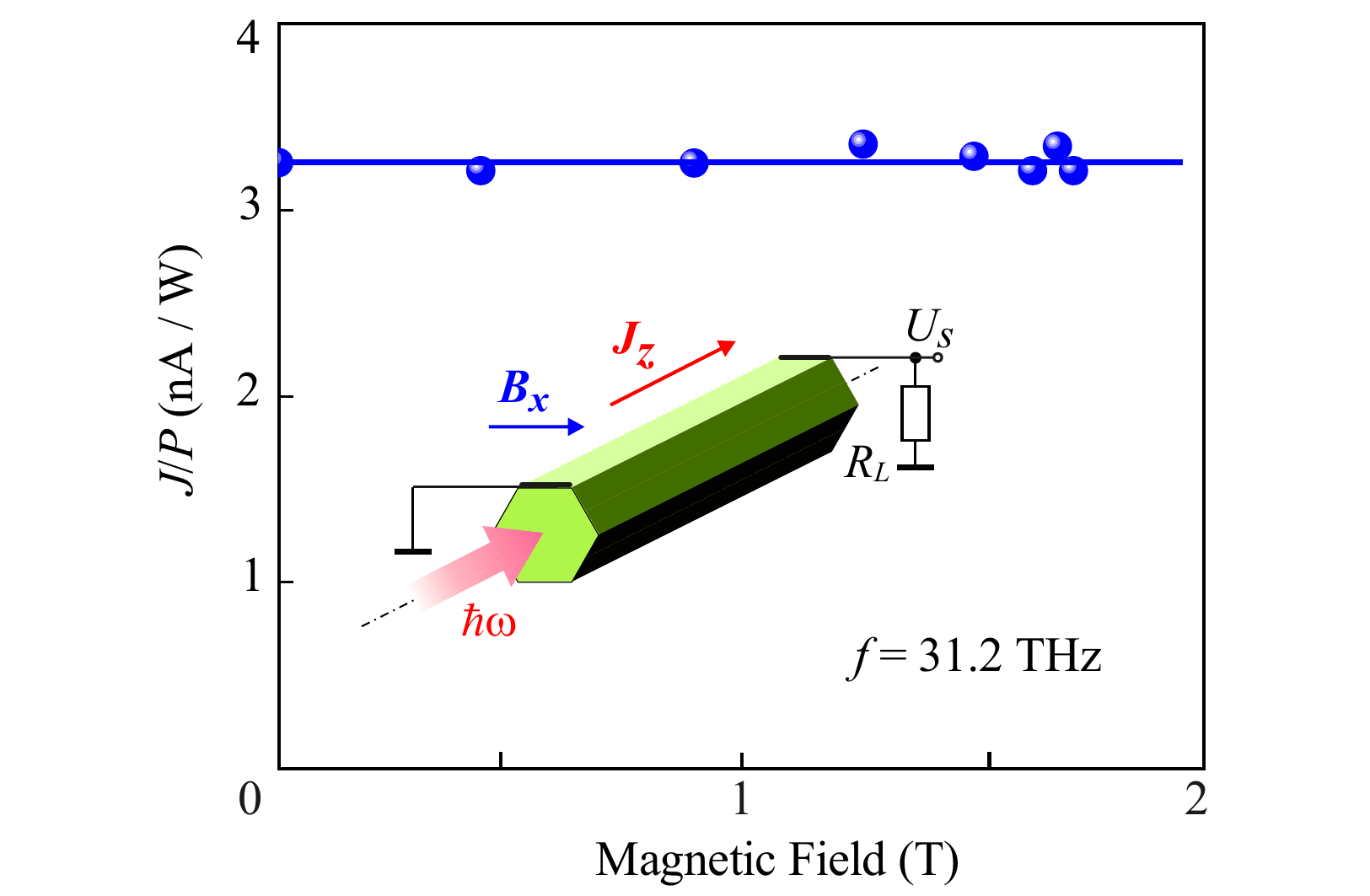}
	\caption{{Magnetic field dependence of the amplitude of the longitudinal circular photocurrent normalized on the radiation power $P$. The inset shows the experimental setup.}
	}
	\label{FigLong}
\end{figure}

{At first glance, one can attribute the magnetic field induced circular photocurrent $J_y$ to the CPGE current excited along $c$-axis~\cite{Ivchenko78p640,Asnin1978}, which is turned towards the sample plane due to the Lorentz force. To examine this possibility, 
we performed additional measurements. Using a tellurium  sample with a length of 25~mm and the cross-section of about 20~mm$^2$ we detected the longitudinal circular photocurrent only.  Application of magnetic field $B_x \leq 2$~T neither change the magnitude of the CPGE current nor causes other photocurrents, {see Fig.~\ref{FigLong}.}
This contradicts with the scenario addressed at the beginning of the paragraph, because that should result in the suppression of the longitudinal photocurrent due to the Lorentz force.  This proves that the measured current Eq.~\eqref{J_MCPGE}
comes from a so far unknown mechanism of the circular magneto-photocurrent. 
The microscopic sense of the fitting parameter $D^{\rm MCPGE}$ is discussed in Sec.~\ref{discussion}.
}

\section{Phenomenological theory and identification of individual photocurrent contributions}
\label{Phenomenological_theory}

Tellurium is a chiral semiconductor with two enantiomorphs  in nature, dextrarotatory and levorotatory, which are mirror images of each other. They can be visualized as two screws with opposite threads. 
We used levorotatory tellurium, which was determined by measuring its natural optical activity~\cite{Shalygin2016}. The point symmetry group of tellurium is $D_3$. This group has a three-fold rotation axis $C_3$, the so-called $c$-axis. We denote this direction as $z$. There are also three $C_2$ rotation axes in the perpendicular plane $(xy)$.

A complete phenomenological analysis of the photocurrents excited by a light propagating uniformly along the $c$-axis ($\bm q \parallel +z$) in a homogeneous Te crystal is presented in Appendix~\ref{appendix2}. In experiments the {polarization-dependent} photocurrent is measured in the $y$ direction for either zero magnetic field or a magnetic field applied along the $x$-axis parallel to one of the $C_2$ axes. For these conditions the phenomenological theory yields:
\begin{align}
	\label{jyphenomen}
j_y =&\qty[ -\chi \tilde{P}_{\rm lin}  + \Phi_l\tilde{P}_{\rm lin}B_x - \Phi_c  P_{\rm circ}B_x]_{\rm PGE}\abs{\bm E}^2 +\nonumber\\
 & \qty[ \tilde{T} P_{\rm lin}q_z - S_l B_x P_{\rm lin} q_z ]_{\rm PDE} \abs{\bm E}^2 \,\,\,,
\end{align}
where $\bm j$ is the photocurrent density, $\bm E$ is the complex amplitude of the radiation electric field
\begin{equation}
\label{E_field}
\bm E(t)=\bm E \exp(-i\omega t)+\bm E^* \exp(i\omega t)
\end{equation}
and
\begin{align}
	\label{Stokes}
	&P_{\rm lin}\abs{\bm E}^2=\abs{E_x}^2-\abs{E_y}^2, \,\,\\
	&\tilde{P}_{\rm lin}\abs{\bm E}^2=E_xE_y^*+E_x^*E_y,\\
	&P_{\rm circ}\abs{\bm E}^2 = i(E_xE_y^*-E_x^*E_y)
\end{align}
are the Stokes parameters, which describe the  polarization of the radiation~\cite{Saleh2019}. 
The terms in the brackets $[...]_{\rm PGE}$ describe the photogalvanic effects (PGE) caused by the
chiral symmetry of bulk Te. The first term in this bracket proportional to the parameter $\chi$ is the trigonal linear photogalvanic effect (LPGE), the second and the third terms proportional to $\Phi_l$ and $\Phi_c$ describe the linear and circular magneto-photogalvanic currents (MPGE). The terms in the bracket $[...]_{\rm PDE}$ describe the photon drag effect (PDE) caused by the transfer of the linear momentum of light to the charge carriers. The first term in this bracket denoted by  parameter $\tilde{T}$ is the  trigonal PDE
and the second one, proportional to the parameter  $S_l$, is the magneto-photon-drag effect (MPDE).

{The PGE current amplitudes obtained in experiments are related with the theoretical values by $[A^{\rm LPGE},D^{\rm MLPGE},D^{\rm MCPGE}] = 2\pi \mathcal S/(cn_\perp)[\chi, \Phi_l, \Phi_c]$, where $\mathcal S$ is the sample {aspect ratio} and $n_\perp$ is the refraction index of Te for radiation propagating along the $c$-axis.}

\subsection{Photocurrents in response to linearly polarized radiation}
\label{linear_polarization}

Equation~\eqref{jyphenomen} shows that a proper choice of  experimental setup together with variation of the  polarization state can be used to identify the contributions of different photocurrent mechanisms. In our experiments linearly polarized light propagates along $z \parallel \bm c$ and the orientation of the in-plane electric field vector $\bm E$ is varied by a counterclockwise rotation of a $\lambda/2$-plate. A zero azimuth angle, $\alpha=0$, corresponds to $\bm E \parallel y$.
Under these conditions the Stokes parameters are given by~\cite{Belkov2005}
\begin{align}
	\label{Plin}
	&P_{\rm lin} = -\cos2\alpha, \,\,\, \,\,\tilde{P}_{\rm lin} = -\sin2\alpha.
\end{align}
	
Consequently, Eq.~(\ref{jyphenomen}), which describes  the total current in $y$-direction and represents the sum of the linear photogalvanic and photon drag effects takes the form
\begin{align}
	\label{jyA2}
	j_y = &(\chi\sin2\alpha - \tilde{T}q_z\cos2\alpha)\abs{\bm E}^2 + \nonumber\\
	&(-\Phi_l\sin2\alpha + S_lq_z\cos2\alpha)B_x\abs{\bm E}^2  \,\,.
\end{align}
For convenience, we have grouped the zero magnetic field (first bracket) and magnetic field induced (second bracket) PGE and PDE   contributions.

Figure~\ref{Fig1} shows that both zero-field (first bracket) and magneto-photocurrents (second bracket) excited by radiation with $f \approx 30$~THz vary with the rotation of the electric fied vector $\bm E$ sinusoidally with the  double angle $\alpha$. This fact shows that at $B_x=0$ the photocurrents are governed by the trigonal LPGE ($j_y = \chi\sin2\alpha \abs{\bm E}^2$) and the photon drag contribution with a current proportional to $\cos 2\alpha $ is not detectable, see Fig.~\ref{Fig1}. In the presence of a magnetic field $B_x$,  this current is superimposed  the linear MPGE  ($j_y = \Phi_lB_x \sin2\alpha \abs{\bm E}^2$), see  Figs.~\ref{Fig1} and~\ref{Fig3}(a) (squares)~\footnote{Note that the trigonal photocurrents given by coefficients $\chi$ depend on the polarization plane orientation
as the 2nd angular harmonics because projections of
the current onto fixed axes are measured. If one detects a direction of the photocurrent as a function of
the light polarization then it has a form of the 3rd
angular harmonics.}.

At low frequencies, however, we found that the dependence of the zero-magnetic field photocurrent  on the angle $\alpha$ is phase shifted by the angle $\psi$, see Fig.~\ref{Fig4}(a). This shows that at these frequencies the photocurrent is caused by the superposition of the LPGE and linear PDE, which are proportional to the sine  and cosine of  $2\alpha$, respectively.
From the fit obtained for zero magnetic field, where $J_y \propto \sin(2\alpha - 46^0)$, see Fig.~\ref{Fig4}(a), we can conclude that $\chi/q_z\tilde{T} = \tan46^0 \approx 1$, hence $\chi \approx q_z\tilde{T}$.

From measurement of the azimuthal dependence of the photocurrent for different magnetic field strengths we found that the phase shift $\psi(B_x)$ depends on the magnitude and sign of $B_x$, see inset in Fig.~\ref{Fig4}.  Considering that, as shown above, $\chi \approx q_z\tilde{T}$, we obtain the azimuthal angle dependence of the magnetic field induced photocurrent in the form
\begin{align}
\label{phase}
j_y \propto \sin(2\alpha - \arctan\frac{1-B_xS_l/\tilde{T}}{1-B_x\Phi_l/\chi}\,)  \,\,.
\end{align}
The dependence of the phase $\psi$ on the in-plane magnetic field $\psi(B_x)$, see inset in Fig.~\ref{Fig4}, allows us to extract the ratio of the parameters defining the linear MPGE and the trigonal PGE, $\Phi_l/\chi = -0.5$~{T$^{-1}$},  as well as the parameters defining the magneto-photon-drag effect and the trigonal photon drag effect, $S_l/\tilde{T} = -0.2$~{T$^{-1}$}.

\subsection{Photocurrents in response to elliptically polarized radiation}
\label{elliptical_polarization}

Experiments show that the circular photocurrent, whose direction is reversed by changing from $\sigma^-$ to $\sigma^+$ circularly polarized radiation, can only be observed in the presence of an external magnetic field, see Fig.~\ref{Fig2} and Figs.~\ref{Fig3}(a) and (c). This is fully consistent with the phenomenological theory which yields a magneto-induced circular photogalvanic current (MCPGE)
\begin{equation}
	\label{Circ)}
	j_y^{\rm MCPGE} = - \Phi_c B_xP_{\rm circ} \abs{\bm E}^2  \,\,.
\end{equation}
In experiments, the  polarization state of the radiation is varied by rotating the $\lambda/4$-plate by an angle $\varphi$ with respect to the $y$-direction, and the Stokes parameters are given by~\cite{Belkov2005}
\begin{align}
	&P_{\rm lin} = -(\cos4\varphi+1)/2 \,\,\,\,\,\\
		\label{Stokesphi}
	&\tilde{P}_{\rm lin} = -\sin4\varphi / 2\\
	&P_{\rm circ}
	= \sin2\varphi \,\,.
\end{align}
Consequently,  the photocurrent in $y$-direction is given by
\begin{align}
	\label{jycircphi}
	j_y = &[\chi\sin4\varphi/2 - \tilde{T}q_z(\cos4\varphi +1)/2]\abs{\bm E}^2 \nonumber\\
	&- [\Phi_{l}\sin4\varphi/2 - S_{l}q_z(\cos4\varphi +1)/2]B_x\abs{\bm E}^2 \nonumber\\
	& - \Phi_c B_x \sin2\varphi \abs{\bm E}^2 \,\,.
\end{align}
Here, the last term describes the circular MPGE. All other terms are due to the photogalvanic and photon drag effects discussed in the previous sections: they are characterized by the same values of the parameters $\chi, \tilde{T}, \Phi_{l}, S_{l}$ as detected in experiments with linearly polarized radiation and the polarization dependence is simply modified according to the modification of the corresponding Stokes parameters. The fit of the experimental data with this function describes the experimental data well and is shown in Fig.~\ref{Fig2}. 
The magnetic field dependencies of the 
MCPGE current, extracted from the experimental helicity dependencies, or, in some measurements, as the half difference between the photocurrent magnitudes in response to $\sigma^+$- and $\sigma^-$- circularly polarized radiation, are shown in Figs.~\ref{Fig3}(a) and~(c). 

\section{Microscopic theory}
\label{theory}

\begin{figure}[t]
	\centering \includegraphics[width=\linewidth]{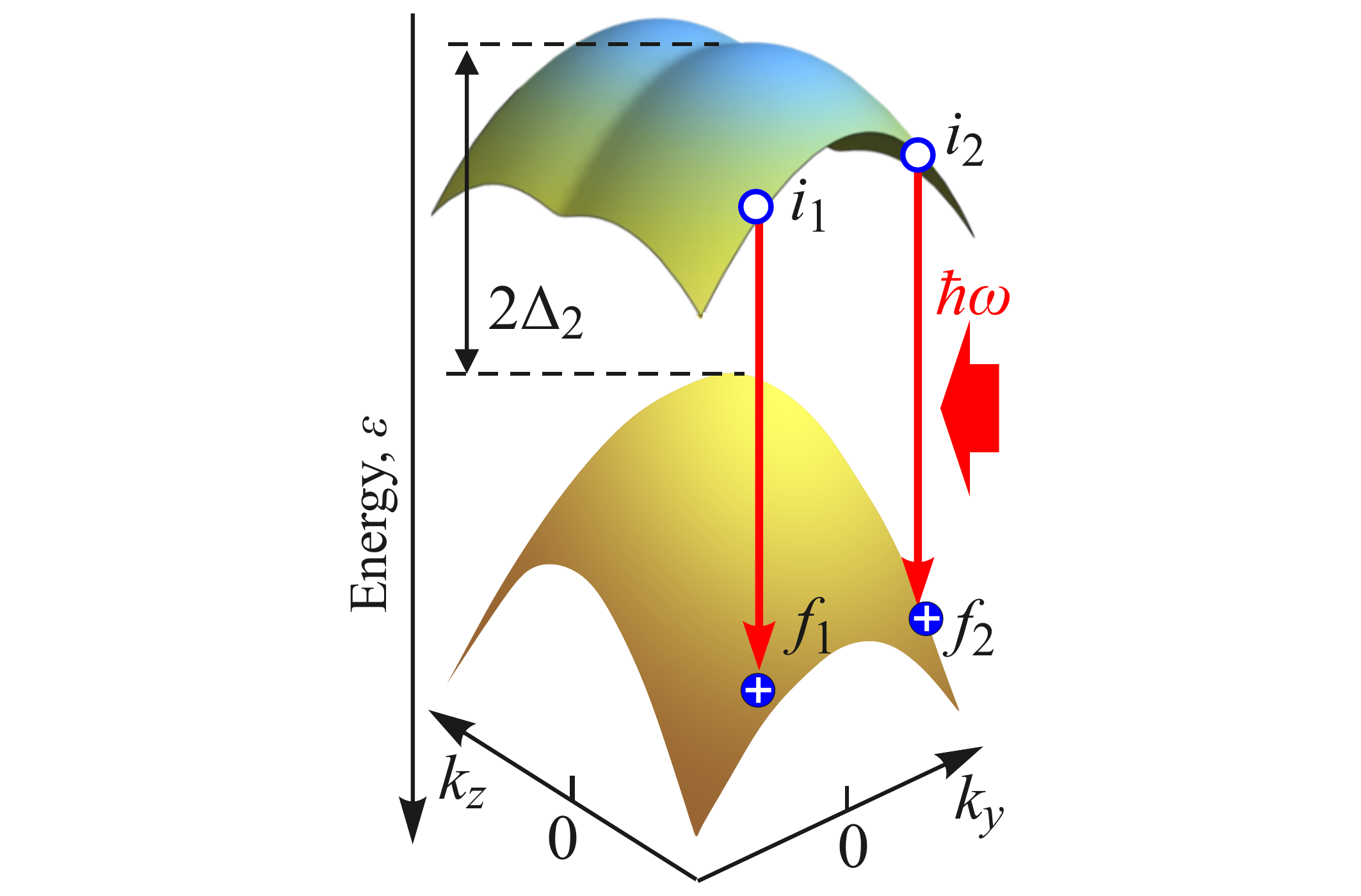}
	\caption{Band structure of the valence band of tellurium. At $\hbar\omega > 2\Delta_2$, direct intersubband optical transitions are allowed.}
	\label{Fig5t}
\end{figure}

Figure~\ref{Fig5t} sketches the valence band structure of bulk Te crystals. 
For the microscopic picture of the observed photocurrents we need to identify optical transitions responsible for their formation. As discussed in Sec.~\ref{Experiment}, the analysis of the photocurrent frequency dependence, see Fig. \ref{Fig3}(b),  shows that the photocurrent excited by infrared radiation with photon energies of the order of 130 meV is driven by the direct intersubband transitions, see red downward arrows in Fig.~\ref{Fig5t}. 
Note, that the photon energies are too low to excite interband transitions ($\hbar \omega \ll E_g=335$~meV). 
For THz radiation with photon energies in the order of 10 meV, i.e. much smaller than the energy separation of the subbands in the valence band $2\Delta_2 \approx 100-125$~meV~\cite{Grosse1,Grosse2,Averkiev_1984_theory,Averkiev_1984_experiment,China2022}, the photocurrents are caused by the indirect Drude-like optical transitions.

We develop a theory of photocurrents in tellurium that are induced by both intersubband optical transitions in the valence band, Fig.~\ref{Fig5t}, and  intrasubband Drude-like transitions. We derive expressions for the trigonal LPGE current and the photocurrents in the presence of a  magnetic field.

In the basis of $\pm3/2$ states, the valence-band Hamiltonian for the $H$ point of the Brillouin zone has the following form
\begin{equation}
	\label{H0}
	\mathcal H =
	\Delta_2 \sigma_x +\beta k_z\sigma_z {+\mathcal A_1 k_z^2 + \mathcal A_2 k_\perp^2}.
\end{equation}
Here $\Delta_2$ is half the gap between the valence subbands at $k=0$, see Fig.~\ref{Fig5t}, and $\beta$ is a constant that is the same for both $H$ and $H'$ valleys but of opposite sign in levorotatory and dextrarotatory tellurium.
The eigenstates of the Hamiltonian $\mathcal H$ have the camel-back dispersion
\begin{equation}
	E_{1,2}=\mathcal A_1 k_z^2 + \mathcal A_2 k_\perp^2\mp\sqrt{\Delta_2^2+(\beta k_z)^2}
\end{equation}
and the envelopes
\begin{equation}
	\psi_{1,2} ={1\over \sqrt{2}}\qty(\sqrt{1\pm \eta},\pm\sqrt{1\mp \eta}), \quad \eta = {\beta k_z\over\sqrt{\Delta_2^2+(\beta k_z)^2}}.
\end{equation}

The direct intersubband optical transition matrix element is  at $k_\perp \neq 0$ nonzero only~\cite{Averkiev_1984_theory}
\begin{equation}
	\label{V_12}
	V_{21} = i{eE\over \hbar \omega} {\Delta_2 \over E_g^2} \abs{L}^2(k_+e_- - k_-e_+) \, .
\end{equation}
Here {$k_\pm = k_x \pm ik_y$,} $E_g$ is the energy gap between the conduction band and 
valence band and $L$ is the interband momentum matrix element~\footnote{In Eq.~\eqref{V_12} the factor $\abs{L}^2/E_g$ absent in Eq.~(25) of
Ref.~\cite{Averkiev_1984_theory} is restored.}.
{The matrix element $V_{21}$ is quadratic in $L$ because the intersubband transitions in the valence band occur via virtual states in the conduction band.}
The absorption coefficient for intersubband transitions $K(\omega)$ calculated according to Fermi's Golden rule with the matrix element~\eqref{V_12}  is given in Appendix~\ref{Abs_intersubband}.

\subsection{Trigonal LPGE current at intersubband transitions}
\label{trigonal_intersubband}

First we derive the trigonal LPGE  which is responsible for the photocurrent $J_1(B=0)$  excited by infrared radiation.
The Hamiltonian~\eqref{H0} describes the uniaxial model of tellurium. In order to account for the trigonality we add an additional term to the valence-band Hamiltonian which has the following form in the basis of the $\pm3/2$ states~\cite{Averkiev_1984_theory}:
\begin{equation}
	\delta \mathcal H = i\gamma' (k_+^3-k_-^3)\sigma_z.
\end{equation}
Here $\gamma'$ is a real constant 
equal at both $H$ and $H'$  Brillouin zone points. 
This correction results in an additional term in the intersubband matrix element $\delta V_{21}(\bm k)=(ie/\hbar \omega)\bm E\cdot \bm \nabla_{\bm k}\delta \mathcal H$.
This results in two competitive contributions to the LPGE: the shift current and the ballistic or injection current. The latter occurs when the interference of light absorption with disorder or phonon scattering is taken into account~\cite{Sturman2020}.  
The shift current is due to holes with a wavevector $\bm k$ undergo a spatial shift  $\bm R_{21}(\bm k)$ during the light absorption process. 
The accumulation of the shifts results in a contribution to the steady-state photocurrent. In general, the ballistic and shift photocurrents have the same order of magnitude and  equally dependent on the system parameters.
Here we estimate the LPGE current by calculating the shift contribution.

The shift value for direct optical transitions is given by~\cite{BelIvchSt_1982,Leppenen2023}
\begin{equation}
	\label{R_shift}
	\bm R_{21}(\bm k) = -\bm \nabla_{\bm k} {\rm arg}(V_{21}+\delta V_{21}) + \bm \Omega_2(\bm k)-\bm \Omega_1(\bm k),
\end{equation}
where ${\rm arg}$ stands for the complex argument, and the Berry connections in the subbands are $\bm \Omega_{1,2}=i\expval{\bm \nabla_{\bm k}}{\psi_{1,2}}$.
The shift photocurrent density reads
\begin{align}
	\label{j_shift}
	&\bm j= 2e
	\\
	& \times \sum_{\bm k} {2\pi\over \hbar}\abs{V_{21}}^2 \bm R_{21} \delta\qty[2\sqrt{\Delta_2^2+(\beta k_z)^2}-\hbar\omega](f_1-f_2). \nonumber
\end{align}
Here the factor 2 accounts for the two valleys of tellurium, and $f_{1,2}$ are occupancies of the initial and final states. 

Calculating the shift photocurrent for intersubband transitions (ib) we obtain in accordance with the phenomenological Eq.~\eqref{jyphenomen}
\begin{align}
	&j_x =  \chi^{\rm ib} {P}_{\rm lin} \abs{\bm E}^2,\\
\label{jshift}
	&j_y =  -\chi^{\rm ib} \tilde{P}_{\rm lin} \abs{\bm E}^2,
\end{align} 
%
%
%
%
where
\begin{equation}
	\label{chi}
	\chi^{\rm ib} = \gamma' K(\omega) {12ec  E_g^2\over \pi \abs{L}^2(\hbar\omega)^2}
\end{equation}
with $K(\omega)$ being the  absorption coefficient for intersubband transitions, see Appendix~\ref{Abs_intersubband}.

\subsection{{Linear and circular} MPGE currents at intersubband transitions}
\label{MPGE_intersubband}

Application of an external magnetic  field results in an additional photocurrent superimposed with the 
trigonal one. This is caused by the change  in the probability of optical transitions for positive and negative wavevectors resulting in an imbalance of the population of states involved in radiation absorption. 
In a magnetic field, 
the following correction to the Hamiltonian appears: 
\begin{equation}
	\label{H_B}
	\delta \mathcal H_B 
	= \gamma_B \sigma_z  (k_+^2 k_- B_- - k_-^2 k_+ B_+) .
\end{equation}
Here, $\gamma_B$ is a real constant that is the same for both $H$ and $H'$ Brillouin-zone points.
{The constant $\gamma_B$ can be obtained in the 5th order of the $\bm k \cdot \bm p$ perturbation theory.}
Accounting for this term results in additional, magnetic field-dependent spatial shifts of the holes under elliptical polarization.
Calculation of the shift $\bm R_{21}(\bm k, \bm B)$ by Eq.~\eqref{R_shift} with $\delta V_{21}=(ie/\hbar \omega)\bm E\cdot \bm \nabla_{\bm k}\delta \mathcal H_B$ and then the photocurrent by Eq.~\eqref{j_shift} we obtain (at $\bm q \parallel +z$) in accordance with Eqs.~\eqref{jyphenomen} the shift MCPGE current
\begin{equation}
	\label{jycircMPGE}
	\bm j =  \Phi_c^{\rm ib}\bm B  \times \hat{\bm z} P_{\rm circ} \abs{\bm E}^2,
\end{equation}
where the shift contribution to $\Phi_c^{\rm ib}$ is given by
\begin{equation}
	\label{Phi_c_MPGE}
	\Phi_c^{\rm ib} = - {\gamma_B\over 4\gamma'}\chi^{\rm ib}
\end{equation}
with $\chi^{\rm ib}$ the zero-field trigonal LPGE constant for intersubband transitions, Eq.~\eqref{chi}.

Accounting for the correction~\eqref{H_B} also changes the selection rules for the intersubband transitions under linearly polarized or at unpolarized excitation. The correction to the matrix element is given by
\begin{align}
	\delta V_{21}^{\rm lin} =& -2{eE\over \hbar \omega}  \sqrt{1-\eta^2} \gamma_B  k_\perp^2 B_\perp
	\\
	&\times \qty[2 \sin(\alpha-\varphi_{\bm B}) + \sin(2\varphi_{\bm k} -\alpha-\varphi_{\bm B})] , \nonumber
\end{align}
where $\varphi_{\bm k}$, $\varphi_{\bm B}$, and $\alpha$  are the azimuthal angles of $\bm k_\perp$, $\bm B_\perp$ and $\bm E_\perp$, respectively.
As a result, the squared matrix element contains a part $\abs{V_{21}+\delta V_{21}^{\rm lin}}^2\propto 2{\Re}(V_{21}^*\delta V_{21}^{\rm lin})$ asymmetrical in $\bm k$. The corresponding asymmetrical part of the intersubband transition probability $W_{21}^{(as)}$ results in a ballistic photocurrent.
Its density is calculated as
\begin{equation}
	\bm j_\perp = 2e\sum_{\bm k} \bm v_\perp W_{21}^{(as)}(f_1-f_2)(\tau_2 -\tau_1).
\end{equation}
Here $\tau_{1,2}$ are the momentum relaxation times in the subbands labeled ``1'' and ``2'', and $\bm v_\perp = 2\mathcal A_2 \bm k_\perp/\hbar$ is the hole velocity equal in both subbands.
Calculation yields in accordance with Eq.~\eqref{jyphenomen} and Appendix~\ref{appendix2}
\begin{equation}
	\label{MPGEinter}
	j_{x,y}= \qty[\Phi_l^{\rm ib}(\pm {P}_{\rm lin} B_{x,y} + \tilde{P}_{\rm lin}B_{y,x}) + {\Lambda}^{\rm ib} B_{x,y}]\abs{\bm E}^2 ,
\end{equation}
where the MLPGE  and the polarization-independent MPGE constants $\Phi_l^{\rm ib}$ and ${\Lambda}^{\rm ib}$ are given by
\begin{equation}
	\label{PhiMPGE}
	\Phi_l^{\rm ib}= - 2 {\Lambda}^{\rm ib} =  {8 k_{\rm B}T (\tau_1 - \tau_2)\over 3\hbar}\Phi_c^{\rm ib}.
\end{equation}
Here $T$ is the temperature, $k_{\rm B}$ is the Boltzmann constant, and $\Phi_c^{\rm ib}$ is given by Eq.~\eqref{Phi_c_MPGE}.

\subsection{Trigonal LPGE current at intraband transitions}
\label{trigonal_intraband}

Now we turn to the photocurrent generated by linearly polarized THz radiation.
For Drude-like intraband optical transitions inside the ground valence subband which are relevant for THz frequencies, the photocurrent is derived from the Boltzmann kinetic equation. In this approach, the low symmetry of tellurium is taken into account in the collision integral. This means that the photocurrent is generated due to the action of polarized radiation and asymmetric hole scattering by disorder or phonons.

In order to obtain the asymmetrical scattering probability,
we consider the following terms of different parity in the interband Hamiltonian:
\begin{equation}
	\label{interband_matr_el}
	\mel{\pm 1/2}{\mathcal H}{\pm 3/2} = Lk_\pm  + Qk_\mp^2.
\end{equation}
Here $\ket{\pm 1/2}$ are the conduction-band states, $L$ is the interband matrix element giving rise to intersubband transitions, Eq.~\eqref{V_12}, and $Q$ fixes the trigonal symmetry of tellurium.
Accordingly, the wavefunction envelope in the ground valence subband  has the form
\begin{equation}
	\label{psi_k}
	\psi_{\bm k} = \psi_0
	+ {(Lk_+ + Qk_-^2)\ket{1/2} + (Lk_- + Qk_+^2)\ket{-1/2}\over E_g\sqrt{2}},
\end{equation}
where $\psi_0$ is the wavefunction 
{caclulated without mixing with the conduction band}.
Then, the matrix element of scattering by the disorder potential $U_{\bm k' \bm k} = U_0\braket{\psi_{\bm k'}}{\psi_{\bm k}}$ 
gets an asymmetric part:
\begin{align}
	&U_{\bm k' \bm k} = U_0 \biggl\{1 +i{{\rm Im}(LQ^*)\over E_g^2}
	\\
	&\times  \qty[k_\perp{k_\perp'}^2\cos(\varphi_{\bm k}+2\varphi_{\bm k'}) - k_\perp^2{k_\perp'}\cos(\varphi_{\bm k'}+2\varphi_{\bm k})] \biggr\}. \nonumber
\end{align}
Here, $U_0$ is the Fourier image of the scattering potential which is assumed to be independent of $\bm k$ and $\bm k'$ corresponding to the scattering by short-range elastic impurities, and the notation `$\perp$' denotes the projection onto the $(xy)$ plane.
This form of $U_{\bm k' \bm k}$ allows us to obtain the asymmetric (skew) scattering probability $W^{a}_{\bm k' \bm k} = -W^{a}_{\bm k \bm k'}$. This can be achieved using the next to Born approximation, which yields~\cite{Belinicher_Sturman_1980,Olbrich2014,Otteneder2020,Hild_PRB_2023}:
\begin{align}
\label{Wa}
	W^{a}_{\bm k' \bm k} = &{(2\pi)^2\over \hbar}\mathcal N \delta(\varepsilon_{\bm k}-\varepsilon_{\bm k'})
	\\
	&\times
	\sum_{\bm p} {\rm Im} (U_{\bm k' \bm p}U_{\bm p \bm k}U_{\bm k \bm k'})\delta(\varepsilon_{\bm k}-\varepsilon_{\bm p}), \nonumber
\end{align}
where $\mathcal N$ is the concentration of scatterers and $\varepsilon_{\bm k}\equiv E_1(\bm k)$ is the hole dispersion in the ground valence subband. 
In the following  we assume an isotropic parabolic dispersion $\varepsilon_k=\hbar^2 k^2/(2m)$.
The calculation yields the asymmetrical scattering probability in the form
\begin{align}
	&W^{a}_{\bm k' \bm k} = {2\pi U_0\over \tau}\delta(\varepsilon_k-\varepsilon_{k'}) {{\rm Im}(LQ^*)\over E_g^2}
	\\
	&\times [k_\perp {k'_\perp}^2 \cos(\varphi_{\bm k}+2\varphi_{\bm k'})-k'_\perp k_\perp^2 \cos(\varphi_{\bm k'}+2\varphi_{\bm k})], \nonumber
\end{align}
where $\tau$ is the relaxation time introduced by $1/ \tau(\varepsilon_k) = (2\pi/ \hbar)\mathcal N U_0^2 g(\varepsilon_k)$ where $g(\varepsilon)$ is the density of states.
Note that the value of ${\rm Im}(LQ^*)$ is the same in the $H$ and $H'$ valleys.

The trigonal PGE constant $\chi^D$ in 3D tellurium which describes the trigonal LPGE current for Drude-like absorption
\begin{align}
	\label{j_x_trigonal}
&j_x =  \chi^D {P}_{\rm lin} \abs{\bm E}^2\\ 
	\label{jtrigonal}
&j_y =  -\chi^D \tilde{P}_{\rm lin} \abs{\bm E}^2,
\end{align}
is calculated analogously to the 2D case in Refs.~\cite{Olbrich2014,Otteneder2020,Hild_PRB_2023}:
\begin{equation}
	\label{chi_1}
	\chi^D = 2e^3 \sum_{\bm k} \xi \tau \qty[\qty({\tau\over 1+\omega^2\tau^2}f_0')' + \tau'f_0'{1-\omega^2\tau^2\over (1+\omega^2\tau^2)^2}].
\end{equation}
Here, $f_0$ is the 
equilibrium distribution function, 
prime means differentiation over energy $\varepsilon_k$, and  the asymmetry parameter is introduced according to
\begin{equation}
	\xi(\varepsilon_k) = \tau \expval{  \sum_{\bm k'}W^{a}_{\bm k \bm k'} {v_x(\bm k)\qty[v_x^2(\bm k')-v_y^2(\bm k')]} },
\end{equation}
where angular brackets indicate averaging over directions of $\bm k$ at a fixed energy $\varepsilon_k$.
The calculation gives
\begin{equation}
\label{xi_parameter}
	\xi(\varepsilon_{k}) = -  {16\pi {\rm Im}(LQ^*)\over 15 \hbar^3 E_g^2}U_0 g(\varepsilon_k)\varepsilon_k^{3}.
\end{equation}
Integrating the first term in Eq.~\eqref{chi_1} by parts we get
\begin{equation}
	\label{chi_2}
	\chi^D = -e^3 \sum_{\bm k}  {\xi \tau^2 {/\varepsilon_k}\over 1+\omega^2\tau^2} f_0'  \qty(7 + {1-\omega^2\tau^2\over 1+\omega^2\tau^2}),
\end{equation}
where we used the energy dependencies $g(\varepsilon_k)\propto \sqrt{\varepsilon_k}$, $\tau \propto 1/\sqrt{\varepsilon_k}$, and $\xi \propto \varepsilon_k^{7/2}$.

Equation~\eqref{chi_2} is valid for any temperature and any relationship between frequency and relaxation rate. For Boltzmann statistics and high frequency ${\omega\tau \gg 1}$ we get
\begin{equation}
	\label{chi_Boltzman}
	\chi{^D}= -{{96} pe^3 k_{\rm B}T {\rm Im}(LQ^*)\over 5\sqrt{\pi}E_g^2\hbar^3 \omega^2}S_{\rm nB}(T).
\end{equation}
Here $p$ is the hole concentration, and we introduced the non-Born dimensionless parameter $S_{\rm nB}(T)=2\pi U_0 g(k_{\rm B}T)$. The temperature dependence is ${\chi^{{D}} \propto T^{{3}/2}}$.

\subsection{{Linear and circular} MPGE current at intraband transitions}
\label{MPGE_intraband}

The experiment shows that also at THz frequencies the application of an external magnetic field $B_x$ results in magnetic field induced linear and circular photocurrents, see Figs.~\ref{Fig4}, \ref{Fig2}, and \ref{Fig3}(b). The MPGE current {caused by the Drude-like intraband optical transitions} is calculated analogously to Eqs.~\eqref{j_x_trigonal}--\eqref{chi_1} but taking into account the magnetic field in the interband mixing. The magnetic field changes the interband matrix elements Eq.~\eqref{interband_matr_el} according to:
\begin{equation}
\label{interband_m_field}
	\mel{\pm 1/2}{\mathcal H}{\pm 3/2} = Lk_\pm + M B_\pm.
\end{equation}
We see that the forbidden interband transitions are allowed by the magnetic field.
Using the same approach as in Eq.~\eqref{psi_k} we obtain the the $\bm B_\perp$-dependence of the disorder scattering matrix element $U_{\bm k' \bm k}$. It contains a part responsible for MPGE, given by
\begin{equation}
	U_{\bm k' \bm k} = U_0\qty[1 +{{\rm Re}(LM^*)\over E_g^2} \bm B_\perp \cdot (\bm k_\perp+\bm k'_\perp)].
\end{equation}
This part of the scattering matrix element leads to the gyrotropic terms in the scattering probability already in the Born approximation:
\begin{equation}
\label{W_kB}
	W_{\bm k' \bm k} 
	=W_{\bm k' \bm k}^{(0)}\qty[1 +2{{\rm Re}(LM^*)\over E_g^2} \bm B_\perp \cdot (\bm k_\perp+\bm k'_\perp)],
\end{equation}
where $W_{\bm k' \bm k}^{(0)}$ is the zero-field symmetrical part.
Note that the value of ${\rm Re}(LM^*)$ is the same in the $H$ and $H'$ valleys.

To calculate the MPGE current, one has to iterate the kinetic equation for the hole distribution function $f_{\bm k}$
\begin{equation}
	\pdv{f_{\bm k}}{t} + {e\over \hbar}\bm E(t) \cdot \pdv{f_{\bm k}}{\bm k} = \sum_{\bm k'} (W_{\bm k \bm k'}f_{\bm k'}-W_{\bm k' \bm k}f_{\bm k})
\end{equation}
with $\bm E(t)$ given by Eq.~\eqref{E_field},
in the small parameters $\bm E$, $\bm E^*$ and $\bm B$. At the first step, we account for $\bm E$ and obtain the correction to the distribution function in the form
\begin{equation}
	f^{(E)}_{\bm k} = -e\tau_{\omega} f_0' \bm E \cdot \bm v_{\bm k}.
\end{equation}
In the following 
we use the notation $\tau_{\omega}=\tau/(1-i\omega \tau)$.

Then there are two ways to get the current carrying distribution. One is an account for $\bm E^*$ at the next step and get the correction $f^{(EE)} \propto \abs{\bm E}^2$, and account for the $\bm{kB}$ terms in the end to  obtain the correction $f^{(EEB)}\propto \abs{\bm E}^2B_\perp$. In the 2nd step we get the correction to the distribution function describing an alignment of electron momenta:
\begin{equation}
	f^{(EE)}_{\bm k} = e^2 \abs{\bm E}^2 \tau v_\perp^2 {\rm Re}(\tau_{\omega} f_0')' 
	(P_{\rm lin}\cos{2\varphi_{\bm k}} + \tilde{P}_{\rm lin}\sin{2\varphi_{\bm k}}).
\end{equation}
Then we include the magnetic field and get the correction $f^{(EEB)}_{\bm k}$ from the equation
\begin{equation}
	{f^{(EEB)}_{\bm k} \over \tau} + \sum_{\bm k'}W_{\bm k' \bm k} \qty(f^{(EE)}_{\bm k} - f^{(EE)}_{\bm k'}) =0,
\end{equation}
which yields
\begin{equation}
	f^{(EEB)}_{\bm k} = -2{{\rm Re}(LM^*)\over E_g^2}(\bm B_\perp \cdot \bm k_\perp)  f^{(EE)}_{\bm k}.
\end{equation}
It contributes to the MPGE current which is calculated as follows:
\begin{equation}
	\label{j_MPGE}
	\bm j = 2e\sum_{\bm k} \bm v_{\bm k} f^{(EEB)}_{\bm k}.
\end{equation}
Substituting $f^{(EEB)}_{\bm k}$ we obtain in accordance with the phenomenological Eq.~\eqref{jyphenomen} 
\begin{equation}
	\label{jLMPGE}
	j_{x,y}=\Phi_l^D(\pm {P}_{\rm lin} B_{x,y} + \tilde{P}_{\rm lin}B_{y,x}) \abs{\bm E}^2,
\end{equation} 
where the MLPGE  constant $\Phi_l^D$ is given by
\begin{equation}
		\label{Phi_l}
	\Phi_l^D = - e^3 {8{\rm Re}(LM^*)\over 5m\hbar E_g^2}\sum_{\bm k} \varepsilon_k^2 \tau \qty({\tau\over 1+ \omega^2\tau^2} f_0')'.
\end{equation}

Now we calculate an additional contribution to the distribution function that takes into account  the $\bm{kB}$ terms in the 2nd step. In doing so, we find a time-dependent correction to the distribution function $f^{(EB)}_{\bm k}\propto EB$ that satisfies the equation
\begin{equation}
	{f^{(EB)}_{\bm k} \over \tau_{\omega}} + \sum_{\bm k'}W_{\bm k' \bm k} \qty(f^{(E)}_{\bm k} - f^{(E)}_{\bm k'}) =0.
\end{equation}
The solution is given by
\begin{align}
	&f^{(EB)}_{\bm k} =  2{{\rm Re}(LM^*)\over E_g^2}
	\\
	&
	\times{\tau_{\omega}^2 \over \tau} e f_0' \qty[(\bm B_\perp \cdot \bm k_\perp)(\bm E \cdot \bm v_{\bm k}) - (\bm B_\perp \cdot \bm E){k_\perp v_\perp \over 2}]. \nonumber
\end{align}
Then we find the quadratic in $E$ and linear in $B$ correction $f^{(EBE)}$ which satisfies the following equation:
\begin{equation}
	{e\over \hbar} \bm E^* \cdot \pdv{f^{(EB)}_{\bm k}}{\bm k} + c.c. = -{f^{(EBE)}_{\bm k}\over \tau}.
\end{equation}
Substituting $f^{(EBE)}_{\bm k}$ to Eq.~\eqref{j_MPGE} (instead of $f^{(EEB)}_{\bm k}$) and integrating by parts we get a contribution to the MPGE current in the form
\begin{equation}
	\delta \bm j = 2{e^2\over \hbar}  \bm E^* \cdot \sum_{\bm k} f^{(EB)}_{\bm k} \pdv{(\tau \bm v_{\bm k})}{\bm k} + c.c.
\end{equation}
Since $f^{(EB)}_{\bm k}$ is zero on average, we differentiate here $\tau$ only. This yields
\begin{equation}
	\delta \bm j = 2e^2  \sum_{\bm k} \tau' \qty[\bm v (\bm E^* \cdot \bm v)  - {v_\perp^2\over 2}\bm E^*]  f^{(EB)}_{\bm k} + c.c.,
\end{equation}
where $\tau'=\dd \tau/\dd \varepsilon_k {=-\tau/(2\varepsilon_k)}$.
Substituting $f^{(EB)}_{\bm k}$ we finally obtain in accordance with Eq.~\eqref{jyphenomen}
\begin{equation}
\label{jMCPGE}
\delta j_{x,y}=\qty(\pm \Phi_c^D {P}_{\rm circ} B_{y,x} + {\Lambda}^D B_{x,y})\abs{\bm E}^2,
\end{equation}
where the MCPGE  and the polarization-independent MPGE constants $\Phi_c^D$ and ${\Lambda}^D$ for Drude-like absorption are given by
\begin{equation}
		\label{Phi_c}
			\Phi_c^D= e^3 {16\omega{\rm Re}(LM^*)\over5m\hbar E_g^2}\sum_{\bm k}
	{\tau^2  \tau' \over (1+ \omega^2\tau^2)^2}\varepsilon_k^2 f_0',
\end{equation}
\begin{equation}
	{\Lambda}^D= e^3  {8{\rm Re}(LM^*)\over 5m\hbar E_g^2}\sum_{\bm k}
	{\tau \tau'(1-\omega^2\tau^2)\over (1+ \omega^2\tau^2)^2} \varepsilon_k^2 f_0'.
\end{equation}
For Boltzmann statistics, short-range scattering potential and high frequency $\omega\tau \gg 1$ we get
\begin{equation}
	\label{Phi_l_c_Boltzman}
	\Phi_l^D = -{24p e^3 {\rm Re}(LM^*) \over 5m\hbar\omega^2 E_g^2}, 
	\quad
	\Phi_c^D = -{4\over 3 \sqrt{\pi} \omega \tau_T}\Phi_l^D,
\end{equation}
and ${\Lambda}^D=-{\Phi_l^D/ 4}$, where $\tau_T =\tau(k_{\rm B}T)$.

\subsection{Chiral 
PDE in tellurium}
\label{PDE_intraband}

We consider linearly-polarized  light propagation along $z \parallel \bm c$. In this setup, the following PDE current is allowed by symmetry in Te, see Eq.~\eqref{jyphenomen} and Appendix~\ref{appendix2}:
\begin{equation}
\label{j_PDE}
j_{x} = \tilde{T} q_z \abs{\bm E}^2 \tilde{P}_{\rm lin}, \qquad j_{y} = \tilde{T} q_z \abs{\bm E}^2  {P}_{\rm lin}.
\end{equation}
The constant $\tilde{T}$ is chiral, i.e. it has opposite sign in two enantiomorphic modifications of tellurium.
This PDE current is different from that in $C_{3v}$ symmetric systems where $q_z$ is invariant,  and we have $j_x \propto q_z {P}_{\rm lin}$, $j_y \propto q_z \tilde{P}_{\rm lin}$.
This results in different dependencies on the light polarization in Te and in previously studied $C_{3v}$ systems such as surface states in topological insulators~\cite{Olbrich2014,Plank2016,Plank2018}.

We calculate the constant $\tilde{T}$ for intraband absorption in the ground valence subband of Te assuming an isotropic energy dispersion for holes $\varepsilon_k=\hbar^2 k^2/(2m)$.
To account for $D_3$ symmetry for intraband transitions, one has to include asymmetric hole scattering in the kinetic theory, which can be described as scattering by wedge-shape defects. We denote the corresponding part of the scattering probability as $W^w_{\bm k' \bm k}$. It is asymmetric with respect to an exchange of the initial and final wavevectors: $W^w_{\bm k' \bm k}=-W^w_{\bm k \bm k'}$~\cite{Belinicher_Sturman_1980}.
This probability is similar to $W^a_{\bm k \bm k'}$, Eq.~\eqref{Wa} but describes skew scattering of holes with a non-zero $z$-component of the velocity.

In order to calculate the PDE current, one has to take into account either the finite wavevector of light $q_z$ or the magnetic field $\tilde{\bm B} \perp z$ of the radiation.  In the first approach, the PDE current is the sum of various contributions obtained by iteration of the Boltzmann kinetic equation in the small parameters $E$, $q_z$, $E^*$ and $W^w$, which we also denote as `$w$'. The analysis shows that the following three iteration sequences can contribute to the PDE current, $EqEw$, $EwqE$, and $EqwE$.
In the second approach, there are two contributions, $Ew\tilde{B}$ and $E\tilde{B}w$.

Qualitatively, the PDE current is formed 
by skew scattering from  wedges of nonequilibrium
holes, which have an anisotropic momentum distribution due to the action of the spatially-dispersive polarized radiation.

We consider a geometry with $\bm q \parallel z$, when the electric and magnetic fields of the radiation, $\bm E_\perp$ and $\tilde{\bm B}_\perp$ are perpendicular to the $z$ axis.
The hole distribution function $f(\bm k)$ satisfies the Boltzmann kinetic equation where the field term contains the forces of the  electric field and the Lorentz force of  magnetic field $\tilde{\bm B}_\perp$ of the  radiation:
\begin{align}
\label{kin_eq_PDE}
\pdv{f_{\bm k}}{t} + iq_z v_z f_{\bm k} &+ {e\over \hbar}\bm E_\perp(t) \cdot \pdv{f_{\bm k}}{\bm k_\perp} \nonumber
\\ 
&+ {e\over \hbar c} [\bm v \times \tilde{\bm B}_\perp(t)]\cdot \pdv{f_{\bm k}}{\bm k} =  {\rm St}[f]. 
\end{align}
Here $\tilde{\bm B}_\perp(t) = \tilde{\bm B}_\perp \exp(-i\omega t) + c.c.$, 
{the second term comes from the ${\bm v}\cdot{\bm \nabla}$ term taking into account that the coordinate dependence of $f_{\bm k}$ is $\propto \exp(iq_zz)$,}
and ${\rm St}[f]$ stands for the elastic collision integral. It describes the isotropization of the distribution over the isoenergetic surface $\varepsilon_k={\rm const}$ and skew scattering by wedges:
\begin{equation}
{\rm St}[f] = -{f_{\bm k} - \expval{f} \over \tau} + \sum_{\bm k'}W^w_{\bm k \bm k'}f_{\bm k'}.
\end{equation}
In the following, we consider either $q_z \neq 0$ or $\tilde{B}\neq 0$ in the kinetic equation because they give contributions to the  PDE currents of the same order.

The microscopic theory developed in Appendix~\ref{PDE_micro} yields the PDE current described by the phenomenological Eq.~\eqref{j_PDE} where the constant $\tilde{T}$ is given by
\begin{equation}
\label{tilde_T}
\tilde{T}=2e^3\sum_{\bm k} \qty[ \tau' (\xi_w-\tilde{\xi}_w) {{\rm Im}(\tau_\omega^3)\over \tau} 
- {(\xi_w\tau \sqrt{\varepsilon_k})'\over\sqrt{\varepsilon_k}} {\rm Im}(\tau_\omega^2)]f_0'.
\end{equation}
The imaginary parts read
\begin{equation}
{\rm Im}(\tau_\omega^3) = {-}{\omega \tau^4 [(\omega\tau)^2-3]\over [1+(\omega\tau)^2]^3},
\quad
{\rm Im}(\tau_\omega^2) = {2\omega \tau^3\over [1+(\omega\tau)^2]^2}.
\end{equation}
We introduced two dimensionless parameters $\xi_w$ and $\tilde{\xi}_w$ which are nonzero due to the $D_3$ point symmetry of Te and describe the wedge-like character of hole scattering:
\begin{equation}
\xi_w
= \tau \expval{\sum_{\bm k'} W^w_{\bm k \bm k'} v_z(\bm k') v_y(\bm k) [v_x^2(\bm k')-v_y^2(\bm k')]},
\end{equation}
\begin{equation}
\tilde{\xi}_w =  \tau \expval{\sum_{\bm k'} W^w_{\bm k \bm k'} v_z(\bm k') v_y(\bm k') [v_x^2(\bm k)-v_y^2(\bm k)]}.
\end{equation}
In contrast to the parameter $\xi$ given by Eq.~\eqref{xi_parameter}, the values $\xi_w$ and $\tilde{\xi}_w$ describe skew scattering of holes propagating oblique to the $(xy)$ plane.

At high frequencies $\omega\tau \gg 1$ we have 
\begin{equation}
\tilde{T} = {2e^3\over \omega^3}\sum_{\bm k} \qty[ {-}{\tau' \over \tau} (\xi_w-\tilde{\xi}_w)  
- 2 {(\xi_w\tau \sqrt{\varepsilon_k})'\over\tau\sqrt{\varepsilon_k}}]f_0'.
\end{equation}
The wedge scattering efficiencies $\xi_w$, $\tilde{\xi}_w$ are obtained in the 4th order of $\bm k\cdot \bm p$ perturbation theory, therefore they $\propto {\varepsilon_k^{9/2}}$. {Using $\tau \propto \varepsilon_k^{-1/2}$} we have for Boltzmann statistics 
\begin{equation}
\label{eq_tilde_T}
\tilde{T} = {{24} pe^3\over\sqrt{\pi} \omega^3 (k_{\rm B}T){^2}} \qty({17} \xi_w + \tilde{\xi}_w)_{\varepsilon_k=k_{\rm B}T},
\end{equation}
where $p$ is the hole concentration. The temperature dependence is $\tilde{T} \propto (k_{\rm B}T)^{{5/2}}$.
%

\subsection{Magnetic field induced photon drag effect at intraband transitions}
\label{Drag_intraband}

For the considered linearly polarized light propagating along $z \parallel \bm c$ and an external magnetic field $\bm B \parallel x$,
the following MPDE current is allowed, see Eq.~\eqref{jyphenomen} and Appendix~\ref{appendix2}:
\begin{equation}
	\label{j_MLPDE}
	j_{x} = B_x q_z \abs{\bm E}^2 S_l \tilde{P}_{\rm lin}, \quad j_{y} = - B_x q_z \abs{\bm E}^2 S_l {P}_{\rm lin}.
\end{equation}
This photocurrent which can also be rewritten in the form  $j_+=S_l iq_z B_-E_+^2$ is allowed in systems of any symmetry.
We calculate the constant $S_l$ for intraband absorption in the ground valence subband of Te. 

We discuss the qualitative picture of the MLPDE assuming, for brevity, a degenerate statistics. Under the action of the radiation electric field, an ac electric current in the $(xy)$ plane $\bm j_\perp(z,t) =\bm j_\perp(\omega)\exp[i(q_zz-\omega t)]+c.c.$ appears. It oscillates in space and time with the amplitude
\begin{equation}
	\bm j_\perp(\omega)= \sigma_\omega \bm E_\perp,
\end{equation}
where $\sigma_\omega = pe^2\tau_\omega/m$ is the ac conductivity.
The magnetic field $B_x$ leads to cyclotron motion, which causes the Hall component of this ac current to flow in the $z$-direction:
\begin{equation}
\label{jz_MPDE}
	j_z(\omega) = -\omega_c\tau_\omega j_y(\omega),
\end{equation}
where $\omega_c=eB_x/(m c)$ is the cyclotron frequency.
This ac current is accompanied by oscillations of the carrier density $\delta p(z,t) = \delta p_\omega \exp[i(q_zz-\omega t)]+c.c.$ Its amplitude is related to the ac current by the continuity equation:
\begin{equation}
	{-} ei\omega \delta p_\omega + iq_zj_z(\omega)=0.
\end{equation}
These density oscillations in the presence of the radiation' electric field result in a  dc current due to rectification~\cite{PerelPinskii,Glazov2014}:
\begin{equation}
	\label{MPDE1}
	\bm j_\perp = {e^2\tau\over m} \overline{\delta p(z,t) \bm E_\perp(z,t)} = {e^2\tau\over m}\delta p(\omega)\bm E_\perp^*+ c.c.,
\end{equation}
where the bar denotes averaging over time and $z$ coordinate.
This approach gives the MLPDE current Eq.~\eqref{j_MLPDE} with the constant $S_l$ given by 
{
\begin{equation}
S_l = -{pe^4\tau^3  (1-\omega^2\tau^2) \over m^3\omega c(1+\omega^2\tau^2)^2}.
\end{equation}
}

There is a competing contribution $\Delta \bm j$ of the same order which stems from the radiation magnetic field. It appears as follows: in the presence of  radiation with amplitude $E_y$ and the Lorentz force from $\bm B \parallel x$, an ac electric current appears which oscillates along and opposite to the $z$ axis
with an amplitude {$j_z(\omega)$, Eq.~\eqref{jz_MPDE}.}
Then, this current is rotated due to the action of the ac magnetic field $\tilde{\bm B}_\perp$. This gives rise to the emergence of a  dc `Hall' component 
\begin{equation}
	\label{MPDE2}
	\Delta j_y = {e\tau\over m c}\tilde{B}_x^* j_z(\omega) + c.c.
	\propto E_y\tilde{B}_x^* + c.c. 
\end{equation}
Noting that $\tilde{\bm B}=c \bm q \times \bm E/\omega$, we obtain the contribution to 
Eq.~\eqref{j_MLPDE} with
\begin{equation}
{\Delta S_l = {pe^4\tau^3  (1-\omega^2\tau^2) \over m^3\omega c(1+\omega^2\tau^2)^2}.}
\end{equation}

This qualitative consideration gives the correct magnitude of the MLPDE current
{at a constant relaxation time $\tau$ independent of electron energy}. However, a microscopic theory is needed 
because, in the  qualitative consideration above, the two contributions cancel each other out. 

The microscopic theory developed in Appendix~\ref{App_MPDE} shows
that the resulting coefficient {$S_l$}
 in Eq.~\eqref{j_MLPDE} is nonzero and depends on the dominating scattering mechanism. 
In order to calculate the MPDE current one has to take into account either the finite wavevector of light $q_z$ or the  magnetic field $\tilde{\bm B} \perp z$ of the radiation. In the first approach, the MLPDE current is a sum of several contributions obtained by iterating the Boltzmann kinetic equation in the small parameters $E$, $q_z$, $E^*$ and $B_x$. The analysis shows that the following three iteration sequences can contribute to the MLPDE current, $EqEB$, $EBqE$, and $EqBE$.
In the second approach, there are two contributions, $EB\tilde{B}$ and $E\tilde{B}B$.
Microscopic calculations show that all the contributions
give rise to  the MLPDE constant $S_l$ in Eq.~\eqref{j_MLPDE} 
{for high frequencies ${\omega\tau \gg 1}$} 
in the following form
{
\begin{equation}
\label{MLPDE_constant}
S_l = a {4e^4 \over 3m^3 c\omega^3} \sum_{\bm k} \varepsilon_k  (-f_0')\tau
\end{equation}
} 
with different pre-factors $a$. All contributions to 
the pre-factor 
$a$ are given in the Table~\ref{tab_MLPDE}.

\begin{table}[h]
	\begin{tabular}{|c|c|}
		\hline
		Contribution                         & Pre-factor $a$         \\ \hline
		$EqEB$                         & 0         \\ 
		$EBqE$                         & $ {1+2r/3}$    \\ 
		$EqBE$                         & $-2r/5$         \\ 
		$EB\tilde{B}$ & ${-}1$        \\ 
		$E\tilde{B}B$ & pol./indep. \\ \hline
	\end{tabular}
	\caption{Various contributions to the MLPDE constant pre-factor $a$ in 
	Eq.~\eqref{MLPDE_constant}. The coefficient $r=\dd \ln\tau/\dd \ln \varepsilon$ is determined by the dominating elastic scattering mechanism.}
		\label{tab_MLPDE}
\end{table}


With Boltzmann statistics and $\tau \propto \varepsilon_k^r$
we obtain {summing all contributions} Eq.~\eqref{MLPDE_constant}
\begin{equation}
S_l = {{16r} \Gamma\qty(r+5/2)p e^4 \tau_T \over 45 \sqrt{\pi}m^3 c\omega^3} \propto T^r,
\end{equation}
where $\tau_T = \tau(\varepsilon_k=k_{\rm B}T)$. 
{At arbitrary frequency,} we obtained for the MLPDE constant $S_l$ the following expression,
{see  Appendix~\ref{App_MPDE}
\begin{equation}
\label{S_l_arb_freq}
S_l = {8 r e^4 \over 45m^3 c\omega} \sum_{\bm k} \varepsilon_k  f_0'\tau \qty[5{\rm Re}(\tau_\omega^2)-3\omega{\rm Im}(\tau_\omega^3)].
\end{equation}
}

\section{discussion}
\label{discussion} 


Below we discuss the experimental data in the context of the
the theory developed.
We compare experimental and theoretical results and present microscopic models that illustrate the generation of the different photocurrents described in the previous section.  The photocurrents were obtained for radiation with strongly different photon energies of about 130~meV in experiments with infrared CO$_2$-laser radiation and  on the order of several meV in experiments with THz radiation. 

The magnitude of the photocurrent excited by infrared radiation
is large enough to enable direct intersubband transitions  sketched in Fig.~\ref{Fig5t} but about three times smaller than the forbidden gap, which excludes interband absorption.

As discussed in Secs.~\ref{Experiment} and~\ref{theory}, the  photon energy in the THz range is much smaller than the energy required for any kind of direct optical transitions (intersubband or interband)  and the photocurrent is due to Drude absorption. 

The microscopic mechanisms responsible for the photocurrents for both intersubband and Drude absorption which summarize the results of Sec.~\ref{theory} are given in Table~\ref{table}.
We begin with the photocurrents and magneto-photocurrents excited by infrared radiation and then consider the photoresponse to THz radiation.

\begin{table}[h]
	\begin{tabular}{|c|c|c|}
		\hline
		& Intersubband & Intraband Drude    \\ \hline 
		LPGE  & shift        & skew scattering by triangles     \\ 
		LPDE  & $-$            & skew scattering by wedges        \\ 
		MCPGE & shift        & $\bm k \bm B$ terms in scattering   \\ 
		MLPGE & $\tau_1\neq \tau_2$      & $\bm k \bm B$ terms in scattering   \\ 
		MLPDE & $-$            & Lorentz force\\
		 & $$            & and photon momentum 
		\\ \hline
	\end{tabular}
	\caption{Microscopic mechanisms responsible for the observed photocurrents for infrared and THz ranges at inter- and intrasubband absorption, respectively. The notation `shift'  indicates the sum of the shift contribution and the ballistic contribution is due to the interference of electron-photon interaction with scattering. {Note that the MLPGE current at intersubband transitions is caused by  ballistic propagation of photoholes in both subbands limited by scattering with different transport times $\tau_1 \neq \tau_2$.}}
		\label{table}
\end{table}

\subsection{Infrared radiation induced photocurrent at zero magnetic field}
\label{dis_trigonal_intersubband}

For all the infrared frequencies used in our experiments we observed that the photocurrent is excited only when the degree of linear polarization is non-zero. This is seen in experiments with linearly polarized radiation, see Fig.~\ref{Fig1}(a), and elliptically polarized radiation, see Fig.~\ref{Fig2}(a). The latter figure clearly shows that the response to circularly polarized radiation ($\sigma^+$ and $\sigma^-$) is  zero. The photocurrent detected in the $y$-direction is proportional to the degree of linear polarization $j_y\propto\tilde{P}_{\rm lin}$, see Eq.~\eqref{jyphenomen} and Figs.~\ref{Fig1}(a) and~\ref{Fig2}(a). Consequently, the azimuthal angle dependence of the current in response to linearly polarized radiation is given by $j~\propto \sin(2\alpha)$ or, in experiments using a $\lambda/4-$plate, $j~\propto \sin(4\varphi)$, see Eqs.~\eqref{Stokes}, \eqref{Plin} and~\eqref{Stokesphi}. This functional behavior corresponds to the trigonal LPGE current obtained in Sec.~\ref{theory}, see the terms proportional to the parameter $\chi$ in Eqs.~\eqref{jyphenomen}, \eqref{jyA2}, \eqref{jycircphi} and~\eqref{jshift}. We emphasize that all other photocurrents either depend differently on the degree of linear polarization, $j_y\propto P_{\rm lin} \propto \cos(2\alpha)$ (contributions proportional to the parameter $\tilde{T}$) or vanish at zero magnetic field (contributions proportional to $\Phi_l$, $\Phi_c$ and $S_l$). We also note that the theory shows that the circular photocurrent at normal incidence and zero magnetic field  is symmetry forbidden, see Eq.~\eqref{jyphenomen} and Appendix~\ref{appendix2}. 

\begin{figure*}[t]
	\centering \includegraphics[width=\linewidth]{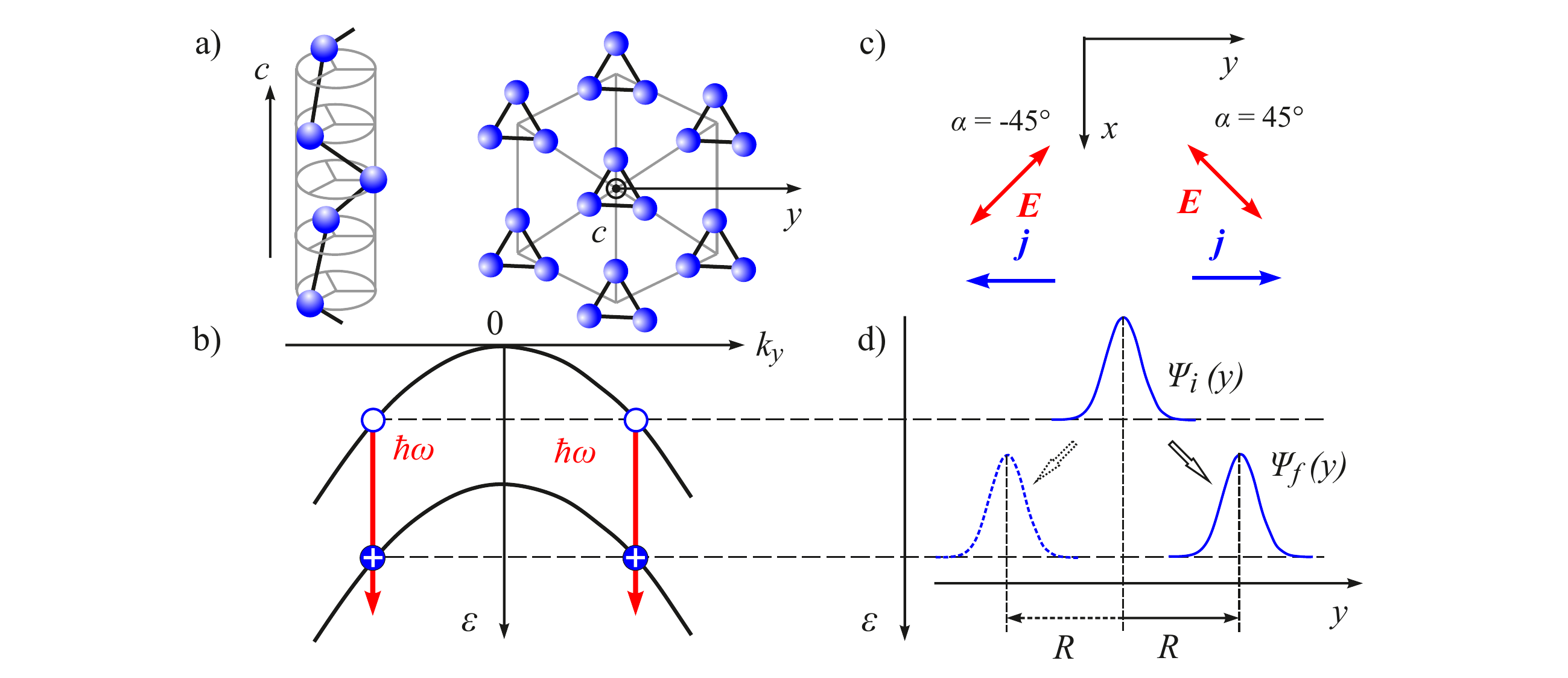}
	\caption{Illustration of the shift contribution to the trigonal LPGE current at intersubband transitions.}
	\label{fig2t}
\end{figure*}

The microscopic theory of the observed LPGE current is developed in Sec.~\ref{trigonal_intersubband}. Taking into account  the shift mechanism of the LPGE we obtained the trigonal LPGE photocurrent and derived the parameter $\chi$, see Eqs.~\eqref{jshift} and~\eqref{chi}. 
The generation of the LPGE current caused by the shifting of the hole wave packets in real space is illustrated in Fig.~\ref{fig2t}. Panel (a) shows the crystallographic structure of the Te crystal with the Te atoms at the corners of the triangles when viewed in the direction of the $c$-axis. Optical transitions between the subbands are sketched in Fig.~\ref{fig2t} by downward vertical arrows. As can be seen  from Eqs.~\eqref{R_shift} and~\eqref{j_shift}, depending on the  orientation of the radiation' electric field vector with respect to the $y$-axis, these transitions lead to shifts of the photoexcited holes by $+R$ ($\alpha = 45^\circ$) or  $-R$ ($\alpha = -45^\circ$), which, consequently, causes a $dc$ electric current (blue horizontal arrows). For vertical or horizontal polarization ($\alpha=0$ or $90^\circ$) the shift along the $y$ axis is zero. 
The frequency dependence of this mechanism is $\chi \propto K(\omega)/\omega^2$.
In the frequency range studied the frequency dependence is weak, which agrees with the experimental data, see curve~1 in the inset of Fig.~\ref{Fig3}(a).

\begin{figure}[t]
	\centering \includegraphics[width=\linewidth]{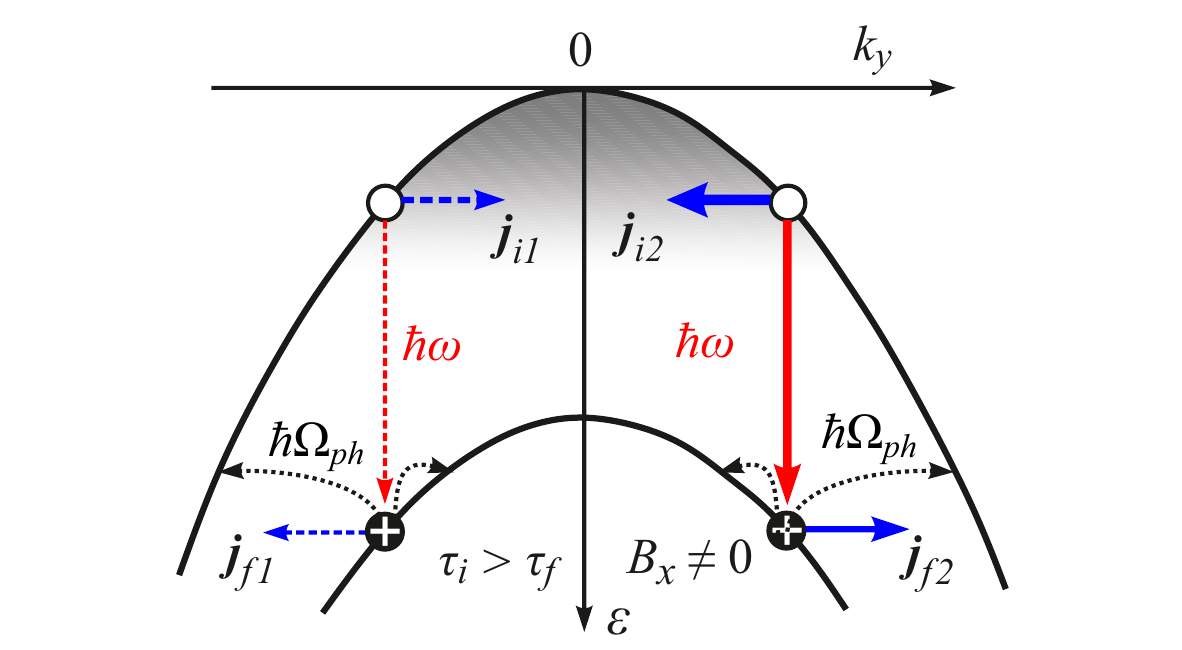}
	\caption{
	Illustration of the MLPGE current $j_y \propto B_x \tilde{P}_{\rm lin}$ formation at intersubband transitions. $\Omega_{ph}$ is the optical phonon frequency.
	}
	\label{fig3t}
\end{figure}

\subsection{Infrared radiation induced  {linear and circular}  MPGE currents}
\label{dis_MPGE_intersubband}

{The application of an in-plane magnetic field ${\bm B \parallel x}$ leads to new photocurrent contributions, which belong to the class of magneto-gyrotropic photogalvanic effects~\cite{35,Belkov2005}. Figures~\ref{Fig1},~\ref{Fig3}(a), and \ref{Fig2} demonstrate that the magnetic field results in the photocurrents being odd in magnetic field $\bm B$ and excited by linearly as well as circularly polarized infrared radiation. 
}

Figure~\ref{Fig1} (blue and olive squares at $\pm B_x$) shows that the magnetic field induced current in the $y$-direction excited by a linearly polarized infrared radiation is characterized by the same azimuthal angle dependence as the photocurrent at zero magnetic field: $j_y(B_x) \propto \tilde{P}_{\rm lin} \propto \sin(2\alpha)$. Analyzing Eq.~\eqref{jyphenomen}  we find  that this  indicates that the magnetic field induced current is caused by the linear MPGE. Note that another possible magnetic field induced photocurrent caused by the photon drag is not detected in the experiments discussed, because it would lead to $j_y(B) \propto P_{\rm lin} \propto \cos(2\alpha)$, and, consequently, to a phase shift in the azimuthal angle dependence. As follows from the microscopic theory presented in Sec.~\ref{MPGE_intersubband}, the linear MPGE is described by Eqs.~\eqref{MPGEinter} and~\eqref{PhiMPGE}.
The generation of the linear MPGE current at intersubband transitions is illustrated in Fig.~\ref{fig3t}. 
For any value of the photon energy $\hbar\omega > 2\Delta_2$, optical transitions are possible for hole states in the upper subband with wavevectors $\pm k_z$
that satisfy the energy conservation law
\[\hbar\omega = 2\sqrt{\Delta_2^2 + (\beta k_z)^2}.
\]
The in-plane wavevector can be arbitrary, and there are pairs of hole states with a fixed $k_x$ and $\abs{k_y}$ that differ by the sign of $k_y$.
These transitions result in the depopulation of initial states $i1$ and $i2$ and the population of the final states $f1$ and $f2$  by photoexcited holes. As a result, four elementary currents are generated, two in the upper subband and two in the lower subband. This is indicated by horizontal blue arrows.  As shown in Sec.~\ref{MPGE_intersubband},  the transitions $i1 \to f1$ and $i2 \to f2$ have different probabilities, which depend on the magnetic field strength and direction. In Fig.~\ref{fig3t} this difference is illustrated by a thick red downward arrow for transitions from state $i2$ and a thin dashed arrow for transitions from state $i1$. As a result the balance between the population of states $i1$ and $i2$ and $f1$ and $f2$ is violated and the corresponding currents have different strength, which is illustrated by sketching the dominating currents by solid arrows. Furthermore, the magnitude of the	contribution of each state to the total current is determined by the momentum relaxation time of the corresponding states. 
Holes in the final states of optical transitions have more effective momentum relaxation due to the emission of optical phonons ($\hbar\Omega_{\rm ph}=11$~meV in Te \cite{Averkiev_1984_theory}), see bent dotted curves in Fig.~\ref{fig3t}, than in the initial states ($\tau_i > \tau_f$). As a result, the contributions of photoexcited holes to the
current is smaller,
and the current in Fig.~\ref{fig3t} is dominated by the contribution $j_{i2}$ caused by the depopulation of the initial state $i2$.
The frequency dependence of this mechanism is the same as for LPGE, Eq.~\eqref{chi}: $\Phi_{l,c} \propto K(\omega)/\omega^2$.
In the investigated frequency range this gives rise to a weak frequency dependence, which agrees with the experimental data, see curves~2 and~3 in the inset of Fig.~\ref{Fig3}(a), respectively.

{The experimentally observed  magnetic field induced circular photocurrent $J_y\propto P_{\rm circ} B_x$, see Figs.~\ref{Fig2}(b,c) and \ref{Fig3}(a), can also be explained by the shift mechanism in the developed theory. 
It shows that such photocurrent is caused by the photogalvanic effect, see Eqs.~\eqref{jyphenomen},~\eqref{jycircMPGE}, and~\eqref{Phi_c_MPGE}. The mechanism of its formation is similar to that of the linear PGE. The only difference is that the probability for optical transitions at positive and negative $k_y$, see Fig.~\ref{fig2t}(b), 
depends on the radiation helicity. Consequently, the resulting current reverses its sign upon reversal of the magnetic field or/and switching polarization from $\sigma^+$ to $\sigma^-$ and vice versa. The latter is described by Eqs.~\eqref{jycircMPGE} and~\eqref{Phi_c_MPGE} and detected experimentally, see Figs.~\ref{Fig2}(b) and~(c).}

\subsection{Terahertz radiation induced trigonal LPGE and photon drag currents}
\label{dis_trigonal_intraband} 

As discussed above the photocurrents excited by infrared radiation are caused by intersubband optical transitions. At terahertz frequencies the corresponding mechanisms are inapplicable, because the photon energies are too small to fulfill the energy conservation law and only indirect optical transitions can be responsible for the radiation absorption (Drude-like mechanism) and the photocurrent generation. In this spectral range the azimuthal angle dependence of the zero-$B$ photocurrent is observed to be phase shifted as compared to the one discussed for the LPGE, see Fig.~\ref{Fig4}(a). This observation indicates that the THz radiation induced photocurrent is caused by the superposition of the linear photogalvanic ($j_y \propto \tilde{P}_{\rm lin} \propto \sin(2\alpha)$) and linear photon drag effects ($j_y \propto P_{\rm lin} \propto \cos(2\alpha)$), see Eq.~\eqref{jyphenomen}. As shown in Sec.~\ref{linear_polarization} both effects contribute almost equally to the total current. 

The microscopic theory developed in Sec.~\ref{trigonal_intraband} shows that the linear PGE is caused by asymmetric scattering of holes driven by the THz electric field. The corresponding current is described by Eqs.~\eqref{jtrigonal} 
and~\eqref{chi_1} which, for the case relevant to our experimental conditions, Boltzmann statistics, and $\omega\tau \gg 1$, takes the form of Eq.~\eqref{chi_Boltzman}. The frequency dependence of the trigonal PGE follows that of the Drude absorption and is for $\omega \tau > 1$  given by $j_y\propto 1/\omega^2$. This is in full agreement with the experiment, see fits in Fig.~\ref{Fig3}(b). The model of the LPGE due to intraband transition is similar to that presented previously for bulk or surface states of other materials having  trigonal symmetry. It has been discussed in detail in works aimed at THz radiation induced LPGE in 
GaN quantum wells~\cite{Weber2007} and surface states of BiSbTe-based 3D topological insulators~\cite{Olbrich2014}.  
Therefore, it is applicable to the description of photocurrent in Te, and will not be presented here. 

The microscopic theory of the chiral linear photon drag effect is developed in Sec.~\ref{PDE_intraband}. It confirms that the PGE and PDE photocurrents in Te are characterized by $\pi/2$-shifted azimuthal angle dependencies. The chiral PDE current is described by Eqs.~\eqref{j_PDE}. Similar to PGE its frequency dependence is given by that of Drude absorption, $q_z\tilde{T} \propto 1/\omega^2$ at $\omega\tau\gg 1$, see Eq.~\eqref{eq_tilde_T}, which is in agreement with the experiment, see fits in Fig.~\ref{Fig3}(b).  The trigonal PDE current has previously been demonstrated for 2D surface  states of BiSbTe 3D topological insulators~\cite{Plank2016,Plank2018}. The model developed to describe it is applicable to the chiral PDE in Te, the only difference with respect to the trigonal PDE in 2D systems is that it is based on scattering by 2D triangle wedges whereas in Te the wedges are three dimensional (top view triangle but atoms are shifted along $z$-axis).  
To avoid repetition this model will not be presented in this paper.

{We estimate the chiral PDE current: $J_{\rm PDE}/P=(2\pi \mathcal S/cn_\perp)\tilde{T}q_z$, where $\mathcal S$ is the sample aspect ratio. From Eq.~\eqref{eq_tilde_T} we obtain $\tilde{T}\approx 10^2 pe^3\xi_w/[\omega^3(k_{\rm B}T)^2]$. The wedge scattering probability is obtained in the 4th order of the $\bm k \cdot \bm p$ perturbation theory and beyond the Born approximation, therefore we have an estimate ${\xi_w \sim S_{\rm nB}\abs{LL'}^2(k_{\rm B}T/\hbar E_g)^4}$. Here $L'$ is the interband momentum matrix element fixing the chirality of Te ($\abs{L'}\ll \abs{L}$), and $S_{\rm nB}$ is the non-Born dimensionless parameter, cf. Eq.~\eqref{chi_Boltzman}. This gives an estimate
\begin{equation}
{J_{\rm PDE}\over P} \approx 10^2{2  \pi\mathcal S pe^3\over c^2 \omega^2}S_{\rm nB} \abs{LL'}^2 {(k_{\rm B}T)^2\over (\hbar E_g)^4}.
\end{equation}
For $\mathcal S = 0.15$, room temperature, $f=1$~THz,
$E_g=335$~meV,  $L=3.3$~eV\AA~\cite{Glazov2022}, and taking $\abs{L'}=0.1\abs{L}$ and $S_{\rm nB}=10^{-2}$, we obtain 
$J_{\rm PDE}/P \sim 1$~nA/W which coincides by an order of magnitude with the measured values. 
}

\subsection{Terahertz radiation induced {linear and circular} MP\textit{G}E and linear MP\textit{D}E currents}
\label{dis_MPGE_intraband}

Finally, we discuss the magnetic field induced photocurrent excited by THz radiation. The analysis of the experimental data in the light of phenomenological theory shows that the magneto-photocurrent induced by linearly polarized radiation is given by 
$$j_y = (\Phi_l \tilde P_{\rm lin} - S_lq_z P_{\rm lin})B_x\abs{\bm E}^2,$$ 
see Sec.~\ref{linear_polarization}, Fig.~\ref{Fig3}, and Eq.~\eqref{jyphenomen}. As a result, the MPGE has  about 2.5 larger magnitude ($\Phi_l$) as compared to the MPDE one ($S_lq_z$).

MPGE  is shown microscopically to be  based on the terms in the scattering probability, which are linear in both the wavevector $\bm k$ and the magnetic field $\bm B$, Eq.~\eqref{W_kB}. These terms caused by the change of the scattering rates by the magnetic field, result in the photocurrent driven by  linear polarizations, see 
 Eqs.~\eqref{jLMPGE},~\eqref{Phi_l}, and the photocurrent driven by circularly polarized radiation, see Eqs.~\eqref{jMCPGE},~\eqref{Phi_c}, also detected in THz experiments, see Fig. \ref{Fig3}(c). The magneto-photogalvanic current is caused by the interband mixing of states by the magnetic field, see the interband matrix element, Eq.~\eqref{interband_m_field}. The $\bm k_\perp \cdot \bm B_\perp$ terms~\eqref{W_kB} are specific for the $D_3$ symmetry of Te because they result in the MPGE currents $j_y \propto B_x \tilde{P}_l$, $j_y \propto B_x P_{\rm circ}$ perpendicular to the magnetic field, while in structure inversion-asymmetric 2D systems (quantum wells, graphene on a substrate etc.)~\cite{Drexler2013,GanichevWeissEromsAnnPhys2017}, similar mechanism gives MPGE photocurrents parallel to the magnetic field at the same polarizations.

The magneto-photon drag effect driven by linearly polarized radiation (linear MPDE) is caused by the simultaneous action of the spatially-varying electric field
of radiation and the Lorentz force, see Eqs.~\eqref{MPDE1} and \eqref{MPDE2}. A major contribution of this photocurrent is manifested in THz experiments as a phase shift in the azimuthal angle dependence, see Fig.~\ref{Fig4}(b) and (c)  and the inset in this figure. The linear MPDE current is allowed in systems of any symmetry. Note that the circular magneto-photocurrent solely due to the magneto-photogalvanic 
effect and the circular MPDE are forbidden by symmetry in the investigated
experimental arrangement, see Eq.~\eqref{jyphenomen}.

Finally, we note that the frequency dependence of all magnetophotocurrents induced by THz radiation is given by that of the Drude absorption, see Eqs.~\eqref{j_MLPDE},~\eqref{MLPDE_constant}, which is in agreement with experimental results (not shown).

\section{summary}
\label{summary} 

Our studies of photocurrents excited by infrared/THz radiation show that depending on the experimental conditions, the radiation induces a series of phenomena caused by photogalvanics and, in THz range, the transfer of linear photon momentum to free carriers. A rich palette of the microscopic mechanisms of the observed photocurrents  is due to the different photocurrent roots depending on the kind of the optical transitions responsible for the radiation absorption,  polarization state, the presence of an external magnetic field, and the transfer of linear photon momentum during absorption of radiation. All observed photocurrents can be described in terms of the developed phenomenological and microscopic theories, taking into account a shift of the electron wavepacket in real space or/and asymmetric scattering in a system without inversion symmetry. The main mechanisms resulting in the observed photocurrents are summarized in Table~\ref{table}. We believe that the results of this study will be useful in future studies of novel Te based materials such as tellurene and Weyl fermions with surface Fermi arcs, which are expected in Te crystals under high pressure.

\section{Acknowledgments}
\label{acknow} 
We thank V. A. Shalygin, K. v.~Klitzing,  I.~Gronwald and I. V.  Sedova for fruitful discussions. The support from the Deutsche Forschungsgemeinschaft (DFG, 
project Ga501/19), and the Volkswagen Stiftung Program (97738) is gratefully acknowledged.  

\appendix

\section{Phenomenology of PGE and PDE currents at normal incidence}
\label{appendix2}

Tellurium is a chiral semiconductor with the point symmetry group $D_3$. We use the coordinate system $(x,y)$ where $x$ is parallel to one of three $C_2$ rotation axes.
We study photocurrents at light propagation along $z \parallel \bm c$ when the unit vector in the light propagation direction $\hat{\bm e} \parallel z$.
The PGE current components in the presence of the magnetic field $\bm B$ 
are given by
\begin{widetext}
\begin{align}
& j_x =\qty[ \chi P_{\rm lin} + \Phi B_z\tilde{P}_{\rm lin}  + \Phi_l (B_xP_{\rm lin} + B_y\tilde{P}_{\rm lin})+ \Phi_c B_y P_{\rm circ}\hat{e}_z +\Lambda B_x]\abs{\bm E}^2,
\\ & j_y =\qty[ -\chi \tilde{P}_{\rm lin} + \Phi B_z{P}_{\rm lin} + \Phi_l(B_x\tilde{P}_{\rm lin} - B_y{P}_{\rm lin}) - \Phi_c B_x P_{\rm circ}\hat{e}_z +\Lambda B_y]\abs{\bm E}^2,
\\& j_z = \qty[\gamma P_{\rm circ}\hat{e}_z + \Phi' (B_y P_{\rm lin} + B_x \tilde{P}_{\rm lin})+\tilde{\Lambda} B_z]\abs{\bm E}^2 .
\end{align}
\end{widetext}
Here the zero-field current $j_z= \gamma P_{\rm circ}\abs{\bm E}^2$ is the longitudinal CPGE and  $j_+ =\chi E_-^2$ is the trigonal LPGE. 
The polarization-independent components along the magnetic field $j_{x,y}\propto {\Lambda} B_{x,y}$, $j_z\propto \tilde{\Lambda} B_z$ and polarization-dependent currents $j_+=\Phi_l B_- E_+^2$, $\bm j_\perp=\Phi_c (\bm B \times \bm \varkappa) \abs{\bm E}^2$ are due to an absence of reflection planes in the $D_3$ group ($\bm \varkappa = P_{\rm circ}\hat{\bm e}$ is the photon angular momentum).
The currents $j_+=\Phi i B_z E_-^2$ and $j_z=\Phi' i(B_-E_-^2 - B_+E_+^2)/2$ describe the trigonal MPGE current which requires both absence of reflections and trigonality.

The PDE current which, by definition, accounts for $q_z \neq 0$ reads
\begin{widetext}
\begin{align}
& j_x =\qty[ \tilde{T}\tilde{P}_{\rm lin} + S B_z P_{\rm lin}  + S_l (B_x\tilde{P}_{\rm lin}+B_yP_{\rm lin})+ S_c B_x P_{\rm circ}\hat{e}_z + RB_y]q_z\abs{\bm E}^2,
\\ & j_y =\qty[ \tilde{T} P_{\rm lin} - S B_z \tilde{P}_{\rm lin} - S_l(B_x P_{\rm lin} + B_y\tilde{P}_{\rm lin}) + S_c B_y P_{\rm circ}\hat{e}_z  - RB_x]q_z \abs{\bm E}^2,
\\ & j_z = \qty[T + \tilde{S}B_z P_{\rm circ}\hat{e}_z + S' (B_x P_{\rm lin} - B_y \tilde{P}_{\rm lin})]q_z \abs{\bm E}^2.
\end{align}
\end{widetext}
The polarization-independent PDE currents $j_z=Tq_z\abs{\bm E}^2$, $\bm j_\perp = R(\bm B \times \bm q) \abs{\bm E}^2$ and helicity-dependent magneto-induced currents $j_z=\tilde{S}B_z\varkappa_zq_z\abs{\bm E}^2$, $\bm j_\perp = S_c \bm B_\perp \varkappa_z q_z\abs{\bm E}^2$, $j_+=S_l i q_z B_- E_+^2$ are allowed in any symmetry.
The currents $j_+=\tilde{T} i q_z E_-^2$,
$j_+=S B_zE_-^2$ and
$j_z=S'(B_+ E_+^2 + B_-E_-^2)q_z\abs{\bm E}^2/2$  
 are allowed in $D_3$ symmetry due to both trigonality and absence of reflection planes.

\section{
Intersubband absorption coefficient}
\label{Abs_intersubband}

The absorption coefficient for direct optical transitions $K(\omega)$ is defined as follows
\begin{align}
	& {K(\omega){I} \over \hbar \omega} 
	\\ 
	& = 2 \sum_{\bm k}  {2\pi\over \hbar} \abs{V_{21}}^2 \delta\qty[2\sqrt{\Delta_2^2+(\beta k_z)^2}-\hbar\omega](f_1-f_2). \nonumber
\end{align}
Here the factor 2 accounts for two valleys of tellurium, ${I}=c n_\perp \abs{\bm E}^2/(2\pi)$  is the light intensity, $f_{1,2}$ are occupations of the initial and final states, and the optical matrix element is given by Eq.~\eqref{V_12}.
For Boltzmann statistics we have $f_1-f_2=[1-\exp(-\hbar\omega/k_{\rm B}T)]f_1$. Since the hole concentration $p=2\sum_{\bm k}f_1$, we obtain
\begin{multline}
	K(\omega) ={8\pi^2 e^2\abs{L}^4\Delta_2^2 \sqrt{\abs{\mathcal A_1} k_{\rm B}T} \over \hbar c E_g^4 \hbar \omega  \abs{\mathcal A_2 \beta}}  p 
	{1-\exp(-\hbar\omega/k_{\rm B}T) \over J_1\sqrt{1-(2\Delta_2/\hbar\omega)^2}}
	\\
	\times {\exp{-{\abs{\mathcal A_1}[(\hbar\omega)^2-(2\Delta_2)^2] \over 4\abs{\beta} k_{\rm B}T} + {\hbar\omega/2-\Delta_2 \over k_{\rm B}T}} 
	},
\end{multline}
where $J_1$ was introduced in Ref.~\cite{Averkiev_1984_theory}:
\begin{equation}
	J_1=\int\limits_0^\infty \dd x \exp[-x^2-{\Delta_2 \over k_{\rm B}T}+\sqrt{\qty({\Delta_2 \over k_{\rm B}T})^2+{x^2\beta^2 \over \abs{\mathcal A_1} k_{\rm B}T}}].
\end{equation}

\section{Microscopic calculation of PDE current}
\label{PDE_micro}

We solve iteratively the kinetic Eq.~\eqref{kin_eq_PDE}.
First, we find the linear in $\bm E$ correction to the distribution function:
\begin{equation}
\label{f_E}
f^{(E)} = e\tau_\omega(-f_0') (\bm E_\perp \cdot \bm v_\perp).
\end{equation}
Then, depending on the contribution under study, we iterate the kinetic equation in small parameters $q_z$, $\bm E_\perp^*$ or $\tilde{\bm B}_\perp$.

Account for $q_z$ yields the correction $f^{(Eq)}$ which satisfies the following kinetic equation:
\begin{equation}
-i\omega f^{(Eq)} + iq_z v_z f^{(E)} = - {f^{(Eq)}\over \tau}.
\end{equation}
It yields
\begin{equation}
\label{f_E_q}
f^{(Eq)} = -iq_z v_z \tau_\omega f^{(E)} = i e q_z \tau_\omega^2 f_0' v_z (\bm E_\perp \cdot \bm v_\perp).
\end{equation}

If we then account for the second power of $\bm E_\perp$, we find the time-independent correction $f^{(EqE)}$:
\begin{equation}
{e\over \hbar}\bm E_\perp^* \cdot \pdv{f^{(Eq)}}{\bm k_\perp} + c.c. = 
-{f^{(EqE)} \over \tilde{\tau}}.
\end{equation}
Here, depending on the anisotropic/isotropic part of the filed term in the left-hand side of this equation, the relaxation time $\tilde{\tau}$ equals to either $\tau$ or to the energy relaxation time $\tau_\varepsilon$.
Splitting the field term to the isotropic and anisotropic in $\bm k$ parts, we obtain
\begin{align}
\label{f_E_q_E}
f^{(EqE)} =  &2 e^2 q_zv_z
{\rm Im}\biggl\{ \tau \qty(\tau_\omega^2 f_0')' \qty(\abs{\bm E_\perp \cdot \bm v_\perp}^2 - \abs{\bm E_\perp}^2 {v_\perp^2\over 2}) \nonumber
\\ & + \tau_\varepsilon \abs{\bm E_\perp}^2 \qty[{v_\perp^2\over 2}\qty(\tau_\omega^2 f_0')'  + {\tau_\omega^2 f_0' \over m}]
\biggr\}.
\end{align}
For linearly polarized radiation we can rewrite this expression as follows
\begin{align}
&f^{(EqE)} =e^2 q_zv_z\abs{\bm E_\perp}^2
\nonumber
\\ &
\times {\rm Im}\biggl\{ \tau \qty(\tau_\omega^2 f_0')' \qty[2v_xv_y P_{\rm lin} + (v_x^2-v_y^2)\tilde{P}_{\rm lin}] \nonumber
\\ & + \tau_\varepsilon \qty[{v_\perp^2\over 2}\qty(\tau_\omega^2 f_0')'  + {\tau_\omega^2 f_0' \over m}]
\biggr\}.
\end{align}

Let us now take into account the last perturbation, the scattering by wedges, and find the current-carrying part of the distribution function $f^{(EqEw)}$. The asymmetric scattering enters the kinetic equation as an incoming term:
\begin{equation}
\sum_{\bm k'} W^w_{\bm k \bm k'} f^{(EqE)}_{\bm k'} = {f^{(EqEw)}_{\bm k}\over \tau}.
\end{equation}
Solving this algebraic equation, we calculate the $EqEw$ contribution to the PDE current which flows in the $(xy)$ plane in this geometry:
\begin{align}
&\bm j_\perp^{(EqEw)} = 2e^3q_z\abs{\bm E_\perp}^2 \sum_{\bm k} \bm v_\perp \tau^2 \qty[{\rm Im}(\tau_\omega^2) f_0']'
 \sum_{\bm k'} W^w_{\bm k \bm k'}\nonumber \\ 
&\times v_z(\bm k') \qty{2v_x(\bm k')v_y(\bm k') P_{\rm lin} + [v_x^2(\bm k')-v_y^2(\bm k')]\tilde{P}_{\rm lin} }.
\end{align}
Here we took into account that the isotropic part of $f^{(EqE)} \propto \tau_\varepsilon$ does not contribute to the current because in $D_3$ symmetry
\begin{equation}
\label{zero_xi}
\expval{\sum_{\bm k'} W^w_{\bm k \bm k'} v_z(\bm k') \bm v_\perp(\bm k)}=0.
\end{equation}
The nonzero average allowed in $D_3$ point symmetry group is the following dimensionless value
\begin{align}
\xi_w
= \tau \expval{\sum_{\bm k'} W^w_{\bm k \bm k'} v_z(\bm k') v_y(\bm k) [v_x^2(\bm k')-v_y^2(\bm k')]}  \nonumber 
\\= 2\tau \expval{\sum_{\bm k'} W^w_{\bm k \bm k'} v_z(\bm k') v_x(\bm k) v_x(\bm k')v_y(\bm k')}.
\end{align}
Then we obtain a contribution to the PDE constant $\tilde{T}$, Eq.~\eqref{j_PDE}, which reads
\begin{equation}
\tilde{T}_{EqEw} = 2e^3\sum_{\bm k} \tau \xi_w \qty[{\rm Im}(\tau_\omega^2) f_0']'.
\end{equation}
Since the density of states is $\propto \sqrt{\varepsilon_k}$, we can rewrite this expression as
\begin{equation}
\tilde{T}_{EqEw} = -2e^3\sum_{\bm k} {(\xi_w\tau \sqrt{\varepsilon_k})'\over\sqrt{\varepsilon_k}} {\rm Im}(\tau_\omega^2) f_0'.
\end{equation}

Now we calculate $EqwE$ contribution to the PDE current. We find the time-dependent correction $f^{(Eqw)}$:
\begin{equation}
\sum_{\bm k'} W^w_{\bm k \bm k'} f^{(Eq)}_{\bm k'} = {f^{(Eqw)}_{\bm k} \over \tau_\omega}.
\end{equation}
The solution of this equation reads
\begin{equation}
f^{(Eqw)}_{\bm k} = i e q_z \tau_\omega^3 f_0' \tau \sum_{\bm k'} W^w_{\bm k \bm k'} v_z(\bm k') [\bm E_\perp \cdot \bm v_\perp(\bm k')].
\end{equation}
The static correction $f^{(EqwE)}$ yielding the contribution to the current satisfies the equation
\begin{equation}
{e\over \hbar}\bm E_\perp^* \cdot \pdv{f^{(Eqw)}}{\bm k_\perp} + c.c. = 
-{f^{(EqwE)} \over \tau}.
\end{equation}
Solving this equation we get 
the current 
\begin{equation}
\bm j^{(EqwE)} = 2e^2\sum_{\bm k} f^{(Eqw)}\qty[{\tau \over m}\bm E^* + \tau' \bm v \qty(\bm E_\perp^* \cdot \bm v_\perp)] + c.c.
\end{equation}
Substituting here $f^{(Eqw)}$
we see that the first term here does not contribute because $f^{(Eqw)}$ is zero on average, but the second term yields a contribution to the current~\eqref{j_PDE} with
\begin{equation}
\tilde{T}_{EqwE} = -2e^3\sum_{\bm k} \tau' \tilde{\xi}_w {{\rm Im}(\tau_\omega^3)\over \tau} f_0'.
\end{equation}
Here we introduced another wedge-scattering efficiency constant $\tilde{\xi}_w$ which is linearly-independent of $\xi_w$:
\begin{align}
\tilde{\xi}_w =  \tau \expval{\sum_{\bm k'} W^w_{\bm k \bm k'} v_z(\bm k') v_y(\bm k') [v_x^2(\bm k)-v_y^2(\bm k)]} \nonumber \\
=2\tau \expval{\sum_{\bm k'} W^w_{\bm k \bm k'} v_z(\bm k') v_x(\bm k') v_x(\bm k)v_y(\bm k)}.
\end{align}

Let us turn now to the $EwqE$ contribution to the PDE current. For its calculation, we account for wedge scattering in the kinetic equation at the second step and find the correction $f^{(Ew)}$:
\begin{equation}
\sum_{\bm k'} W^w_{\bm k \bm k'}f^{(E)}_{\bm k'} = {f^{(Ew)} \over \tau_\omega}.
\end{equation}
Solution of this equation reads
\begin{equation}
f^{(Ew)} = -e\tau_\omega^2 f_0' \sum_{\bm k'} W^w_{\bm k \bm k'}[\bm E \cdot \bm v(\bm k')].
\end{equation}
Then we should take into account $q_z$ and then $\bm E$. In the next step we find $f^{(Ewq)}$ which satisfies
\begin{equation}
iq_zv_zf^{(Ew)} = - {f^{(Ewq)}\over \tau_\omega},
\end{equation} 
and we get
\begin{equation}
f^{(Ewq)} = iq_zv_ze\tau_\omega^3 f_0'\sum_{\bm k'} W^w_{\bm k \bm k'}[\bm E \cdot \bm v(\bm k')].
\end{equation}
Then we search for the correction $f^{(EwqE)}$:
\begin{equation}
{e\over \hbar}\bm E_\perp^* \cdot \pdv{f^{(Ewq)}}{\bm k_\perp} + c.c. = 
-{f^{(EwqE)} \over \tau}.
\end{equation}
It allows for calculation of the corresponding contribution to the PGE current:
\begin{equation}
\bm j_\perp^{(EwqE)} = 2e\sum_{\bm k} \bm v_\perp f^{(EwqE)}.
\end{equation}
Substituting $f^{(EwqE)}$ and integrating by parts we obtain
\begin{equation}
\bm j_\perp^{(EwqE)} = 2e^2 \sum_{\bm k} \bm v \tau' (\bm E^*\cdot \bm v)  f^{(Ewq)} + c.c.
\end{equation}
Substituting here $f^{(Ewq)}$ we get
\begin{align}
&\bm j_\perp^{(EwqE)} \nonumber
\\
&= 2iq_ze^3 \sum_{\bm k \bm k'}  \tau' \tau_\omega^3 f_0'\bm v(\bm E^*\cdot \bm v) v_z  W^w_{\bm k \bm k'}[\bm E \cdot \bm v(\bm k')]+ c.c.
\end{align}
This yields a contribution to the PDE constant $\tilde{T}$:
\begin{equation}
\tilde{T}_{EwqE} = 2e^3\sum_{\bm k} \tau' \xi_w {{\rm Im}(\tau_\omega^3)\over \tau} f_0'.
\end{equation}

Now we put $q_z=0$ and take into account the radiation magnetic field $\tilde{\bm B}$.
If we do this after accounting for $E$ and $w$, then we get a steady-state correction $f^{(Ew\tilde{B})}$ which satisfies the equation
\begin{equation}
{e\over \hbar c} (\bm v \times \tilde{\bm B}^*)\cdot \pdv{f^{(Ew)}}{\bm k} + c.c.= -{f^{(Ew\tilde{B})} \over \tau}.
\end{equation}
Solving this algebraic equation and calculating the current $\bm j^{(Ew\tilde{B})} = 2e\sum_{\bm k} \bm v_\perp f^{(Ew\tilde{B})}$, integrating by parts we get
\begin{equation}
\bm j^{(Ew\tilde{B})} = {2e^2\over mc}\sum_{\bm k} \tau (\bm v \times \tilde{\bm B}^*) f^{(Ew)}.
\end{equation}
Substituting $f^{(Ew)}$ we see that this contribution is zero because it is proportional to an average~\eqref{zero_xi}.

Finally, we have one more contribution to the PDE current coming from the correction to the distribution function $f^{(E\tilde{B}w)}$ obtained by account for $E$, $\tilde{\bm B}_\perp$ and then $w$. 
It is found from the equation
\begin{equation}
\sum_{\bm k'}W^w_{\bm k \bm k'}f^{(E\tilde{B})}_{\bm k'} = {f^{(E\tilde{B}w)} \over \tau},
\end{equation}
where the steady-state correction $f^{(E\tilde{B})}$ is given by
\begin{equation}
f^{(E\tilde{B})} = {e^2\over m c} v_z \tau \tau_\omega f_0' (\tilde{\bm B}^*\times \bm E) + c.c.
\end{equation}
We see that $f^{(E\tilde{B})} \propto q_z\abs{\bm E}^2$, i.e. it is polarization-independent. Moreover, this contribution to the PDE current is zero because it is also proportional to the value~\eqref{zero_xi}.

We obtained that the PDE current is described by the phenomenological Eq.~\eqref{j_PDE} where the constant $\tilde{T}$ is given by
\begin{equation}
\tilde{T}=\tilde{T}_{EqEw}+\tilde{T}_{EqwE}+\tilde{T}_{EwqE},
\end{equation}
which yields Eq.~\eqref{tilde_T} from the main text.

\section{Microscopic calculation of  MLPDE current}
\label{App_MPDE}

We consider the geometry with $\bm q \parallel z$, when the electric and magnetic fields of the radiation, $\bm E_\perp$ and $\tilde{\bm B}_\perp$ are perpendicular to the $z$ axis, and the external magnetic field  $\bm B \parallel x$.
The hole distribution function $f(\bm k)$ satisfies the Boltzmann kinetic equation where the field term contains the forces of the radiation electric field and two Lorentz forces -- of the static magnetic field $B_x$ and the ac radiation magnetic field $\tilde{\bm B}_\perp$:
\begin{align}
\pdv{f}{t} + iq_z v_z f &+ {e\over \hbar}\bm E_\perp(t) \cdot \pdv{f}{\bm k_\perp} \nonumber
\\ 
&+ {e\over \hbar c} [\bm v \times (\bm B+\tilde{\bm B}_\perp(t))]\cdot \pdv{f}{\bm k} =  {\rm St}[f]. 
\end{align}
Here $\tilde{\bm B}_\perp(t) = \tilde{\bm B}_\perp \exp(-i\omega t) + c.c.$, ${\rm St}[f]$ stands for the elastic collision integral describing isotropization of the distribution over the isoenergetic surface $\varepsilon_k={\rm const}$.
In what follows we account either for $q_z \neq 0$ or for $\tilde{B}\neq 0$ in the kinetic equation because they give contributions to the the PDE currents of the same order.

First, we find the linear in $\bm E$ correction to the distribution function~\eqref{f_E}.
Then, depending on the contribution under study, we iterate the kinetic equation in small parameters $q_z$, $\bm E_\perp^*$ or $\tilde{\bm B}_\perp$.

Account for $q_z$ yields the correction $f^{(Eq)}$~\eqref{f_E_q}. 
If we then account for the second power of $\bm E_\perp$, we find the time-independent correction $f^{(EqE)}$~\eqref{f_E_q_E}.
%
Let us now take into account the last perturbation, the magnetic field $B_x$, and find the current-carrying part of the distribution function $f^{(EqEB)}$. The magnetic field enters the kinetic equation via the Lorentz force:
\begin{equation}
{e\over \hbar c} (\bm v \times \bm B)\cdot \pdv{f^{(EqE)}}{\bm k} = -{f^{(EqEB)} \over \tau}.
\end{equation}
The solution reads
\begin{equation}
\label{f_EqEB}
f^{(EqEB)} = -{m \omega_c \tau \over \hbar} \qty(v_z \partial_{k_y} -v_y \partial_{k_z})f^{(EqE)},
\end{equation}
where $\omega_c = eB_x/(m c)$ is the cyclotron frequency.
Then we calculate the $EqEB$ contribution to the MLPDE current which flows in the $(xy)$ plane in this geometry:
\begin{equation}
\bm j_\perp^{(EqEB)} = 2e\sum_{\bm k} \bm v_\perp f^{(EqEB)}.
\end{equation}
Substituting $f^{(EqEB)}$ from Eq.~\eqref{f_EqEB} and integrating by parts we obtain that this contribution is zero:
\begin{equation}
\bm j_\perp^{(EqEB)} =0
\end{equation}
(the simplest way to see this is to calculate the $x$ component of the current).

Now we calculate $EqBE$ contribution to the MLPDE current. We find the time-dependent correction $f^{(EqB)}$:
\begin{equation}
{e\over \hbar c} (\bm v \times \bm B)\cdot \pdv{f^{(Eq)}}{\bm k} = -{f^{(EqB)} \over \tau_\omega}.
\end{equation}
The solution of this equation reads
\begin{equation}
f^{(EqB)} =-   i e q_z \omega_c \tau_\omega^3 f_0'\qty[(v_z^2-v_y^2) E_y- v_x v_y E_x] .
\end{equation}
The static correction $f^{(EqBE)}$ yielding the contribution to the current satisfies the equation
\begin{equation}
{e\over \hbar}\bm E_\perp^* \cdot \pdv{f^{(EqB)}}{\bm k_\perp} + c.c. = 
-{f^{(EqBE)} \over \tau}.
\end{equation}
Solving this equation we get for the $y$ component of the current 
\begin{equation}
j_y^{(EqBE)} = 2e^2\sum_{\bm k} f^{(EqB)}\qty[{\tau \over m}E_y^* + \tau' v_y \qty(\bm E_\perp^* \cdot \bm v_\perp)] + c.c.
\end{equation}
Substituting here $f^{(EqB)}$ and averaging over directions of $\bm k$ 
we obtain
\begin{align}
&j_y^{(EqBE)} = 2ie^3q_z \omega_c \\
&\times \sum_{\bm k}\tau_\omega^3 f_0'   \tau'  {v^4\over 15} (2\abs{E_y}^2+\abs{E_x}^2)+ c.c. \nonumber
\end{align}
Noting that 
$\abs{E_{x,y}}^2=\abs{\bm E}^2(1\pm P_{\rm lin})/2$,
and introducing $r=\dd \ln\tau/\dd \ln \varepsilon_k$,
we get 
a contribution to the polarization-dependent MLPGE current Eq.~\eqref{j_MLPDE}, $j_y^{(EqBE)}= - B_x q_z \abs{\bm E}^2 S_l^{(EqBE)} {P}_{\rm lin}$ where $S_l^{(EqBE)}$ is given by 
\begin{equation}
S_l^{(EqBE)} = -{8re^4\over 15m^3c} \sum_{\bm k} {\rm Im}(\tau_\omega^3) f_0'   \tau  \varepsilon_k.
\end{equation}
At $\omega \tau \gg 1$ we have ${\rm Im}(\tau_\omega^3)=-1/\omega^3$. This gives
\begin{equation}
S_l^{(EqBE)} = {8re^4 \over 15m^3 c \omega^3} \sum_{\bm k} f_0'   \tau  \varepsilon_k
= -{8r\Gamma(r+5/2) p e^4 \tau_T\over 15\sqrt{\pi}m^3 c \omega^3}.
\end{equation}

Let us turn now to the $EBqE$ contribution to the MLPDE current. For its calculation, we account for the Lorentz force in the kinetic equation at the second step and find the correction $f^{(EB)} \propto E_\perp B_x$:
\begin{equation}
{e\over \hbar c} (\bm v \times \bm B)\cdot \pdv{f^{(E)}}{\bm k} = -{f^{(EB)} \over \tau_\omega}.
\end{equation}
Solution of this equation reads
\begin{equation}
f^{(EB)} = e\omega_c\tau_\omega^2 f_0' E_y v_z.
\end{equation}
Then we should take into account $q_z$ and then $\bm E$. In the next step we find $f^{(EBq)}$ which satisfies
\begin{equation}
iq_zv_zf^{(EB)} = - {f^{(EBq)}\over \tau_*},
\end{equation} 
where $\tau_*$ equals to $\tau_\omega$ or $i/\omega$ for the anisotropic/isotropic part of the left-hand side,
and we get
\begin{equation}
f^{(EBq)} = e\omega_c q_z E_y f_0' \qty[-i\tau_\omega^3 \qty(v_z^2 - {v^2 \over 3}) + {\tau_\omega^2\over \omega} {v^2\over 3}].
\end{equation}
Then we search for the correction $f^{(EBqE)}$:
\begin{equation}
{e\over \hbar}\bm E_\perp^* \cdot \pdv{f^{(EBq)}}{\bm k_\perp} + c.c. = 
-{f^{(EBqE)} \over \tau}.
\end{equation}
It allows for calculation of the corresponding contribution to the MLPGE current:
\begin{equation}
\bm j_\perp^{(EBqE)} = 2e\sum_{\bm k} \bm v_\perp f^{(EBqE)}.
\end{equation}
Integration by parts and averaging over directions of $\bm k$ yields
\begin{equation}
\bm j_\perp^{(EBqE)} = 2e^2 \bm E^*\sum_{\bm k} \qty({\tau\over m}+\tau' {v^2\over 3})  f^{(EBq)} + c.c.
\end{equation}
We see that only the isotropic part of $f^{(EBq)}$ which contains a factor $1/\omega$ contributes to the current. 
Using $\tau'=r\tau/\varepsilon_k$, we obtain
\begin{equation}
\bm j_\perp^{(EBqE)} = {4e^3 \omega_c q_z \over 3 m^2 \omega} \qty(1+{2r\over 3})\bm E^*E_y\sum_{\bm k} f_0' \varepsilon_k \tau \tau_\omega^2 + c.c.
\end{equation}
In particular, it means that the $EBqE$ contribution to $S_l$ is given by  
\begin{equation}
S_l^{(EBqE)} = {4e^4 \over 3 m^3 c \omega} \qty(1+{2r\over 3})\sum_{\bm k} f_0' \varepsilon_k \tau {\rm Re}(\tau_\omega^2).
\end{equation}

At $\omega\tau \gg 1$ when ${\rm Re}(\tau_\omega^2)=-1/\omega^2$ we have
\begin{equation}
S_l^{(EBqE)} = {4e^4 \over 3 m^3 c \omega^3} \qty(1+{2r\over 3})\sum_{\bm k} (-f_0') \varepsilon_k \tau .
\end{equation}
For Boltzmann statistics we get
\begin{equation}
S_l^{(EBqE)} = {4 p e^4 \tau_T \over 3 \sqrt{\pi} m^3 c \omega^3} \qty(1+{2r\over 3})\Gamma\qty(r+{5\over 2}).
\end{equation}

Now we put $q_z=0$ and take into account the radiation magnetic field $\tilde{\bm B}$.
If we do this after accounting for $E$ and $B_x$, then we get a steady-state correction $f^{(EB\tilde{B})}$ which satisfies the equation
\begin{equation}
{e\over \hbar c} (\bm v \times \tilde{\bm B}^*)\cdot \pdv{f^{(EB)}}{\bm k} + c.c.= -{f^{(EB\tilde{B})} \over \tau}.
\end{equation}
It solution reads
\begin{equation}
f^{(EB\tilde{B})} = {e^2\omega_c \tau \over mc}\tau_\omega^2 f_0' E_y \qty(\tilde{\bm B}^* \times \bm v)_z + c.c.
\end{equation}
Calculating the current $\bm j^{(EB\tilde{B})} = 2e\sum_{\bm k} \bm v_\perp f^{(EB\tilde{B})}$,
we obtain 
\begin{equation}
j_y^{(EB\tilde{B})} = {4e^3\omega_c \over 3m^2c} E_y\tilde{B_x^*}\sum_{\bm k} \varepsilon_k \tau_\omega^2 f_0'\tau+c.c.
\end{equation}
Using the relation between the radiation magnetic and electric fields $\tilde{\bm B}=c\bm q \times \bm E/\omega$ 
we obtain that this contribution yields $S_l$ in the form 
\begin{equation}
S_l^{(EB\tilde{B})} =- {4e^4 \over 3m^3 c\omega} \sum_{\bm k} \varepsilon_k {\rm Re}(\tau_\omega^2) f_0'\tau.
\end{equation}
At $\omega\tau \gg 1$
we have
\begin{equation}
S_l^{(EB\tilde{B})} = {4e^4 \over 3m^3 c\omega^3} \sum_{\bm k} \varepsilon_k  f_0'\tau.
\end{equation}
For Boltzmann statistics and $\tau(\varepsilon_k) \propto \varepsilon_k^r$ we get
\begin{equation}
S_l^{(EB\tilde{B})} = {4 p e^4 \over 3 \sqrt{\pi}m^3 c\omega^3} \Gamma\qty(r+{5\over 2}) \propto T^r.
\end{equation}

Finally, we have one more contribution to the MPDE current coming from the correction to the distribution function $f^{(E\tilde{B}B)}$ obtained by account for $E$, $\tilde{\bm B}_\perp$ and then $B_x$. 
It is found from the equation
\begin{equation}
{e\over \hbar c} (\bm v \times \bm B)\cdot \pdv{f^{(E\tilde{B})}}{\bm k} = -{f^{(E\tilde{B}B)} \over \tau},
\end{equation}
where the steady-state correction $f^{(E\tilde{B})}$ is given by
\begin{equation}
f^{(E\tilde{B})} = {e^2\over m c} v_z \tau \tau_\omega f_0' (\tilde{\bm B}^*\times \bm E)_z + c.c.
\end{equation}
Calculating the current density we get
\begin{equation}
 j_y^{(E\tilde{B}B)} = 2e\sum_{\bm k} v_y f^{(E\tilde{B}B)} = 2e\omega_c  \sum_{\bm k} v_z \tau f^{(E\tilde{B})}.
\end{equation}
Substitution of $f^{(E\tilde{B})}$ results in a polarization-independent contribution only.

For high frequencies $\omega\tau \gg 1$ we obtained Eq.~\eqref{MLPDE_constant}  where the pre-factor $a$ is a sum of various contributions summarized in Table~\ref{tab_MLPDE} of the main text.
For arbitrary frequencies, the MLPDE constant $S_l =S_l^{(EqBE)} +S_l^{(EBqE)} + S_l^{(EB\tilde{B})}$ is given by Eq.~\eqref{S_l_arb_freq} of the main text.

\bibliography{savedrecs}

\begin{thebibliography}{71}%
\makeatletter
\providecommand \@ifxundefined [1]{%
 \@ifx{#1\undefined}
}%
\providecommand \@ifnum [1]{%
 \ifnum #1\expandafter \@firstoftwo
 \else \expandafter \@secondoftwo
 \fi
}%
\providecommand \@ifx [1]{%
 \ifx #1\expandafter \@firstoftwo
 \else \expandafter \@secondoftwo
 \fi
}%
\providecommand \natexlab [1]{#1}%
\providecommand \enquote  [1]{``#1''}%
\providecommand \bibnamefont  [1]{#1}%
\providecommand \bibfnamefont [1]{#1}%
\providecommand \citenamefont [1]{#1}%
\providecommand \href@noop [0]{\@secondoftwo}%
\providecommand \href [0]{\begingroup \@sanitize@url \@href}%
\providecommand \@href[1]{\@@startlink{#1}\@@href}%
\providecommand \@@href[1]{\endgroup#1\@@endlink}%
\providecommand \@sanitize@url [0]{\catcode `\\12\catcode `\$12\catcode
  `\&12\catcode `\#12\catcode `\^12\catcode `\_12\catcode `\%12\relax}%
\providecommand \@@startlink[1]{}%
\providecommand \@@endlink[0]{}%
\providecommand \url  [0]{\begingroup\@sanitize@url \@url }%
\providecommand \@url [1]{\endgroup\@href {#1}{\urlprefix }}%
\providecommand \urlprefix  [0]{URL }%
\providecommand \Eprint [0]{\href }%
\providecommand \doibase [0]{https://doi.org/}%
\providecommand \selectlanguage [0]{\@gobble}%
\providecommand \bibinfo  [0]{\@secondoftwo}%
\providecommand \bibfield  [0]{\@secondoftwo}%
\providecommand \translation [1]{[#1]}%
\providecommand \BibitemOpen [0]{}%
\providecommand \bibitemStop [0]{}%
\providecommand \bibitemNoStop [0]{.\EOS\space}%
\providecommand \EOS [0]{\spacefactor3000\relax}%
\providecommand \BibitemShut  [1]{\csname bibitem#1\endcsname}%
\let\auto@bib@innerbib\@empty
\bibitem [{\citenamefont {Button}\ \emph {et~al.}(1969)\citenamefont {Button},
  \citenamefont {Landwehr}, \citenamefont {Bradley}, \citenamefont {Grosse},\
  and\ \citenamefont {Lax}}]{Button1969}%
  \BibitemOpen
  \bibfield  {author} {\bibinfo {author} {\bibfnamefont {K.~J.}\ \bibnamefont
  {Button}}, \bibinfo {author} {\bibfnamefont {G.}~\bibnamefont {Landwehr}},
  \bibinfo {author} {\bibfnamefont {C.~C.}\ \bibnamefont {Bradley}}, \bibinfo
  {author} {\bibfnamefont {P.}~\bibnamefont {Grosse}},\ and\ \bibinfo {author}
  {\bibfnamefont {B.}~\bibnamefont {Lax}},\ }\bibfield  {title} {\bibinfo
  {title} {Quantum effects in cyclotron resonance in $p$-type tellurium},\
  }\href {https://doi.org/10.1103/physrevlett.23.14} {\bibfield  {journal}
  {\bibinfo  {journal} {Physical Review Letters}\ }\textbf {\bibinfo {volume}
  {23}},\ \bibinfo {pages} {14} (\bibinfo {year} {1969})}\BibitemShut {NoStop}%
\bibitem [{\citenamefont {Bresler}\ \emph {et~al.}(1970)\citenamefont
  {Bresler}, \citenamefont {Veselago}, \citenamefont {Kosichkin}, \citenamefont
  {Pikus}, \citenamefont {Farbshtein},\ and\ \citenamefont
  {Shalyt}}]{Bresler1970}%
  \BibitemOpen
  \bibfield  {author} {\bibinfo {author} {\bibfnamefont {M.~S.}\ \bibnamefont
  {Bresler}}, \bibinfo {author} {\bibfnamefont {V.~G.}\ \bibnamefont
  {Veselago}}, \bibinfo {author} {\bibfnamefont {Y.~V.}\ \bibnamefont
  {Kosichkin}}, \bibinfo {author} {\bibfnamefont {G.~E.}\ \bibnamefont
  {Pikus}}, \bibinfo {author} {\bibfnamefont {I.~I.}\ \bibnamefont
  {Farbshtein}},\ and\ \bibinfo {author} {\bibfnamefont {S.~S.}\ \bibnamefont
  {Shalyt}},\ }\bibfield  {title} {\bibinfo {title} {Energy scheme of the
  tellurium valence band},\ }\href
  {http://www.jetp.ras.ru/cgi-bin/dn/e_030_05_0799.pdf} {\bibfield  {journal}
  {\bibinfo  {journal} {Sov. JETP}\ }\textbf {\bibinfo {volume} {30}},\
  \bibinfo {pages} {799} (\bibinfo {year} {1970})}\BibitemShut {NoStop}%
\bibitem [{\citenamefont {Saffert}\ \emph {et~al.}(1974)\citenamefont
  {Saffert}, \citenamefont {Schapawalow}, \citenamefont {Landwehr},\ and\
  \citenamefont {Gmelin}}]{Saffert1974}%
  \BibitemOpen
  \bibfield  {author} {\bibinfo {author} {\bibfnamefont {R.}~\bibnamefont
  {Saffert}}, \bibinfo {author} {\bibfnamefont {J.}~\bibnamefont
  {Schapawalow}}, \bibinfo {author} {\bibfnamefont {G.}~\bibnamefont
  {Landwehr}},\ and\ \bibinfo {author} {\bibfnamefont {E.}~\bibnamefont
  {Gmelin}},\ }\bibfield  {title} {\bibinfo {title} {Nernst-ettingshausen and
  seebeck effect of pure and electron-irradiated tellurium at
  low-temperatures},\ }\href {https://doi.org/10.1002/pssb.2220610216}
  {\bibfield  {journal} {\bibinfo  {journal} {phys. stat. sol. (b)}\ }\textbf
  {\bibinfo {volume} {61}},\ \bibinfo {pages} {509} (\bibinfo {year}
  {1974})}\BibitemShut {NoStop}%
\bibitem [{\citenamefont {{von Klitzing}}\ and\ \citenamefont
  {Landwehr}(1971)}]{Klitzing1971b}%
  \BibitemOpen
  \bibfield  {author} {\bibinfo {author} {\bibfnamefont {K.}~\bibnamefont {{von
  Klitzing}}}\ and\ \bibinfo {author} {\bibfnamefont {G.}~\bibnamefont
  {Landwehr}},\ }\bibfield  {title} {\bibinfo {title} {Surface quantum states
  in tellurium},\ }\href {https://doi.org/10.1016/0038-1098(71)90630-2}
  {\bibfield  {journal} {\bibinfo  {journal} {Solid State Communications}\
  }\textbf {\bibinfo {volume} {9}},\ \bibinfo {pages} {2201} (\bibinfo {year}
  {1971})}\BibitemShut {NoStop}%
\bibitem [{\citenamefont {Englert}\ \emph {et~al.}(1977)\citenamefont
  {Englert}, \citenamefont {von Klitzing}, \citenamefont {Silbermann},\ and\
  \citenamefont {Landwehr}}]{Englert1977}%
  \BibitemOpen
  \bibfield  {author} {\bibinfo {author} {\bibfnamefont {T.}~\bibnamefont
  {Englert}}, \bibinfo {author} {\bibfnamefont {K.}~\bibnamefont {von
  Klitzing}}, \bibinfo {author} {\bibfnamefont {R.}~\bibnamefont
  {Silbermann}},\ and\ \bibinfo {author} {\bibfnamefont {G.}~\bibnamefont
  {Landwehr}},\ }\bibfield  {title} {\bibinfo {title} {Influence of surface on
  galvanomagnetic properties of tellurium},\ }\href
  {https://doi.org/10.1002/pssb.2220810110} {\bibfield  {journal} {\bibinfo
  {journal} {phys. stat. sol. (b)}\ }\textbf {\bibinfo {volume} {81}},\
  \bibinfo {pages} {119} (\bibinfo {year} {1977})}\BibitemShut {NoStop}%
\bibitem [{\citenamefont {Nomura}(1960)}]{Nomura1960}%
  \BibitemOpen
  \bibfield  {author} {\bibinfo {author} {\bibfnamefont {K.~C.}\ \bibnamefont
  {Nomura}},\ }\bibfield  {title} {\bibinfo {title} {Optical activity in
  tellurium},\ }\href {https://doi.org/10.1103/PhysRevLett.5.500} {\bibfield
  {journal} {\bibinfo  {journal} {Phys. Rev. Lett.}\ }\textbf {\bibinfo
  {volume} {5}},\ \bibinfo {pages} {500} (\bibinfo {year} {1960})}\BibitemShut
  {NoStop}%
\bibitem [{\citenamefont {Ivchenko}\ and\ \citenamefont
  {Pikus}(1974)}]{Ivchenko1974}%
  \BibitemOpen
  \bibfield  {author} {\bibinfo {author} {\bibfnamefont {E.}~\bibnamefont
  {Ivchenko}}\ and\ \bibinfo {author} {\bibfnamefont {G.~E.}\ \bibnamefont
  {Pikus}},\ }\bibfield  {title} {\bibinfo {title} {Natural optical activity of
  semiconductor (tellurium)},\ }\href@noop {} {\bibfield  {journal} {\bibinfo
  {journal} {Sov. Phys. Solid State}\ }\textbf {\bibinfo {volume} {16}},\
  \bibinfo {pages} {1933} (\bibinfo {year} {1974})}\BibitemShut {NoStop}%
\bibitem [{\citenamefont {Fukuda}\ \emph {et~al.}(1975)\citenamefont {Fukuda},
  \citenamefont {Shiosaki},\ and\ \citenamefont {Kawabata}}]{Fukuda1975}%
  \BibitemOpen
  \bibfield  {author} {\bibinfo {author} {\bibfnamefont {S.}~\bibnamefont
  {Fukuda}}, \bibinfo {author} {\bibfnamefont {T.}~\bibnamefont {Shiosaki}},\
  and\ \bibinfo {author} {\bibfnamefont {A.}~\bibnamefont {Kawabata}},\
  }\bibfield  {title} {\bibinfo {title} {Infrared optical-activity in
  tellurium},\ }\href {https://doi.org/10.1002/pssb.2220680247} {\bibfield
  {journal} {\bibinfo  {journal} {phys. stat. sol. (b)}\ }\textbf {\bibinfo
  {volume} {68}},\ \bibinfo {pages} {K107} (\bibinfo {year}
  {1975})}\BibitemShut {NoStop}%
\bibitem [{\citenamefont {Stolze}\ \emph {et~al.}(1977)\citenamefont {Stolze},
  \citenamefont {Lutz},\ and\ \citenamefont {Grosse}}]{Stolze1977}%
  \BibitemOpen
  \bibfield  {author} {\bibinfo {author} {\bibfnamefont {H.}~\bibnamefont
  {Stolze}}, \bibinfo {author} {\bibfnamefont {M.}~\bibnamefont {Lutz}},\ and\
  \bibinfo {author} {\bibfnamefont {P.}~\bibnamefont {Grosse}},\ }\bibfield
  {title} {\bibinfo {title} {Optical-activity of tellurium},\ }\href
  {https://doi.org/10.1002/pssb.2220820206} {\bibfield  {journal} {\bibinfo
  {journal} {phys. stat. sol. (b)}\ }\textbf {\bibinfo {volume} {82}},\
  \bibinfo {pages} {457} (\bibinfo {year} {1977})}\BibitemShut {NoStop}%
\bibitem [{\citenamefont {Ivchenko}\ and\ \citenamefont
  {Pikus}(1978)}]{Ivchenko78p640}%
  \BibitemOpen
  \bibfield  {author} {\bibinfo {author} {\bibfnamefont {E.~L.}\ \bibnamefont
  {Ivchenko}}\ and\ \bibinfo {author} {\bibfnamefont {G.~E.}\ \bibnamefont
  {Pikus}},\ }\bibfield  {title} {\bibinfo {title} {{New photogalvanic effect
  in gyrotropic crystals}},\ }\href
  {http://jetpletters.ru/ps/1554/article_23792.shtml} {\bibfield  {journal}
  {\bibinfo  {journal} {JETP Lett.}\ }\textbf {\bibinfo {volume} {27}},\
  \bibinfo {pages} {604} (\bibinfo {year} {1978})}\BibitemShut {NoStop}%
\bibitem [{\citenamefont {Asnin}\ \emph {et~al.}(1978)\citenamefont {Asnin},
  \citenamefont {Bakun}, \citenamefont {Danishevskii}, \citenamefont
  {Ivchenko}, \citenamefont {Pikus},\ and\ \citenamefont
  {Rogachev}}]{Asnin1978}%
  \BibitemOpen
  \bibfield  {author} {\bibinfo {author} {\bibfnamefont {V.~M.}\ \bibnamefont
  {Asnin}}, \bibinfo {author} {\bibfnamefont {A.~A.}\ \bibnamefont {Bakun}},
  \bibinfo {author} {\bibfnamefont {A.~M.}\ \bibnamefont {Danishevskii}},
  \bibinfo {author} {\bibfnamefont {E.~L.}\ \bibnamefont {Ivchenko}}, \bibinfo
  {author} {\bibfnamefont {G.~E.}\ \bibnamefont {Pikus}},\ and\ \bibinfo
  {author} {\bibfnamefont {A.~A.}\ \bibnamefont {Rogachev}},\ }\bibfield
  {title} {\bibinfo {title} {Observation of a photo-emf that depends on the
  sign of the circular-polarization of the light},\ }\href
  {http://jetpletters.ru/ps/1557/article_23830.shtml} {\bibfield  {journal}
  {\bibinfo  {journal} {JETP Lett.}\ }\textbf {\bibinfo {volume} {28}},\
  \bibinfo {pages} {74} (\bibinfo {year} {1978})}\BibitemShut {NoStop}%
\bibitem [{\citenamefont {Shalygin}\ \emph {et~al.}(2016)\citenamefont
  {Shalygin}, \citenamefont {Moldavskaya}, \citenamefont {Danilov},
  \citenamefont {Farbshtein},\ and\ \citenamefont {Golub}}]{Shalygin2016}%
  \BibitemOpen
  \bibfield  {author} {\bibinfo {author} {\bibfnamefont {V.~A.}\ \bibnamefont
  {Shalygin}}, \bibinfo {author} {\bibfnamefont {M.~D.}\ \bibnamefont
  {Moldavskaya}}, \bibinfo {author} {\bibfnamefont {S.~N.}\ \bibnamefont
  {Danilov}}, \bibinfo {author} {\bibfnamefont {I.~I.}\ \bibnamefont
  {Farbshtein}},\ and\ \bibinfo {author} {\bibfnamefont {L.~E.}\ \bibnamefont
  {Golub}},\ }\bibfield  {title} {\bibinfo {title} {Circular photon drag effect
  in bulk tellurium},\ }\href {https://doi.org/10.1103/PhysRevB.93.045207}
  {\bibfield  {journal} {\bibinfo  {journal} {Phys. Rev. B}\ }\textbf {\bibinfo
  {volume} {93}},\ \bibinfo {pages} {045207} (\bibinfo {year}
  {2016})}\BibitemShut {NoStop}%
\bibitem [{\citenamefont {Vorob'ev}\ \emph {et~al.}(1979)\citenamefont
  {Vorob'ev}, \citenamefont {Ivchenko}, \citenamefont {Pikus}, \citenamefont
  {Farbshtein}, \citenamefont {Shalygin},\ and\ \citenamefont
  {Shturbin}}]{Vorobjev79p441}%
  \BibitemOpen
  \bibfield  {author} {\bibinfo {author} {\bibfnamefont {L.~E.}\ \bibnamefont
  {Vorob'ev}}, \bibinfo {author} {\bibfnamefont {E.~L.}\ \bibnamefont
  {Ivchenko}}, \bibinfo {author} {\bibfnamefont {G.~E.}\ \bibnamefont {Pikus}},
  \bibinfo {author} {\bibfnamefont {I.~I.}\ \bibnamefont {Farbshtein}},
  \bibinfo {author} {\bibfnamefont {V.~A.}\ \bibnamefont {Shalygin}},\ and\
  \bibinfo {author} {\bibfnamefont {A.~I.}\ \bibnamefont {Shturbin}},\
  }\bibfield  {title} {\bibinfo {title} {Optical-activity in tellurium induced
  by a current},\ }\href {http://jetpletters.ru/ps/1454/article_22128.shtml}
  {\bibfield  {journal} {\bibinfo  {journal} {JETP Lett.}\ }\textbf {\bibinfo
  {volume} {29}},\ \bibinfo {pages} {441} (\bibinfo {year} {1979})}\BibitemShut
  {NoStop}%
\bibitem [{\citenamefont {Shalygin}(2022)}]{Shalygin2023}%
  \BibitemOpen
  \bibfield  {author} {\bibinfo {author} {\bibfnamefont {V.~A.}\ \bibnamefont
  {Shalygin}},\ }\bibfield  {title} {\bibinfo {title} {Current-induced optical
  activity: First observation and comprehensive study},\ }in\ \href
  {https://doi.org/https://doi.org/10.1007/978-3-031-11287-4_1} {\emph
  {\bibinfo {booktitle} {Optics and Its Applications}}},\ \bibinfo {editor}
  {edited by\ \bibinfo {editor} {\bibfnamefont {D.}~\bibnamefont {Blaschke}},
  \bibinfo {editor} {\bibfnamefont {D.}~\bibnamefont {Firsov}}, \bibinfo
  {editor} {\bibfnamefont {A.}~\bibnamefont {Papoyan}},\ and\ \bibinfo {editor}
  {\bibfnamefont {H.~A.}\ \bibnamefont {Sarkisyan}}}\ (\bibinfo  {publisher}
  {Springer International Publishing},\ \bibinfo {address} {Cham},\ \bibinfo
  {year} {2022})\ pp.\ \bibinfo {pages} {1--19}\BibitemShut {NoStop}%
\bibitem [{\citenamefont {Wang}\ \emph {et~al.}(2018)\citenamefont {Wang},
  \citenamefont {Qiu}, \citenamefont {Wang}, \citenamefont {Huang},
  \citenamefont {Wang}, \citenamefont {Liu}, \citenamefont {Du}, \citenamefont
  {Goddard}, \citenamefont {Kim}, \citenamefont {Xu}, \citenamefont {Ye},\ and\
  \citenamefont {Wu}}]{Wang2018}%
  \BibitemOpen
  \bibfield  {author} {\bibinfo {author} {\bibfnamefont {Y.}~\bibnamefont
  {Wang}}, \bibinfo {author} {\bibfnamefont {G.}~\bibnamefont {Qiu}}, \bibinfo
  {author} {\bibfnamefont {R.}~\bibnamefont {Wang}}, \bibinfo {author}
  {\bibfnamefont {S.}~\bibnamefont {Huang}}, \bibinfo {author} {\bibfnamefont
  {Q.}~\bibnamefont {Wang}}, \bibinfo {author} {\bibfnamefont {Y.}~\bibnamefont
  {Liu}}, \bibinfo {author} {\bibfnamefont {Y.}~\bibnamefont {Du}}, \bibinfo
  {author} {\bibfnamefont {W.~A.}\ \bibnamefont {Goddard}, \bibfnamefont
  {III}}, \bibinfo {author} {\bibfnamefont {M.~J.}\ \bibnamefont {Kim}},
  \bibinfo {author} {\bibfnamefont {X.}~\bibnamefont {Xu}}, \bibinfo {author}
  {\bibfnamefont {P.~D.}\ \bibnamefont {Ye}},\ and\ \bibinfo {author}
  {\bibfnamefont {W.}~\bibnamefont {Wu}},\ }\bibfield  {title} {\bibinfo
  {title} {Field-effect transistors made from solution-grown two-dimensional
  tellurene},\ }\href {https://doi.org/10.1038/s41928-018-0058-4} {\bibfield
  {journal} {\bibinfo  {journal} {Nat. Electron.}\ }\textbf {\bibinfo {volume}
  {1}},\ \bibinfo {pages} {228} (\bibinfo {year} {2018})}\BibitemShut {NoStop}%
\bibitem [{\citenamefont {Wu}\ \emph {et~al.}(2018)\citenamefont {Wu},
  \citenamefont {Qiu}, \citenamefont {Wang}, \citenamefont {Wang},\ and\
  \citenamefont {Ye}}]{Wu2018b}%
  \BibitemOpen
  \bibfield  {author} {\bibinfo {author} {\bibfnamefont {W.}~\bibnamefont
  {Wu}}, \bibinfo {author} {\bibfnamefont {G.}~\bibnamefont {Qiu}}, \bibinfo
  {author} {\bibfnamefont {Y.}~\bibnamefont {Wang}}, \bibinfo {author}
  {\bibfnamefont {R.}~\bibnamefont {Wang}},\ and\ \bibinfo {author}
  {\bibfnamefont {P.}~\bibnamefont {Ye}},\ }\bibfield  {title} {\bibinfo
  {title} {Tellurene: its physical properties, scalable nanomanufacturing, and
  device applications},\ }\href {https://doi.org/10.1039/c8cs00598b} {\bibfield
   {journal} {\bibinfo  {journal} {Chem. Soc. Rev.}\ }\textbf {\bibinfo
  {volume} {47}},\ \bibinfo {pages} {7203} (\bibinfo {year}
  {2018})}\BibitemShut {NoStop}%
\bibitem [{\citenamefont {Shi}\ \emph {et~al.}(2020)\citenamefont {Shi},
  \citenamefont {Cao}, \citenamefont {Khan}, \citenamefont {Tareen},
  \citenamefont {Liu}, \citenamefont {Liang}, \citenamefont {Zhang},
  \citenamefont {Ma}, \citenamefont {Guo}, \citenamefont {Luo},\ and\
  \citenamefont {Zhang}}]{Shi2020}%
  \BibitemOpen
  \bibfield  {author} {\bibinfo {author} {\bibfnamefont {Z.}~\bibnamefont
  {Shi}}, \bibinfo {author} {\bibfnamefont {R.}~\bibnamefont {Cao}}, \bibinfo
  {author} {\bibfnamefont {K.}~\bibnamefont {Khan}}, \bibinfo {author}
  {\bibfnamefont {A.~K.}\ \bibnamefont {Tareen}}, \bibinfo {author}
  {\bibfnamefont {X.}~\bibnamefont {Liu}}, \bibinfo {author} {\bibfnamefont
  {W.}~\bibnamefont {Liang}}, \bibinfo {author} {\bibfnamefont
  {Y.}~\bibnamefont {Zhang}}, \bibinfo {author} {\bibfnamefont
  {C.}~\bibnamefont {Ma}}, \bibinfo {author} {\bibfnamefont {Z.}~\bibnamefont
  {Guo}}, \bibinfo {author} {\bibfnamefont {X.}~\bibnamefont {Luo}},\ and\
  \bibinfo {author} {\bibfnamefont {H.}~\bibnamefont {Zhang}},\ }\bibfield
  {title} {\bibinfo {title} {Two-dimensional tellurium: Progress, challenges,
  and prospects},\ }\href {https://doi.org/10.1007/s40820-020-00427-z}
  {\bibfield  {journal} {\bibinfo  {journal} {Nano-Micro Letters}\ }\textbf
  {\bibinfo {volume} {12}},\ \bibinfo {pages} {99} (\bibinfo {year}
  {2020})}\BibitemShut {NoStop}%
\bibitem [{\citenamefont {Zhang}\ \emph {et~al.}(2021)\citenamefont {Zhang},
  \citenamefont {Gong}, \citenamefont {Yu}, \citenamefont {Dai}, \citenamefont
  {Yang}, \citenamefont {Chen}, \citenamefont {Li}, \citenamefont {Pan},
  \citenamefont {Wang}, \citenamefont {Guo}, \citenamefont {Zhang},\ and\
  \citenamefont {Fu}}]{Zhang2020t}%
  \BibitemOpen
  \bibfield  {author} {\bibinfo {author} {\bibfnamefont {L.}~\bibnamefont
  {Zhang}}, \bibinfo {author} {\bibfnamefont {T.}~\bibnamefont {Gong}},
  \bibinfo {author} {\bibfnamefont {Z.}~\bibnamefont {Yu}}, \bibinfo {author}
  {\bibfnamefont {H.}~\bibnamefont {Dai}}, \bibinfo {author} {\bibfnamefont
  {Z.}~\bibnamefont {Yang}}, \bibinfo {author} {\bibfnamefont {G.}~\bibnamefont
  {Chen}}, \bibinfo {author} {\bibfnamefont {J.}~\bibnamefont {Li}}, \bibinfo
  {author} {\bibfnamefont {R.}~\bibnamefont {Pan}}, \bibinfo {author}
  {\bibfnamefont {H.}~\bibnamefont {Wang}}, \bibinfo {author} {\bibfnamefont
  {Z.}~\bibnamefont {Guo}}, \bibinfo {author} {\bibfnamefont {H.}~\bibnamefont
  {Zhang}},\ and\ \bibinfo {author} {\bibfnamefont {X.}~\bibnamefont {Fu}},\
  }\bibfield  {title} {\bibinfo {title} {Recent advances in hybridization,
  doping, and functionalization of 2d xenes},\ }\href
  {https://doi.org/10.1002/adfm.202005471} {\bibfield  {journal} {\bibinfo
  {journal} {Adv. Funct. Mater.}\ }\textbf {\bibinfo {volume} {31}},\ \bibinfo
  {pages} {2005471} (\bibinfo {year} {2021})}\BibitemShut {NoStop}%
\bibitem [{\citenamefont {Xu}\ \emph {et~al.}(2020)\citenamefont {Xu},
  \citenamefont {Ma}, \citenamefont {Fu}, \citenamefont {Shang}, \citenamefont
  {Jiang}, \citenamefont {Wang}, \citenamefont {Li},\ and\ \citenamefont
  {Zhang}}]{Xu2020}%
  \BibitemOpen
  \bibfield  {author} {\bibinfo {author} {\bibfnamefont {N.}~\bibnamefont
  {Xu}}, \bibinfo {author} {\bibfnamefont {P.}~\bibnamefont {Ma}}, \bibinfo
  {author} {\bibfnamefont {S.}~\bibnamefont {Fu}}, \bibinfo {author}
  {\bibfnamefont {X.}~\bibnamefont {Shang}}, \bibinfo {author} {\bibfnamefont
  {S.}~\bibnamefont {Jiang}}, \bibinfo {author} {\bibfnamefont
  {S.}~\bibnamefont {Wang}}, \bibinfo {author} {\bibfnamefont {D.}~\bibnamefont
  {Li}},\ and\ \bibinfo {author} {\bibfnamefont {H.}~\bibnamefont {Zhang}},\
  }\bibfield  {title} {\bibinfo {title} {Tellurene-based saturable absorber to
  demonstrate large-energy dissipative soliton and noise-like pulse
  generations},\ }\href {https://doi.org/10.1515/nanoph-2019-0545} {\bibfield
  {journal} {\bibinfo  {journal} {Nanophotonics}\ }\textbf {\bibinfo {volume}
  {9}},\ \bibinfo {pages} {2783} (\bibinfo {year} {2020})}\BibitemShut
  {NoStop}%
\bibitem [{\citenamefont {Yan}\ \emph {et~al.}(2022)\citenamefont {Yan},
  \citenamefont {Yang}, \citenamefont {Yang}, \citenamefont {Ji}, \citenamefont
  {Zhang}, \citenamefont {Tu}, \citenamefont {Du}, \citenamefont {Cai},\ and\
  \citenamefont {Lin}}]{Yan2022}%
  \BibitemOpen
  \bibfield  {author} {\bibinfo {author} {\bibfnamefont {Z.}~\bibnamefont
  {Yan}}, \bibinfo {author} {\bibfnamefont {H.}~\bibnamefont {Yang}}, \bibinfo
  {author} {\bibfnamefont {Z.}~\bibnamefont {Yang}}, \bibinfo {author}
  {\bibfnamefont {C.}~\bibnamefont {Ji}}, \bibinfo {author} {\bibfnamefont
  {G.}~\bibnamefont {Zhang}}, \bibinfo {author} {\bibfnamefont
  {Y.}~\bibnamefont {Tu}}, \bibinfo {author} {\bibfnamefont {G.}~\bibnamefont
  {Du}}, \bibinfo {author} {\bibfnamefont {S.}~\bibnamefont {Cai}},\ and\
  \bibinfo {author} {\bibfnamefont {S.}~\bibnamefont {Lin}},\ }\bibfield
  {title} {\bibinfo {title} {Emerging two-dimensional tellurene and tellurides
  for broadband photodetectors},\ }\href
  {https://doi.org/10.1002/smll.202200016} {\bibfield  {journal} {\bibinfo
  {journal} {Small}\ }\textbf {\bibinfo {volume} {18}},\ \bibinfo {pages}
  {2200016} (\bibinfo {year} {2022})}\BibitemShut {NoStop}%
\bibitem [{\citenamefont {Agapito}\ \emph {et~al.}(2013)\citenamefont
  {Agapito}, \citenamefont {Kioussis}, \citenamefont {Goddard},\ and\
  \citenamefont {Ong}}]{Agapito2013}%
  \BibitemOpen
  \bibfield  {author} {\bibinfo {author} {\bibfnamefont {L.~A.}\ \bibnamefont
  {Agapito}}, \bibinfo {author} {\bibfnamefont {N.}~\bibnamefont {Kioussis}},
  \bibinfo {author} {\bibfnamefont {W.~A.}\ \bibnamefont {Goddard},
  \bibfnamefont {III}},\ and\ \bibinfo {author} {\bibfnamefont {N.~P.}\
  \bibnamefont {Ong}},\ }\bibfield  {title} {\bibinfo {title} {Novel family of
  chiral-based topological insulators: Elemental tellurium under strain},\
  }\href {https://doi.org/10.1103/PhysRevLett.110.176401} {\bibfield  {journal}
  {\bibinfo  {journal} {Phys. Rev. Lett.}\ }\textbf {\bibinfo {volume} {110}},\
  \bibinfo {pages} {176401} (\bibinfo {year} {2013})}\BibitemShut {NoStop}%
\bibitem [{\citenamefont {Hirayama}\ \emph {et~al.}(2015)\citenamefont
  {Hirayama}, \citenamefont {Okugawa}, \citenamefont {Ishibashi}, \citenamefont
  {Murakami},\ and\ \citenamefont {Miyake}}]{Hirayama2015}%
  \BibitemOpen
  \bibfield  {author} {\bibinfo {author} {\bibfnamefont {M.}~\bibnamefont
  {Hirayama}}, \bibinfo {author} {\bibfnamefont {R.}~\bibnamefont {Okugawa}},
  \bibinfo {author} {\bibfnamefont {S.}~\bibnamefont {Ishibashi}}, \bibinfo
  {author} {\bibfnamefont {S.}~\bibnamefont {Murakami}},\ and\ \bibinfo
  {author} {\bibfnamefont {T.}~\bibnamefont {Miyake}},\ }\bibfield  {title}
  {\bibinfo {title} {Weyl node and spin texture in trigonal tellurium and
  selenium},\ }\href {https://doi.org/10.1103/PhysRevLett.114.206401}
  {\bibfield  {journal} {\bibinfo  {journal} {Phys. Rev. Lett.}\ }\textbf
  {\bibinfo {volume} {114}},\ \bibinfo {pages} {206401} (\bibinfo {year}
  {2015})}\BibitemShut {NoStop}%
\bibitem [{\citenamefont {Murakami}\ \emph {et~al.}(2017)\citenamefont
  {Murakami}, \citenamefont {Hirayama}, \citenamefont {Okugawa},\ and\
  \citenamefont {Miyake}}]{Murakami2017}%
  \BibitemOpen
  \bibfield  {author} {\bibinfo {author} {\bibfnamefont {S.}~\bibnamefont
  {Murakami}}, \bibinfo {author} {\bibfnamefont {M.}~\bibnamefont {Hirayama}},
  \bibinfo {author} {\bibfnamefont {R.}~\bibnamefont {Okugawa}},\ and\ \bibinfo
  {author} {\bibfnamefont {T.}~\bibnamefont {Miyake}},\ }\bibfield  {title}
  {\bibinfo {title} {Emergence of topological semimetals in gap closing in
  semiconductors without inversion symmetry},\ }\href
  {https://doi.org/10.1126/sciadv.1602680} {\bibfield  {journal} {\bibinfo
  {journal} {Science Advances}\ }\textbf {\bibinfo {volume} {3}},\ \bibinfo
  {pages} {e1602680} (\bibinfo {year} {2017})}\BibitemShut {NoStop}%
\bibitem [{\citenamefont {Ideue}\ \emph {et~al.}(2019)\citenamefont {Ideue},
  \citenamefont {Hirayama}, \citenamefont {Taiko}, \citenamefont {Takahashi},
  \citenamefont {Murase}, \citenamefont {Miyake}, \citenamefont {Murakami},
  \citenamefont {Sasagawa},\ and\ \citenamefont {Iwasa}}]{Ideue2019}%
  \BibitemOpen
  \bibfield  {author} {\bibinfo {author} {\bibfnamefont {T.}~\bibnamefont
  {Ideue}}, \bibinfo {author} {\bibfnamefont {M.}~\bibnamefont {Hirayama}},
  \bibinfo {author} {\bibfnamefont {H.}~\bibnamefont {Taiko}}, \bibinfo
  {author} {\bibfnamefont {T.}~\bibnamefont {Takahashi}}, \bibinfo {author}
  {\bibfnamefont {M.}~\bibnamefont {Murase}}, \bibinfo {author} {\bibfnamefont
  {T.}~\bibnamefont {Miyake}}, \bibinfo {author} {\bibfnamefont
  {S.}~\bibnamefont {Murakami}}, \bibinfo {author} {\bibfnamefont
  {T.}~\bibnamefont {Sasagawa}},\ and\ \bibinfo {author} {\bibfnamefont
  {Y.}~\bibnamefont {Iwasa}},\ }\bibfield  {title} {\bibinfo {title}
  {Pressure-induced topological phase transition in noncentrosymmetric
  elemental tellurium},\ }\href {https://doi.org/10.1073/pnas.1905524116}
  {\bibfield  {journal} {\bibinfo  {journal} {Proceedings of the National
  Academy of Sciences}\ }\textbf {\bibinfo {volume} {116}},\ \bibinfo {pages}
  {25530} (\bibinfo {year} {2019})}\BibitemShut {NoStop}%
\bibitem [{\citenamefont {Zhang}\ \emph {et~al.}(2020)\citenamefont {Zhang},
  \citenamefont {Zhao}, \citenamefont {Li}, \citenamefont {Wang}, \citenamefont
  {Xie}, \citenamefont {Cheng}, \citenamefont {Li}, \citenamefont {Lin},
  \citenamefont {Xi}, \citenamefont {Ke}, \citenamefont {Yang}, \citenamefont
  {He}, \citenamefont {Sun}, \citenamefont {Wang}, \citenamefont {Zhang},\ and\
  \citenamefont {Zeng}}]{Zhang2020}%
  \BibitemOpen
  \bibfield  {author} {\bibinfo {author} {\bibfnamefont {N.}~\bibnamefont
  {Zhang}}, \bibinfo {author} {\bibfnamefont {G.}~\bibnamefont {Zhao}},
  \bibinfo {author} {\bibfnamefont {L.}~\bibnamefont {Li}}, \bibinfo {author}
  {\bibfnamefont {P.}~\bibnamefont {Wang}}, \bibinfo {author} {\bibfnamefont
  {L.}~\bibnamefont {Xie}}, \bibinfo {author} {\bibfnamefont {B.}~\bibnamefont
  {Cheng}}, \bibinfo {author} {\bibfnamefont {H.}~\bibnamefont {Li}}, \bibinfo
  {author} {\bibfnamefont {Z.}~\bibnamefont {Lin}}, \bibinfo {author}
  {\bibfnamefont {C.}~\bibnamefont {Xi}}, \bibinfo {author} {\bibfnamefont
  {J.}~\bibnamefont {Ke}}, \bibinfo {author} {\bibfnamefont {M.}~\bibnamefont
  {Yang}}, \bibinfo {author} {\bibfnamefont {J.}~\bibnamefont {He}}, \bibinfo
  {author} {\bibfnamefont {Z.}~\bibnamefont {Sun}}, \bibinfo {author}
  {\bibfnamefont {Z.}~\bibnamefont {Wang}}, \bibinfo {author} {\bibfnamefont
  {Z.}~\bibnamefont {Zhang}},\ and\ \bibinfo {author} {\bibfnamefont
  {C.}~\bibnamefont {Zeng}},\ }\bibfield  {title} {\bibinfo {title}
  {Magnetotransport signatures of {Weyl} physics and discrete scale invariance
  in the elemental semiconductor tellurium},\ }\href
  {https://doi.org/10.1073/pnas.2002913117} {\bibfield  {journal} {\bibinfo
  {journal} {Proceedings of the National Academy of Sciences}\ }\textbf
  {\bibinfo {volume} {117}},\ \bibinfo {pages} {11337} (\bibinfo {year}
  {2020})}\BibitemShut {NoStop}%
\bibitem [{\citenamefont {Oliveira}\ \emph {et~al.}(2021)\citenamefont
  {Oliveira}, \citenamefont {Fontes}, \citenamefont {Moutinho}, \citenamefont
  {Rowley}, \citenamefont {Baggio-Saitovitch}, \citenamefont {Silva~Neto},\
  and\ \citenamefont {Enderlein}}]{Oliveira2021}%
  \BibitemOpen
  \bibfield  {author} {\bibinfo {author} {\bibfnamefont {J.~F.}\ \bibnamefont
  {Oliveira}}, \bibinfo {author} {\bibfnamefont {M.~B.}\ \bibnamefont
  {Fontes}}, \bibinfo {author} {\bibfnamefont {M.}~\bibnamefont {Moutinho}},
  \bibinfo {author} {\bibfnamefont {S.~E.}\ \bibnamefont {Rowley}}, \bibinfo
  {author} {\bibfnamefont {E.}~\bibnamefont {Baggio-Saitovitch}}, \bibinfo
  {author} {\bibfnamefont {M.~B.}\ \bibnamefont {Silva~Neto}},\ and\ \bibinfo
  {author} {\bibfnamefont {C.}~\bibnamefont {Enderlein}},\ }\bibfield  {title}
  {\bibinfo {title} {Pressure-induced anderson-mott transition in elemental
  tellurium},\ }\href {https://doi.org/10.1038/s43246-020-00110-1} {\bibfield
  {journal} {\bibinfo  {journal} {Communications Materials}\ }\textbf {\bibinfo
  {volume} {2}},\ \bibinfo {pages} {1} (\bibinfo {year} {2021})}\BibitemShut
  {NoStop}%
\bibitem [{\citenamefont {Glazov}\ \emph {et~al.}(2022)\citenamefont {Glazov},
  \citenamefont {Ivchenko},\ and\ \citenamefont {Nestoklon}}]{Glazov2022}%
  \BibitemOpen
  \bibfield  {author} {\bibinfo {author} {\bibfnamefont {M.~M.}\ \bibnamefont
  {Glazov}}, \bibinfo {author} {\bibfnamefont {E.~L.}\ \bibnamefont
  {Ivchenko}},\ and\ \bibinfo {author} {\bibfnamefont {M.~O.}\ \bibnamefont
  {Nestoklon}},\ }\bibfield  {title} {\bibinfo {title} {Effect of pressure on
  the electronic band structure and circular photocurrent in tellurium},\
  }\href {https://doi.org/10.1134/S1063776122100053} {\bibfield  {journal}
  {\bibinfo  {journal} {JETP}\ }\textbf {\bibinfo {volume} {135}},\ \bibinfo
  {pages} {575} (\bibinfo {year} {2022})}\BibitemShut {NoStop}%
\bibitem [{\citenamefont {Glazov}\ and\ \citenamefont
  {Ganichev}(2014)}]{Glazov2014}%
  \BibitemOpen
  \bibfield  {author} {\bibinfo {author} {\bibfnamefont {M.~M.}\ \bibnamefont
  {Glazov}}\ and\ \bibinfo {author} {\bibfnamefont {S.~D.}\ \bibnamefont
  {Ganichev}},\ }\bibfield  {title} {\bibinfo {title} {High frequency electric
  field induced nonlinear effects in graphene},\ }\href
  {https://doi.org/10.1016/j.physrep.2013.10.003} {\bibfield  {journal}
  {\bibinfo  {journal} {Phys. Rep.}\ }\textbf {\bibinfo {volume} {535}},\
  \bibinfo {pages} {101} (\bibinfo {year} {2014})}\BibitemShut {NoStop}%
\bibitem [{\citenamefont {Otteneder}\ \emph {et~al.}(2020)\citenamefont
  {Otteneder}, \citenamefont {Hubmann}, \citenamefont {Lu}, \citenamefont
  {Kozlov}, \citenamefont {Golub}, \citenamefont {Watanabe}, \citenamefont
  {Taniguchi}, \citenamefont {Efetov},\ and\ \citenamefont
  {Ganichev}}]{Otteneder2020}%
  \BibitemOpen
  \bibfield  {author} {\bibinfo {author} {\bibfnamefont {M.}~\bibnamefont
  {Otteneder}}, \bibinfo {author} {\bibfnamefont {S.}~\bibnamefont {Hubmann}},
  \bibinfo {author} {\bibfnamefont {X.}~\bibnamefont {Lu}}, \bibinfo {author}
  {\bibfnamefont {D.~A.}\ \bibnamefont {Kozlov}}, \bibinfo {author}
  {\bibfnamefont {L.~E.}\ \bibnamefont {Golub}}, \bibinfo {author}
  {\bibfnamefont {K.}~\bibnamefont {Watanabe}}, \bibinfo {author}
  {\bibfnamefont {T.}~\bibnamefont {Taniguchi}}, \bibinfo {author}
  {\bibfnamefont {D.~K.}\ \bibnamefont {Efetov}},\ and\ \bibinfo {author}
  {\bibfnamefont {S.~D.}\ \bibnamefont {Ganichev}},\ }\bibfield  {title}
  {\bibinfo {title} {Terahertz photogalvanics in twisted bilayer graphene close
  to the second magic angle},\ }\href
  {https://doi.org/10.1021/acs.nanolett.0c02474} {\bibfield  {journal}
  {\bibinfo  {journal} {Nano Lett.}\ }\textbf {\bibinfo {volume} {20}},\
  \bibinfo {pages} {7152} (\bibinfo {year} {2020})}\BibitemShut {NoStop}%
\bibitem [{\citenamefont {Ivchenko}\ and\ \citenamefont {Ganichev}(2017)}]{35}%
  \BibitemOpen
  \bibfield  {author} {\bibinfo {author} {\bibfnamefont {E.~L.}\ \bibnamefont
  {Ivchenko}}\ and\ \bibinfo {author} {\bibfnamefont {S.~D.}\ \bibnamefont
  {Ganichev}},\ }\bibinfo {title} {Spin--photogalvanics},\ in\ \href@noop {}
  {\emph {\bibinfo {booktitle} {Spin physics in semiconductors}}},\ \bibinfo
  {editor} {edited by\ \bibinfo {editor} {\bibfnamefont {M.~I.}\ \bibnamefont
  {Dyakonov}}}\ (\bibinfo  {publisher} {Springer},\ \bibinfo {year}
  {2017})\BibitemShut {NoStop}%
\bibitem [{\citenamefont {Plank}\ and\ \citenamefont
  {Ganichev}(2018)}]{Plank2018}%
  \BibitemOpen
  \bibfield  {author} {\bibinfo {author} {\bibfnamefont {H.}~\bibnamefont
  {Plank}}\ and\ \bibinfo {author} {\bibfnamefont {S.~D.}\ \bibnamefont
  {Ganichev}},\ }\bibfield  {title} {\bibinfo {title} {A review on terahertz
  photogalvanic spectroscopy of {Bi$_2$Te$_3$}- and {Sb$_2$Te$_3$}-based three
  dimensional topological insulators},\ }\href
  {https://doi.org/10.1016/j.sse.2018.06.002} {\bibfield  {journal} {\bibinfo
  {journal} {Solid-State Electron.}\ }\textbf {\bibinfo {volume} {147}},\
  \bibinfo {pages} {44} (\bibinfo {year} {2018})}\BibitemShut {NoStop}%
\bibitem [{\citenamefont {Ishizuka}\ \emph {et~al.}(2016)\citenamefont
  {Ishizuka}, \citenamefont {Hayata}, \citenamefont {Ueda},\ and\ \citenamefont
  {Nagaosa}}]{Ishizuka2016}%
  \BibitemOpen
  \bibfield  {author} {\bibinfo {author} {\bibfnamefont {H.}~\bibnamefont
  {Ishizuka}}, \bibinfo {author} {\bibfnamefont {T.}~\bibnamefont {Hayata}},
  \bibinfo {author} {\bibfnamefont {M.}~\bibnamefont {Ueda}},\ and\ \bibinfo
  {author} {\bibfnamefont {N.}~\bibnamefont {Nagaosa}},\ }\bibfield  {title}
  {\bibinfo {title} {Emergent electromagnetic induction and adiabatic charge
  pumping in noncentrosymmetric {W}eyl semimetals},\ }\href
  {https://doi.org/10.1103/PhysRevLett.117.216601} {\bibfield  {journal}
  {\bibinfo  {journal} {Phys. Rev. Lett.}\ }\textbf {\bibinfo {volume} {117}},\
  \bibinfo {pages} {216601} (\bibinfo {year} {2016})}\BibitemShut {NoStop}%
\bibitem [{\citenamefont {Golub}\ \emph {et~al.}(2017)\citenamefont {Golub},
  \citenamefont {Ivchenko},\ and\ \citenamefont {Spivak}}]{Golub2017}%
  \BibitemOpen
  \bibfield  {author} {\bibinfo {author} {\bibfnamefont {L.~E.}\ \bibnamefont
  {Golub}}, \bibinfo {author} {\bibfnamefont {E.~L.}\ \bibnamefont
  {Ivchenko}},\ and\ \bibinfo {author} {\bibfnamefont {B.~Z.}\ \bibnamefont
  {Spivak}},\ }\bibfield  {title} {\bibinfo {title} {Photocurrent in gyrotropic
  {Weyl} semimetals},\ }\href {https://doi.org/10.1134/S0021364017120062}
  {\bibfield  {journal} {\bibinfo  {journal} {JETP Lett.}\ }\textbf {\bibinfo
  {volume} {105}},\ \bibinfo {pages} {782} (\bibinfo {year}
  {2017})}\BibitemShut {NoStop}%
\bibitem [{\citenamefont {de~Juan}\ \emph {et~al.}(2017)\citenamefont
  {de~Juan}, \citenamefont {Grushin}, \citenamefont {Morimoto},\ and\
  \citenamefont {Moore}}]{Juan2017}%
  \BibitemOpen
  \bibfield  {author} {\bibinfo {author} {\bibfnamefont {F.}~\bibnamefont
  {de~Juan}}, \bibinfo {author} {\bibfnamefont {A.~G.}\ \bibnamefont
  {Grushin}}, \bibinfo {author} {\bibfnamefont {T.}~\bibnamefont {Morimoto}},\
  and\ \bibinfo {author} {\bibfnamefont {J.~E.}\ \bibnamefont {Moore}},\
  }\bibfield  {title} {\bibinfo {title} {Quantized circular photogalvanic
  effect in {W}eyl semimetals},\ }\href {https://doi.org/10.1038/ncomms15995}
  {\bibfield  {journal} {\bibinfo  {journal} {Nat. Comm.}\ }\textbf {\bibinfo
  {volume} {8}},\ \bibinfo {pages} {15995} (\bibinfo {year}
  {2017})}\BibitemShut {NoStop}%
\bibitem [{\citenamefont {Chan}\ \emph {et~al.}(2017)\citenamefont {Chan},
  \citenamefont {Lindner}, \citenamefont {Refael},\ and\ \citenamefont
  {Lee}}]{Chan2017}%
  \BibitemOpen
  \bibfield  {author} {\bibinfo {author} {\bibfnamefont {C.-K.}\ \bibnamefont
  {Chan}}, \bibinfo {author} {\bibfnamefont {N.~H.}\ \bibnamefont {Lindner}},
  \bibinfo {author} {\bibfnamefont {G.}~\bibnamefont {Refael}},\ and\ \bibinfo
  {author} {\bibfnamefont {P.~A.}\ \bibnamefont {Lee}},\ }\bibfield  {title}
  {\bibinfo {title} {Photocurrents in {W}eyl semimetals},\ }\href
  {https://doi.org/10.1103/PhysRevB.95.041104} {\bibfield  {journal} {\bibinfo
  {journal} {Phys. Rev. B}\ }\textbf {\bibinfo {volume} {95}},\ \bibinfo
  {pages} {041104} (\bibinfo {year} {2017})}\BibitemShut {NoStop}%
\bibitem [{\citenamefont {Ma}\ \emph {et~al.}(2017)\citenamefont {Ma},
  \citenamefont {Xu}, \citenamefont {Chan}, \citenamefont {Zhang},
  \citenamefont {Chang}, \citenamefont {Lin}, \citenamefont {Xie},
  \citenamefont {Palacios}, \citenamefont {Lin}, \citenamefont {Jia},
  \citenamefont {Lee}, \citenamefont {Jarillo-Herrero},\ and\ \citenamefont
  {Gedik}}]{Ma2017}%
  \BibitemOpen
  \bibfield  {author} {\bibinfo {author} {\bibfnamefont {Q.}~\bibnamefont
  {Ma}}, \bibinfo {author} {\bibfnamefont {S.-Y.}\ \bibnamefont {Xu}}, \bibinfo
  {author} {\bibfnamefont {C.-K.}\ \bibnamefont {Chan}}, \bibinfo {author}
  {\bibfnamefont {C.-L.}\ \bibnamefont {Zhang}}, \bibinfo {author}
  {\bibfnamefont {G.}~\bibnamefont {Chang}}, \bibinfo {author} {\bibfnamefont
  {Y.}~\bibnamefont {Lin}}, \bibinfo {author} {\bibfnamefont {W.}~\bibnamefont
  {Xie}}, \bibinfo {author} {\bibfnamefont {T.}~\bibnamefont {Palacios}},
  \bibinfo {author} {\bibfnamefont {H.}~\bibnamefont {Lin}}, \bibinfo {author}
  {\bibfnamefont {S.}~\bibnamefont {Jia}}, \bibinfo {author} {\bibfnamefont
  {P.~A.}\ \bibnamefont {Lee}}, \bibinfo {author} {\bibfnamefont
  {P.}~\bibnamefont {Jarillo-Herrero}},\ and\ \bibinfo {author} {\bibfnamefont
  {N.}~\bibnamefont {Gedik}},\ }\bibfield  {title} {\bibinfo {title} {Direct
  optical detection of {Weyl} fermion chirality in a topological semimetal},\
  }\href {https://doi.org/10.1038/NPHYS4146} {\bibfield  {journal} {\bibinfo
  {journal} {Nat. Phys.}\ }\textbf {\bibinfo {volume} {13}},\ \bibinfo {pages}
  {842} (\bibinfo {year} {2017})}\BibitemShut {NoStop}%
\bibitem [{\citenamefont {Ji}\ \emph {et~al.}(2019)\citenamefont {Ji},
  \citenamefont {Liu}, \citenamefont {Addison}, \citenamefont {Liu},
  \citenamefont {Yu}, \citenamefont {Gao}, \citenamefont {Liu}, \citenamefont
  {Rappe}, \citenamefont {Kane}, \citenamefont {Mele},\ and\ \citenamefont
  {Agarwal}}]{Ji2019}%
  \BibitemOpen
  \bibfield  {author} {\bibinfo {author} {\bibfnamefont {Z.}~\bibnamefont
  {Ji}}, \bibinfo {author} {\bibfnamefont {G.}~\bibnamefont {Liu}}, \bibinfo
  {author} {\bibfnamefont {Z.}~\bibnamefont {Addison}}, \bibinfo {author}
  {\bibfnamefont {W.}~\bibnamefont {Liu}}, \bibinfo {author} {\bibfnamefont
  {P.}~\bibnamefont {Yu}}, \bibinfo {author} {\bibfnamefont {H.}~\bibnamefont
  {Gao}}, \bibinfo {author} {\bibfnamefont {Z.}~\bibnamefont {Liu}}, \bibinfo
  {author} {\bibfnamefont {A.~M.}\ \bibnamefont {Rappe}}, \bibinfo {author}
  {\bibfnamefont {C.~L.}\ \bibnamefont {Kane}}, \bibinfo {author}
  {\bibfnamefont {E.~J.}\ \bibnamefont {Mele}},\ and\ \bibinfo {author}
  {\bibfnamefont {R.}~\bibnamefont {Agarwal}},\ }\bibfield  {title} {\bibinfo
  {title} {Spatially dispersive circular photogalvanic effect in a {W}eyl
  semimetal},\ }\href {https://doi.org/10.1038/s41563-019-0421-5} {\bibfield
  {journal} {\bibinfo  {journal} {Nat. Mat.}\ }\textbf {\bibinfo {volume}
  {18}},\ \bibinfo {pages} {955} (\bibinfo {year} {2019})}\BibitemShut
  {NoStop}%
\bibitem [{\citenamefont {Ma}\ \emph {et~al.}(2022)\citenamefont {Ma},
  \citenamefont {Cheng}, \citenamefont {Li}, \citenamefont {Fan}, \citenamefont
  {Mu}, \citenamefont {Lai}, \citenamefont {Song}, \citenamefont {Yang},
  \citenamefont {Cheng}, \citenamefont {Wang}, \citenamefont {Zeng},\ and\
  \citenamefont {Sun}}]{Ma2022}%
  \BibitemOpen
  \bibfield  {author} {\bibinfo {author} {\bibfnamefont {J.}~\bibnamefont
  {Ma}}, \bibinfo {author} {\bibfnamefont {B.}~\bibnamefont {Cheng}}, \bibinfo
  {author} {\bibfnamefont {L.}~\bibnamefont {Li}}, \bibinfo {author}
  {\bibfnamefont {Z.}~\bibnamefont {Fan}}, \bibinfo {author} {\bibfnamefont
  {H.}~\bibnamefont {Mu}}, \bibinfo {author} {\bibfnamefont {J.}~\bibnamefont
  {Lai}}, \bibinfo {author} {\bibfnamefont {X.}~\bibnamefont {Song}}, \bibinfo
  {author} {\bibfnamefont {D.}~\bibnamefont {Yang}}, \bibinfo {author}
  {\bibfnamefont {J.}~\bibnamefont {Cheng}}, \bibinfo {author} {\bibfnamefont
  {Z.}~\bibnamefont {Wang}}, \bibinfo {author} {\bibfnamefont {C.}~\bibnamefont
  {Zeng}},\ and\ \bibinfo {author} {\bibfnamefont {D.}~\bibnamefont {Sun}},\
  }\bibfield  {title} {\bibinfo {title} {Unveiling {Weyl}-related optical
  responses in semiconducting tellurium by mid-infrared circular photogalvanic
  effect},\ }\href {https://doi.org/10.1038/s41467-022-33190-3} {\bibfield
  {journal} {\bibinfo  {journal} {Nat. Comm.}\ }\textbf {\bibinfo {volume}
  {13}},\ \bibinfo {pages} {5425} (\bibinfo {year} {2022})}\BibitemShut
  {NoStop}%
\bibitem [{Note1()}]{Note1}%
  \BibitemOpen
  \bibinfo {note} {Note that due to growth conditions the hexagon was slightly
  distorted.}\BibitemShut {Stop}%
\bibitem [{\citenamefont {Bauer}\ \emph {et~al.}(1973)\citenamefont {Bauer},
  \citenamefont {Kahilert}, \citenamefont {von Klitzing},\ and\ \citenamefont
  {Landwehr}}]{Bauer1973}%
  \BibitemOpen
  \bibfield  {author} {\bibinfo {author} {\bibfnamefont {G.}~\bibnamefont
  {Bauer}}, \bibinfo {author} {\bibfnamefont {H.}~\bibnamefont {Kahilert}},
  \bibinfo {author} {\bibfnamefont {K.}~\bibnamefont {von Klitzing}},\ and\
  \bibinfo {author} {\bibfnamefont {G.}~\bibnamefont {Landwehr}},\ }\bibfield
  {title} {\bibinfo {title} {Time-dependent non-ohmic conductivity and
  magnetoresistance in p-tellurium at 2 k},\ }\href
  {https://doi.org/10.1002/pssb.2220590214} {\bibfield  {journal} {\bibinfo
  {journal} {Physica Status Solidi (b)}\ }\textbf {\bibinfo {volume} {59}},\
  \bibinfo {pages} {479} (\bibinfo {year} {1973})}\BibitemShut {NoStop}%
\bibitem [{\citenamefont {Ribakovs}\ and\ \citenamefont
  {Gundjian}(1977)}]{Ribakovs1977}%
  \BibitemOpen
  \bibfield  {author} {\bibinfo {author} {\bibfnamefont {G.}~\bibnamefont
  {Ribakovs}}\ and\ \bibinfo {author} {\bibfnamefont {A.~A.}\ \bibnamefont
  {Gundjian}},\ }\bibfield  {title} {\bibinfo {title} {Theory of the photon
  drag effect in tellurium},\ }\href {https://doi.org/10.1063/1.323520}
  {\bibfield  {journal} {\bibinfo  {journal} {Journal of Applied Physics}\
  }\textbf {\bibinfo {volume} {48}},\ \bibinfo {pages} {4609} (\bibinfo {year}
  {1977})}\BibitemShut {NoStop}%
\bibitem [{\citenamefont {Ganichev}\ and\ \citenamefont
  {Prettl}(2006)}]{Ganichev2006}%
  \BibitemOpen
  \bibfield  {author} {\bibinfo {author} {\bibfnamefont {S.~D.}\ \bibnamefont
  {Ganichev}}\ and\ \bibinfo {author} {\bibfnamefont {W.}~\bibnamefont
  {Prettl}},\ }\href@noop {} {\emph {\bibinfo {title} {Intense Terahertz
  Excitation of Semiconductors}}}\ (\bibinfo  {publisher} {Oxford University
  Press, Oxford},\ \bibinfo {year} {2006})\BibitemShut {NoStop}%
\bibitem [{\citenamefont {Ganichev}\ \emph
  {et~al.}(2003{\natexlab{a}})\citenamefont {Ganichev}, \citenamefont
  {Bel'kov}, \citenamefont {Schneider}, \citenamefont {Ivchenko}, \citenamefont
  {Tarasenko}, \citenamefont {Wegscheider}, \citenamefont {Weiss},
  \citenamefont {Schuh}, \citenamefont {Beregulin},\ and\ \citenamefont
  {Prettl}}]{Ganichev2003a}%
  \BibitemOpen
  \bibfield  {author} {\bibinfo {author} {\bibfnamefont {S.~D.}\ \bibnamefont
  {Ganichev}}, \bibinfo {author} {\bibfnamefont {V.~V.}\ \bibnamefont
  {Bel'kov}}, \bibinfo {author} {\bibfnamefont {P.}~\bibnamefont {Schneider}},
  \bibinfo {author} {\bibfnamefont {E.~L.}\ \bibnamefont {Ivchenko}}, \bibinfo
  {author} {\bibfnamefont {S.~A.}\ \bibnamefont {Tarasenko}}, \bibinfo {author}
  {\bibfnamefont {W.}~\bibnamefont {Wegscheider}}, \bibinfo {author}
  {\bibfnamefont {D.}~\bibnamefont {Weiss}}, \bibinfo {author} {\bibfnamefont
  {D.}~\bibnamefont {Schuh}}, \bibinfo {author} {\bibfnamefont {E.~V.}\
  \bibnamefont {Beregulin}},\ and\ \bibinfo {author} {\bibfnamefont
  {W.}~\bibnamefont {Prettl}},\ }\bibfield  {title} {\bibinfo {title} {Resonant
  inversion of the circular photogalvanic effect in n-doped quantum wells},\
  }\href {https://doi.org/10.1103/PhysRevB.68.035319} {\bibfield  {journal}
  {\bibinfo  {journal} {Phys. Rev. B}\ }\textbf {\bibinfo {volume} {68}},\
  \bibinfo {pages} {035319} (\bibinfo {year} {2003}{\natexlab{a}})}\BibitemShut
  {NoStop}%
\bibitem [{\citenamefont {Ganichev}\ \emph
  {et~al.}(2003{\natexlab{b}})\citenamefont {Ganichev}, \citenamefont
  {Schneider}, \citenamefont {Bel'kov}, \citenamefont {Ivchenko}, \citenamefont
  {Tarasenko}, \citenamefont {Wegscheider}, \citenamefont {Weiss},
  \citenamefont {Schuh}, \citenamefont {Murdin}, \citenamefont {Phillips},
  \citenamefont {Pidgeon}, \citenamefont {Clarke}, \citenamefont {Merrick},
  \citenamefont {Murzyn}, \citenamefont {Beregulin},\ and\ \citenamefont
  {Prettl}}]{Ganichev2003}%
  \BibitemOpen
  \bibfield  {author} {\bibinfo {author} {\bibfnamefont {S.~D.}\ \bibnamefont
  {Ganichev}}, \bibinfo {author} {\bibfnamefont {P.}~\bibnamefont {Schneider}},
  \bibinfo {author} {\bibfnamefont {V.~V.}\ \bibnamefont {Bel'kov}}, \bibinfo
  {author} {\bibfnamefont {E.~L.}\ \bibnamefont {Ivchenko}}, \bibinfo {author}
  {\bibfnamefont {S.~A.}\ \bibnamefont {Tarasenko}}, \bibinfo {author}
  {\bibfnamefont {W.}~\bibnamefont {Wegscheider}}, \bibinfo {author}
  {\bibfnamefont {D.}~\bibnamefont {Weiss}}, \bibinfo {author} {\bibfnamefont
  {D.}~\bibnamefont {Schuh}}, \bibinfo {author} {\bibfnamefont {B.~N.}\
  \bibnamefont {Murdin}}, \bibinfo {author} {\bibfnamefont {P.~J.}\
  \bibnamefont {Phillips}}, \bibinfo {author} {\bibfnamefont {C.~R.}\
  \bibnamefont {Pidgeon}}, \bibinfo {author} {\bibfnamefont {D.~G.}\
  \bibnamefont {Clarke}}, \bibinfo {author} {\bibfnamefont {M.}~\bibnamefont
  {Merrick}}, \bibinfo {author} {\bibfnamefont {P.}~\bibnamefont {Murzyn}},
  \bibinfo {author} {\bibfnamefont {E.~V.}\ \bibnamefont {Beregulin}},\ and\
  \bibinfo {author} {\bibfnamefont {W.}~\bibnamefont {Prettl}},\ }\bibfield
  {title} {\bibinfo {title} {Spin-galvanic effect due to optical spin
  orientation in n-type gaas quantum well structures},\ }\href
  {https://doi.org/10.1103/PhysRevB.68.081302} {\bibfield  {journal} {\bibinfo
  {journal} {Phys. Rev. B}\ }\textbf {\bibinfo {volume} {68}},\ \bibinfo
  {pages} {081302} (\bibinfo {year} {2003}{\natexlab{b}})}\BibitemShut
  {NoStop}%
\bibitem [{Note2()}]{Note2}%
  \BibitemOpen
  \bibinfo {note} {The number of studied frequencies was defined by the laser
  tunability and availability of the polarizers operating at specific
  frequencies.}\BibitemShut {Stop}%
\bibitem [{\citenamefont {Ganichev}\ \emph {et~al.}(1993)\citenamefont
  {Ganichev}, \citenamefont {Prettl},\ and\ \citenamefont
  {Huggard}}]{Ganichev1993}%
  \BibitemOpen
  \bibfield  {author} {\bibinfo {author} {\bibfnamefont {S.~D.}\ \bibnamefont
  {Ganichev}}, \bibinfo {author} {\bibfnamefont {W.}~\bibnamefont {Prettl}},\
  and\ \bibinfo {author} {\bibfnamefont {P.~G.}\ \bibnamefont {Huggard}},\
  }\bibfield  {title} {\bibinfo {title} {Phonon assisted tunnel ionization of
  deep impurities in the electric field of far-infrared radiation},\ }\href
  {https://doi.org/10.1103/physrevlett.71.3882} {\bibfield  {journal} {\bibinfo
   {journal} {Phys. Rev. Lett.}\ }\textbf {\bibinfo {volume} {71}},\ \bibinfo
  {pages} {3882} (\bibinfo {year} {1993})}\BibitemShut {NoStop}%
\bibitem [{\citenamefont {Ganichev}\ \emph {et~al.}(1995)\citenamefont
  {Ganichev}, \citenamefont {Yassievich}, \citenamefont {Prettl}, \citenamefont
  {Diener}, \citenamefont {Meyer},\ and\ \citenamefont {Benz}}]{Ganichev1995}%
  \BibitemOpen
  \bibfield  {author} {\bibinfo {author} {\bibfnamefont {S.~D.}\ \bibnamefont
  {Ganichev}}, \bibinfo {author} {\bibfnamefont {I.~N.}\ \bibnamefont
  {Yassievich}}, \bibinfo {author} {\bibfnamefont {W.}~\bibnamefont {Prettl}},
  \bibinfo {author} {\bibfnamefont {J.}~\bibnamefont {Diener}}, \bibinfo
  {author} {\bibfnamefont {B.~K.}\ \bibnamefont {Meyer}},\ and\ \bibinfo
  {author} {\bibfnamefont {K.~W.}\ \bibnamefont {Benz}},\ }\bibfield  {title}
  {\bibinfo {title} {Tunneling ionization of {AutolocalizedDX}-centers in
  terahertz fields},\ }\href {https://doi.org/10.1103/physrevlett.75.1590}
  {\bibfield  {journal} {\bibinfo  {journal} {Phys. Rev. Lett.}\ }\textbf
  {\bibinfo {volume} {75}},\ \bibinfo {pages} {1590} (\bibinfo {year}
  {1995})}\BibitemShut {NoStop}%
\bibitem [{\citenamefont {Ganichev}\ \emph {et~al.}(1998)\citenamefont
  {Ganichev}, \citenamefont {Ziemann}, \citenamefont {Gleim}, \citenamefont
  {Prettl}, \citenamefont {Yassievich}, \citenamefont {Perel}, \citenamefont
  {Wilke},\ and\ \citenamefont {Haller}}]{Ganichev1998}%
  \BibitemOpen
  \bibfield  {author} {\bibinfo {author} {\bibfnamefont {S.~D.}\ \bibnamefont
  {Ganichev}}, \bibinfo {author} {\bibfnamefont {E.}~\bibnamefont {Ziemann}},
  \bibinfo {author} {\bibfnamefont {T.}~\bibnamefont {Gleim}}, \bibinfo
  {author} {\bibfnamefont {W.}~\bibnamefont {Prettl}}, \bibinfo {author}
  {\bibfnamefont {I.~N.}\ \bibnamefont {Yassievich}}, \bibinfo {author}
  {\bibfnamefont {V.~I.}\ \bibnamefont {Perel}}, \bibinfo {author}
  {\bibfnamefont {I.}~\bibnamefont {Wilke}},\ and\ \bibinfo {author}
  {\bibfnamefont {E.~E.}\ \bibnamefont {Haller}},\ }\bibfield  {title}
  {\bibinfo {title} {Carrier tunneling in high-frequency electric fields},\
  }\href {https://doi.org/10.1103/physrevlett.80.2409} {\bibfield  {journal}
  {\bibinfo  {journal} {Phys. Rev. Lett.}\ }\textbf {\bibinfo {volume} {80}},\
  \bibinfo {pages} {2409} (\bibinfo {year} {1998})}\BibitemShut {NoStop}%
\bibitem [{\citenamefont {Ganichev}\ \emph {et~al.}(1985)\citenamefont
  {Ganichev}, \citenamefont {Terent'ev},\ and\ \citenamefont
  {Yaroshetskii}}]{Ganichev1985}%
  \BibitemOpen
  \bibfield  {author} {\bibinfo {author} {\bibfnamefont {S.~D.}\ \bibnamefont
  {Ganichev}}, \bibinfo {author} {\bibfnamefont {Y.~V.}\ \bibnamefont
  {Terent'ev}},\ and\ \bibinfo {author} {\bibfnamefont {I.~D.}\ \bibnamefont
  {Yaroshetskii}},\ }\bibfield  {title} {\bibinfo {title} {Photon-drag
  photodetectors for the far-ir and submillimeter regions},\ }\href@noop {}
  {\bibfield  {journal} {\bibinfo  {journal} {Pis'ma Zh. Tekh. Fiz.}\ }\textbf
  {\bibinfo {volume} {11}},\ \bibinfo {pages} {46} (\bibinfo {year} {1985})},\
  \bibinfo {note} {[Sov. Tech. Phys. Lett. \textbf{11}, 20 (1985)]}\BibitemShut
  {NoStop}%
\bibitem [{\citenamefont {Bel'kov}\ \emph {et~al.}(2005)\citenamefont
  {Bel'kov}, \citenamefont {Ganichev}, \citenamefont {Ivchenko}, \citenamefont
  {Tarasenko}, \citenamefont {Weber}, \citenamefont {Giglberger}, \citenamefont
  {Olteanu}, \citenamefont {Tranitz}, \citenamefont {Danilov}, \citenamefont
  {Schneider}, \citenamefont {Wegscheider}, \citenamefont {Weiss},\ and\
  \citenamefont {Prettl}}]{Belkov2005}%
  \BibitemOpen
  \bibfield  {author} {\bibinfo {author} {\bibfnamefont {V.~V.}\ \bibnamefont
  {Bel'kov}}, \bibinfo {author} {\bibfnamefont {S.~D.}\ \bibnamefont
  {Ganichev}}, \bibinfo {author} {\bibfnamefont {E.~L.}\ \bibnamefont
  {Ivchenko}}, \bibinfo {author} {\bibfnamefont {S.~A.}\ \bibnamefont
  {Tarasenko}}, \bibinfo {author} {\bibfnamefont {W.}~\bibnamefont {Weber}},
  \bibinfo {author} {\bibfnamefont {S.}~\bibnamefont {Giglberger}}, \bibinfo
  {author} {\bibfnamefont {M.}~\bibnamefont {Olteanu}}, \bibinfo {author}
  {\bibfnamefont {H.~P.}\ \bibnamefont {Tranitz}}, \bibinfo {author}
  {\bibfnamefont {S.~N.}\ \bibnamefont {Danilov}}, \bibinfo {author}
  {\bibfnamefont {P.}~\bibnamefont {Schneider}}, \bibinfo {author}
  {\bibfnamefont {W.}~\bibnamefont {Wegscheider}}, \bibinfo {author}
  {\bibfnamefont {D.}~\bibnamefont {Weiss}},\ and\ \bibinfo {author}
  {\bibfnamefont {W.}~\bibnamefont {Prettl}},\ }\bibfield  {title} {\bibinfo
  {title} {Magneto-gyrotropic photogalvanic effects in semiconductor quantum
  wells},\ }\href {https://doi.org/10.1088/0953-8984/17/21/032} {\bibfield
  {journal} {\bibinfo  {journal} {J. Phys. Condens. Matter}\ }\textbf {\bibinfo
  {volume} {17}},\ \bibinfo {pages} {3405} (\bibinfo {year}
  {2005})}\BibitemShut {NoStop}%
\bibitem [{Note3()}]{Note3}%
  \BibitemOpen
  \bibinfo {note} {{While in magnetic fields up to about $\pm ~1.2$~T the
  current linearly depends on the magnetic field at higher fields, it possibly
  tends to deviate from this behavior. To justify this tendency experiments at
  higher magnetic fields are needed, which was out of scope of the present
  work.}}\BibitemShut {Stop}%
\bibitem [{Note4()}]{Note4}%
  \BibitemOpen
  \bibinfo {note} {Note that in the geometry applying $\lambda $-quarter plate
  the $\sin 4\varphi /2$ corresponds to the Stokes parameter describing the
  degree of linear polarization and the polarization ellipse orientation, which
  in experiments applying $\lambda $-half plate is given by $\sin 2\alpha $,
  see below and Refs.~\cite {Saleh2019,Belkov2005}.}\BibitemShut {Stop}%
\bibitem [{\citenamefont {Saleh}\ and\ \citenamefont
  {Teich}(2019)}]{Saleh2019}%
  \BibitemOpen
  \bibfield  {author} {\bibinfo {author} {\bibfnamefont {B.~E.~A.}\
  \bibnamefont {Saleh}}\ and\ \bibinfo {author} {\bibfnamefont {M.~C.}\
  \bibnamefont {Teich}},\ }\href@noop {} {\emph {\bibinfo {title} {Fundamentals
  of Photonics}}}\ (\bibinfo  {publisher} {John Wiley and Sons Ltd.},\ \bibinfo
  {year} {2019})\BibitemShut {NoStop}%
\bibitem [{Note5()}]{Note5}%
  \BibitemOpen
  \bibinfo {note} {Note that the trigonal photocurrents given by coefficients
  $\chi $ depend on the polarization plane orientation as the 2nd angular
  harmonics because projections of the current onto fixed axes are measured. If
  one detects a direction of the photocurrent as a function of the light
  polarization then it has a form of the 3rd angular harmonics.}\BibitemShut
  {Stop}%
\bibitem [{\citenamefont {Fischer}\ \emph {et~al.}(1973)\citenamefont
  {Fischer}, \citenamefont {Bangert},\ and\ \citenamefont {Grosse}}]{Grosse1}%
  \BibitemOpen
  \bibfield  {author} {\bibinfo {author} {\bibfnamefont {D.}~\bibnamefont
  {Fischer}}, \bibinfo {author} {\bibfnamefont {E.}~\bibnamefont {Bangert}},\
  and\ \bibinfo {author} {\bibfnamefont {P.}~\bibnamefont {Grosse}},\
  }\bibfield  {title} {\bibinfo {title} {Intervalence band transitions in
  tellurium. {I}. {Polarization} ${E \parallel c}$},\ }\href
  {https://doi.org/https://doi.org/10.1002/pssb.2220550208} {\bibfield
  {journal} {\bibinfo  {journal} {physica status solidi (b)}\ }\textbf
  {\bibinfo {volume} {55}},\ \bibinfo {pages} {527} (\bibinfo {year}
  {1973})}\BibitemShut {NoStop}%
\bibitem [{\citenamefont {Bangert}\ \emph {et~al.}(1973)\citenamefont
  {Bangert}, \citenamefont {Fischer},\ and\ \citenamefont {Grosse}}]{Grosse2}%
  \BibitemOpen
  \bibfield  {author} {\bibinfo {author} {\bibfnamefont {E.}~\bibnamefont
  {Bangert}}, \bibinfo {author} {\bibfnamefont {D.}~\bibnamefont {Fischer}},\
  and\ \bibinfo {author} {\bibfnamefont {P.}~\bibnamefont {Grosse}},\
  }\bibfield  {title} {\bibinfo {title} {Intervalence band transitions in
  tellurium. {II}. {Polarization} ${E \perp c}$},\ }\href
  {https://doi.org/https://doi.org/10.1002/pssb.2220590206} {\bibfield
  {journal} {\bibinfo  {journal} {physica status solidi (b)}\ }\textbf
  {\bibinfo {volume} {59}},\ \bibinfo {pages} {419} (\bibinfo {year}
  {1973})}\BibitemShut {NoStop}%
\bibitem [{\citenamefont {Averkiev}\ \emph
  {et~al.}(1984{\natexlab{a}})\citenamefont {Averkiev}, \citenamefont {Asnin},
  \citenamefont {Bakun}, \citenamefont {Danishevskii}, \citenamefont
  {Ivchenko}, \citenamefont {Pikus},\ and\ \citenamefont
  {Rogachev}}]{Averkiev_1984_theory}%
  \BibitemOpen
  \bibfield  {author} {\bibinfo {author} {\bibfnamefont {N.~S.}\ \bibnamefont
  {Averkiev}}, \bibinfo {author} {\bibfnamefont {V.~M.}\ \bibnamefont {Asnin}},
  \bibinfo {author} {\bibfnamefont {A.~A.}\ \bibnamefont {Bakun}}, \bibinfo
  {author} {\bibfnamefont {A.~M.}\ \bibnamefont {Danishevskii}}, \bibinfo
  {author} {\bibfnamefont {E.~L.}\ \bibnamefont {Ivchenko}}, \bibinfo {author}
  {\bibfnamefont {G.~E.}\ \bibnamefont {Pikus}},\ and\ \bibinfo {author}
  {\bibfnamefont {A.~A.}\ \bibnamefont {Rogachev}},\ }\bibfield  {title}
  {\bibinfo {title} {Circular photogalvanic effect in tellurium. {I}.
  {T}heory},\ }\href@noop {} {\bibfield  {journal} {\bibinfo  {journal} {Sov.
  Phys. Semicond.}\ }\textbf {\bibinfo {volume} {18}},\ \bibinfo {pages} {397}
  (\bibinfo {year} {1984}{\natexlab{a}})}\BibitemShut {NoStop}%
\bibitem [{\citenamefont {Averkiev}\ \emph
  {et~al.}(1984{\natexlab{b}})\citenamefont {Averkiev}, \citenamefont {Asnin},
  \citenamefont {Bakun}, \citenamefont {Danishevskii}, \citenamefont
  {Ivchenko}, \citenamefont {Pikus},\ and\ \citenamefont
  {Rogachev}}]{Averkiev_1984_experiment}%
  \BibitemOpen
  \bibfield  {author} {\bibinfo {author} {\bibfnamefont {N.~S.}\ \bibnamefont
  {Averkiev}}, \bibinfo {author} {\bibfnamefont {V.~M.}\ \bibnamefont {Asnin}},
  \bibinfo {author} {\bibfnamefont {A.~A.}\ \bibnamefont {Bakun}}, \bibinfo
  {author} {\bibfnamefont {A.~M.}\ \bibnamefont {Danishevskii}}, \bibinfo
  {author} {\bibfnamefont {E.~L.}\ \bibnamefont {Ivchenko}}, \bibinfo {author}
  {\bibfnamefont {G.~E.}\ \bibnamefont {Pikus}},\ and\ \bibinfo {author}
  {\bibfnamefont {A.~A.}\ \bibnamefont {Rogachev}},\ }\bibfield  {title}
  {\bibinfo {title} {Circular photogalvanic effect in tellurium. {II}.
  {E}xperiment},\ }\href@noop {} {\bibfield  {journal} {\bibinfo  {journal}
  {Sov. Phys. Semicond.}\ }\textbf {\bibinfo {volume} {18}},\ \bibinfo {pages}
  {402} (\bibinfo {year} {1984}{\natexlab{b}})}\BibitemShut {NoStop}%
\bibitem [{\citenamefont {Yang}\ \emph {et~al.}(2022)\citenamefont {Yang},
  \citenamefont {Zhu}, \citenamefont {Wang}, \citenamefont {Xia}, \citenamefont
  {Zhang}, \citenamefont {Jiang}, \citenamefont {Shang}, \citenamefont {Li},\
  and\ \citenamefont {Hu}}]{China2022}%
  \BibitemOpen
  \bibfield  {author} {\bibinfo {author} {\bibfnamefont {D.-Q.}\ \bibnamefont
  {Yang}}, \bibinfo {author} {\bibfnamefont {L.-Q.}\ \bibnamefont {Zhu}},
  \bibinfo {author} {\bibfnamefont {J.-L.}\ \bibnamefont {Wang}}, \bibinfo
  {author} {\bibfnamefont {W.}~\bibnamefont {Xia}}, \bibinfo {author}
  {\bibfnamefont {J.-Z.}\ \bibnamefont {Zhang}}, \bibinfo {author}
  {\bibfnamefont {K.}~\bibnamefont {Jiang}}, \bibinfo {author} {\bibfnamefont
  {L.-Y.}\ \bibnamefont {Shang}}, \bibinfo {author} {\bibfnamefont {Y.-W.}\
  \bibnamefont {Li}},\ and\ \bibinfo {author} {\bibfnamefont {Z.-G.}\
  \bibnamefont {Hu}},\ }\bibfield  {title} {\bibinfo {title} {Band structure
  and lattice vibration of elemental tellurium investigated by
  temperature-dependent mid-and-far infrared transmission and raman
  spectroscopy},\ }\href
  {https://doi.org/https://doi.org/10.1002/pssb.202100625} {\bibfield
  {journal} {\bibinfo  {journal} {physica status solidi (b)}\ }\textbf
  {\bibinfo {volume} {259}},\ \bibinfo {pages} {2100625} (\bibinfo {year}
  {2022})}\BibitemShut {NoStop}%
\bibitem [{Note6()}]{Note6}%
  \BibitemOpen
  \bibinfo {note} {In Eq.~\protect \textup {\hbox {\mathsurround \z@ \protect
  \normalfont (\ignorespaces \ref {V_12}\unskip \@@italiccorr )}} the factor
  $\abs {L}^2/E_g$ absent in Eq.~(25) of Ref.~\cite {Averkiev_1984_theory} is
  restored.}\BibitemShut {Stop}%
\bibitem [{\citenamefont {Sturman}(2020)}]{Sturman2020}%
  \BibitemOpen
  \bibfield  {author} {\bibinfo {author} {\bibfnamefont {B.~I.}\ \bibnamefont
  {Sturman}},\ }\bibfield  {title} {\bibinfo {title} {Ballistic and shift
  currents in the bulk photovoltaic effect theory},\ }\href
  {https://doi.org/10.3367/ufne.2019.06.038578} {\bibfield  {journal} {\bibinfo
   {journal} {Physics-Uspekhi}\ }\textbf {\bibinfo {volume} {63}},\ \bibinfo
  {pages} {407} (\bibinfo {year} {2020})}\BibitemShut {NoStop}%
\bibitem [{\citenamefont {Belinicher}\ \emph {et~al.}(1982)\citenamefont
  {Belinicher}, \citenamefont {Ivchenko},\ and\ \citenamefont
  {Sturman}}]{BelIvchSt_1982}%
  \BibitemOpen
  \bibfield  {author} {\bibinfo {author} {\bibfnamefont {V.~I.}\ \bibnamefont
  {Belinicher}}, \bibinfo {author} {\bibfnamefont {E.~L.}\ \bibnamefont
  {Ivchenko}},\ and\ \bibinfo {author} {\bibfnamefont {B.~I.}\ \bibnamefont
  {Sturman}},\ }\bibfield  {title} {\bibinfo {title} {Kinetic theory of the
  displacement photovoltaic effect in piezoelectric},\ }\href@noop {}
  {\bibfield  {journal} {\bibinfo  {journal} {Zh. Eksp. Teor. Fiz.}\ }\textbf
  {\bibinfo {volume} {83}},\ \bibinfo {pages} {649} (\bibinfo {year} {1982})},\
  \bibinfo {note} {[JETP \textbf{56}, 359 (1982)]}\BibitemShut {NoStop}%
\bibitem [{\citenamefont {Leppenen}\ and\ \citenamefont
  {Golub}(2023)}]{Leppenen2023}%
  \BibitemOpen
  \bibfield  {author} {\bibinfo {author} {\bibfnamefont {N.~V.}\ \bibnamefont
  {Leppenen}}\ and\ \bibinfo {author} {\bibfnamefont {L.~E.}\ \bibnamefont
  {Golub}},\ }\bibfield  {title} {\bibinfo {title} {Linear photogalvanic effect
  in surface states of topological insulators},\ }\href
  {https://doi.org/10.1103/PhysRevB.107.L161403} {\bibfield  {journal}
  {\bibinfo  {journal} {Phys. Rev. B}\ }\textbf {\bibinfo {volume} {107}},\
  \bibinfo {pages} {L161403} (\bibinfo {year} {2023})}\BibitemShut {NoStop}%
\bibitem [{\citenamefont {Belinicher}\ and\ \citenamefont
  {Sturman}(1980)}]{Belinicher_Sturman_1980}%
  \BibitemOpen
  \bibfield  {author} {\bibinfo {author} {\bibfnamefont {V.~I.}\ \bibnamefont
  {Belinicher}}\ and\ \bibinfo {author} {\bibfnamefont {B.~I.}\ \bibnamefont
  {Sturman}},\ }\bibfield  {title} {\bibinfo {title} {The photogalvanic effect
  in media lacking a center of symmetry},\ }\href
  {https://doi.org/10.1070/PU1980v023n03ABEH004703} {\bibfield  {journal}
  {\bibinfo  {journal} {Sov. Phys. Usp.}\ }\textbf {\bibinfo {volume} {23}},\
  \bibinfo {pages} {199} (\bibinfo {year} {1980})},\ \bibinfo {note} {[Usp.
  Fiz. Nauk \textbf{130}, 415 (1980)]}\BibitemShut {NoStop}%
\bibitem [{\citenamefont {Olbrich}\ \emph {et~al.}(2014)\citenamefont
  {Olbrich}, \citenamefont {Golub}, \citenamefont {Herrmann}, \citenamefont
  {Danilov}, \citenamefont {Plank}, \citenamefont {Bel'kov}, \citenamefont
  {Mussler}, \citenamefont {Weyrich}, \citenamefont {Schneider}, \citenamefont
  {Kampmeier}, \citenamefont {Gr\"utzmacher}, \citenamefont {Plucinski},
  \citenamefont {Eschbach},\ and\ \citenamefont {Ganichev}}]{Olbrich2014}%
  \BibitemOpen
  \bibfield  {author} {\bibinfo {author} {\bibfnamefont {P.}~\bibnamefont
  {Olbrich}}, \bibinfo {author} {\bibfnamefont {L.~E.}\ \bibnamefont {Golub}},
  \bibinfo {author} {\bibfnamefont {T.}~\bibnamefont {Herrmann}}, \bibinfo
  {author} {\bibfnamefont {S.~N.}\ \bibnamefont {Danilov}}, \bibinfo {author}
  {\bibfnamefont {H.}~\bibnamefont {Plank}}, \bibinfo {author} {\bibfnamefont
  {V.~V.}\ \bibnamefont {Bel'kov}}, \bibinfo {author} {\bibfnamefont
  {G.}~\bibnamefont {Mussler}}, \bibinfo {author} {\bibfnamefont
  {C.}~\bibnamefont {Weyrich}}, \bibinfo {author} {\bibfnamefont {C.~M.}\
  \bibnamefont {Schneider}}, \bibinfo {author} {\bibfnamefont {J.}~\bibnamefont
  {Kampmeier}}, \bibinfo {author} {\bibfnamefont {D.}~\bibnamefont
  {Gr\"utzmacher}}, \bibinfo {author} {\bibfnamefont {L.}~\bibnamefont
  {Plucinski}}, \bibinfo {author} {\bibfnamefont {M.}~\bibnamefont
  {Eschbach}},\ and\ \bibinfo {author} {\bibfnamefont {S.~D.}\ \bibnamefont
  {Ganichev}},\ }\bibfield  {title} {\bibinfo {title} {Room-temperature
  high-frequency transport of dirac fermions in epitaxially grown
  {Sb}$_{2}${Te}$_{3}$- and {Bi}$_2${Te}$_{3}$-based topological insulators},\
  }\href {https://doi.org/10.1103/PhysRevLett.113.096601} {\bibfield  {journal}
  {\bibinfo  {journal} {Phys. Rev. Lett.}\ }\textbf {\bibinfo {volume} {113}},\
  \bibinfo {pages} {096601} (\bibinfo {year} {2014})}\BibitemShut {NoStop}%
\bibitem [{\citenamefont {Hild}\ \emph {et~al.}(2023)\citenamefont {Hild},
  \citenamefont {Golub}, \citenamefont {Fuhrmann}, \citenamefont {Otteneder},
  \citenamefont {Kronseder}, \citenamefont {Matsubara}, \citenamefont
  {Kobayashi}, \citenamefont {Oshima}, \citenamefont {Honda}, \citenamefont
  {Kato}, \citenamefont {Wunderlich}, \citenamefont {Back},\ and\ \citenamefont
  {Ganichev}}]{Hild_PRB_2023}%
  \BibitemOpen
  \bibfield  {author} {\bibinfo {author} {\bibfnamefont {M.}~\bibnamefont
  {Hild}}, \bibinfo {author} {\bibfnamefont {L.~E.}\ \bibnamefont {Golub}},
  \bibinfo {author} {\bibfnamefont {A.}~\bibnamefont {Fuhrmann}}, \bibinfo
  {author} {\bibfnamefont {M.}~\bibnamefont {Otteneder}}, \bibinfo {author}
  {\bibfnamefont {M.}~\bibnamefont {Kronseder}}, \bibinfo {author}
  {\bibfnamefont {M.}~\bibnamefont {Matsubara}}, \bibinfo {author}
  {\bibfnamefont {T.}~\bibnamefont {Kobayashi}}, \bibinfo {author}
  {\bibfnamefont {D.}~\bibnamefont {Oshima}}, \bibinfo {author} {\bibfnamefont
  {A.}~\bibnamefont {Honda}}, \bibinfo {author} {\bibfnamefont
  {T.}~\bibnamefont {Kato}}, \bibinfo {author} {\bibfnamefont {J.}~\bibnamefont
  {Wunderlich}}, \bibinfo {author} {\bibfnamefont {C.}~\bibnamefont {Back}},\
  and\ \bibinfo {author} {\bibfnamefont {S.~D.}\ \bibnamefont {Ganichev}},\
  }\bibfield  {title} {\bibinfo {title} {Terahertz spin ratchet effect in
  magnetic metamaterials},\ }\href
  {https://doi.org/10.1103/PhysRevB.107.155419} {\bibfield  {journal} {\bibinfo
   {journal} {Phys. Rev. B}\ }\textbf {\bibinfo {volume} {107}},\ \bibinfo
  {pages} {155419} (\bibinfo {year} {2023})}\BibitemShut {NoStop}%
\bibitem [{\citenamefont {Plank}\ \emph {et~al.}(2016)\citenamefont {Plank},
  \citenamefont {Golub}, \citenamefont {Bauer}, \citenamefont {Bel'kov},
  \citenamefont {Herrmann}, \citenamefont {Olbrich}, \citenamefont {Eschbach},
  \citenamefont {Plucinski}, \citenamefont {Schneider}, \citenamefont
  {Kampmeier}, \citenamefont {Lanius}, \citenamefont {Mussler}, \citenamefont
  {Gr\"utzmacher},\ and\ \citenamefont {Ganichev}}]{Plank2016}%
  \BibitemOpen
  \bibfield  {author} {\bibinfo {author} {\bibfnamefont {H.}~\bibnamefont
  {Plank}}, \bibinfo {author} {\bibfnamefont {L.~E.}\ \bibnamefont {Golub}},
  \bibinfo {author} {\bibfnamefont {S.}~\bibnamefont {Bauer}}, \bibinfo
  {author} {\bibfnamefont {V.~V.}\ \bibnamefont {Bel'kov}}, \bibinfo {author}
  {\bibfnamefont {T.}~\bibnamefont {Herrmann}}, \bibinfo {author}
  {\bibfnamefont {P.}~\bibnamefont {Olbrich}}, \bibinfo {author} {\bibfnamefont
  {M.}~\bibnamefont {Eschbach}}, \bibinfo {author} {\bibfnamefont
  {L.}~\bibnamefont {Plucinski}}, \bibinfo {author} {\bibfnamefont {C.~M.}\
  \bibnamefont {Schneider}}, \bibinfo {author} {\bibfnamefont {J.}~\bibnamefont
  {Kampmeier}}, \bibinfo {author} {\bibfnamefont {M.}~\bibnamefont {Lanius}},
  \bibinfo {author} {\bibfnamefont {G.}~\bibnamefont {Mussler}}, \bibinfo
  {author} {\bibfnamefont {D.}~\bibnamefont {Gr\"utzmacher}},\ and\ \bibinfo
  {author} {\bibfnamefont {S.~D.}\ \bibnamefont {Ganichev}},\ }\bibfield
  {title} {\bibinfo {title} {Photon drag effect in
  {$({\mathrm{Bi}}_{1\ensuremath{-}x}{\mathrm{Sb}}_{x}){}_{2}{\mathrm{Te}}_{3}$}
  three-dimensional topological insulators},\ }\href
  {https://doi.org/10.1103/PhysRevB.93.125434} {\bibfield  {journal} {\bibinfo
  {journal} {Phys. Rev. B}\ }\textbf {\bibinfo {volume} {93}},\ \bibinfo
  {pages} {125434} (\bibinfo {year} {2016})}\BibitemShut {NoStop}%
\bibitem [{\citenamefont {Perel}\ and\ \citenamefont
  {Pinskii}(1973)}]{PerelPinskii}%
  \BibitemOpen
  \bibfield  {author} {\bibinfo {author} {\bibfnamefont {V.~I.}\ \bibnamefont
  {Perel}}\ and\ \bibinfo {author} {\bibfnamefont {Y.~M.}\ \bibnamefont
  {Pinskii}},\ }\bibfield  {title} {\bibinfo {title} {Constant current in
  conducting media due to a high-frequency electron electromagnetic field},\
  }\href@noop {} {\bibfield  {journal} {\bibinfo  {journal} {Sov. Phys. - Solid
  State}\ }\textbf {\bibinfo {volume} {15}},\ \bibinfo {pages} {688} (\bibinfo
  {year} {1973})}\BibitemShut {NoStop}%
\bibitem [{\citenamefont {Weber}\ \emph {et~al.}(2008)\citenamefont {Weber},
  \citenamefont {Golub}, \citenamefont {Danilov}, \citenamefont {Karch},
  \citenamefont {Reitmaier}, \citenamefont {Wittmann}, \citenamefont {Bel'kov},
  \citenamefont {Ivchenko}, \citenamefont {Kvon}, \citenamefont {Vinh},
  \citenamefont {van~der Meer}, \citenamefont {Murdin},\ and\ \citenamefont
  {Ganichev}}]{Weber2007}%
  \BibitemOpen
  \bibfield  {author} {\bibinfo {author} {\bibfnamefont {W.}~\bibnamefont
  {Weber}}, \bibinfo {author} {\bibfnamefont {L.~E.}\ \bibnamefont {Golub}},
  \bibinfo {author} {\bibfnamefont {S.~N.}\ \bibnamefont {Danilov}}, \bibinfo
  {author} {\bibfnamefont {J.}~\bibnamefont {Karch}}, \bibinfo {author}
  {\bibfnamefont {C.}~\bibnamefont {Reitmaier}}, \bibinfo {author}
  {\bibfnamefont {B.}~\bibnamefont {Wittmann}}, \bibinfo {author}
  {\bibfnamefont {V.~V.}\ \bibnamefont {Bel'kov}}, \bibinfo {author}
  {\bibfnamefont {E.~L.}\ \bibnamefont {Ivchenko}}, \bibinfo {author}
  {\bibfnamefont {Z.~D.}\ \bibnamefont {Kvon}}, \bibinfo {author}
  {\bibfnamefont {N.~Q.}\ \bibnamefont {Vinh}}, \bibinfo {author}
  {\bibfnamefont {A.~F.~G.}\ \bibnamefont {van~der Meer}}, \bibinfo {author}
  {\bibfnamefont {B.}~\bibnamefont {Murdin}},\ and\ \bibinfo {author}
  {\bibfnamefont {S.~D.}\ \bibnamefont {Ganichev}},\ }\bibfield  {title}
  {\bibinfo {title} {Quantum ratchet effects induced by terahertz radiation in
  gan-based two-dimensional structures},\ }\href
  {https://doi.org/10.1103/PhysRevB.77.245304} {\bibfield  {journal} {\bibinfo
  {journal} {Phys. Rev. B}\ }\textbf {\bibinfo {volume} {77}},\ \bibinfo
  {pages} {245304} (\bibinfo {year} {2008})}\BibitemShut {NoStop}%
\bibitem [{\citenamefont {Drexler}\ \emph {et~al.}(2013)\citenamefont
  {Drexler}, \citenamefont {Tarasenko}, \citenamefont {Olbrich}, \citenamefont
  {Karch}, \citenamefont {Hirmer}, \citenamefont {Müller}, \citenamefont
  {Gmitra}, \citenamefont {Fabian}, \citenamefont {Yakimova}, \citenamefont
  {Lara-Avila}, \citenamefont {Kubatkin}, \citenamefont {Wang}, \citenamefont
  {Vajtai}, \citenamefont {Ajayan}, \citenamefont {Kono},\ and\ \citenamefont
  {Ganichev}}]{Drexler2013}%
  \BibitemOpen
  \bibfield  {author} {\bibinfo {author} {\bibfnamefont {C.}~\bibnamefont
  {Drexler}}, \bibinfo {author} {\bibfnamefont {S.~A.}\ \bibnamefont
  {Tarasenko}}, \bibinfo {author} {\bibfnamefont {P.}~\bibnamefont {Olbrich}},
  \bibinfo {author} {\bibfnamefont {J.}~\bibnamefont {Karch}}, \bibinfo
  {author} {\bibfnamefont {M.}~\bibnamefont {Hirmer}}, \bibinfo {author}
  {\bibfnamefont {F.}~\bibnamefont {Müller}}, \bibinfo {author} {\bibfnamefont
  {M.}~\bibnamefont {Gmitra}}, \bibinfo {author} {\bibfnamefont
  {J.}~\bibnamefont {Fabian}}, \bibinfo {author} {\bibfnamefont
  {R.}~\bibnamefont {Yakimova}}, \bibinfo {author} {\bibfnamefont
  {S.}~\bibnamefont {Lara-Avila}}, \bibinfo {author} {\bibfnamefont
  {S.}~\bibnamefont {Kubatkin}}, \bibinfo {author} {\bibfnamefont
  {M.}~\bibnamefont {Wang}}, \bibinfo {author} {\bibfnamefont {R.}~\bibnamefont
  {Vajtai}}, \bibinfo {author} {\bibfnamefont {P.~M.}\ \bibnamefont {Ajayan}},
  \bibinfo {author} {\bibfnamefont {J.}~\bibnamefont {Kono}},\ and\ \bibinfo
  {author} {\bibfnamefont {S.~D.}\ \bibnamefont {Ganichev}},\ }\bibfield
  {title} {\bibinfo {title} {Magnetic quantum ratchet effect in graphene},\
  }\href {https://doi.org/10.1038/nnano.2012.231} {\bibfield  {journal}
  {\bibinfo  {journal} {Nature Nanotechnology}\ }\textbf {\bibinfo {volume}
  {8}},\ \bibinfo {pages} {104} (\bibinfo {year} {2013})}\BibitemShut {NoStop}%
\bibitem [{\citenamefont {Ganichev}\ \emph {et~al.}(2017)\citenamefont
  {Ganichev}, \citenamefont {Weiss},\ and\ \citenamefont
  {Eroms}}]{GanichevWeissEromsAnnPhys2017}%
  \BibitemOpen
  \bibfield  {author} {\bibinfo {author} {\bibfnamefont {S.~D.}\ \bibnamefont
  {Ganichev}}, \bibinfo {author} {\bibfnamefont {D.}~\bibnamefont {Weiss}},\
  and\ \bibinfo {author} {\bibfnamefont {J.}~\bibnamefont {Eroms}},\ }\bibfield
   {title} {\bibinfo {title} {Terahertz electric field driven electric currents
  and ratchet effects in graphene},\ }\href
  {https://doi.org/https://doi.org/10.1002/andp.201600406} {\bibfield
  {journal} {\bibinfo  {journal} {Annalen der Physik}\ }\textbf {\bibinfo
  {volume} {529}},\ \bibinfo {pages} {1600406} (\bibinfo {year}
  {2017})}\BibitemShut {NoStop}%
\end{thebibliography}%

\end{document}